\documentclass[a4paper,11pt]{article}
\pdfoutput=1 

\usepackage{jcappub} 

\makeatletter
\gdef\@fpheader{Prepared for submission to JCAP\hfill UCI-HEP-TR-2024-02}
\makeatother

\usepackage[T1]{fontenc} 
\usepackage{soul}

\usepackage{graphicx}
\usepackage{bm}
\usepackage{hyperref}
\usepackage{slashed}
\usepackage{float}
\usepackage{slashed}
\usepackage{lipsum}
\usepackage{subfigure}
\usepackage{multirow}
\usepackage{amsmath}
\usepackage{array} 
\usepackage{varwidth} 
\usepackage{tabularx}
\usepackage{braket}
\usepackage[dvipsnames]{xcolor}
\usepackage{orcidlink}

\newcommand{\be}{\begin{equation}}
\newcommand{\ee}{\end{equation}}
\newcommand{\bea}{\begin{eqnarray}}
\newcommand{\eea}{\end{eqnarray}}

\hypersetup{
     colorlinks   = true,
     citecolor    = Maroon,
     urlcolor     = Maroon,
     linkcolor    = Maroon
}

\usepackage{listings}
\usepackage{color,xcolor}

\title{Indirect Searches for\\ Dark Photon-Photon Tridents\\ in Celestial Objects}

\author[a]{Tim Linden\,\orcidlink{0000-0001-9888-0971},}
\author[a,b,1]{Thong T.Q. Nguyen\,\orcidlink{0000-0002-8460-0219}\note{Corresponding author.},}
\author[c]{Tim M.P. Tait\,\orcidlink{0000-0003-3002-6909}}

\affiliation[a]{Stockholm University and The Oskar Klein Centre for Cosmoparticle Physics,\\ Alba Nova, 10691 Stockholm, Sweden}
\affiliation[b]{Institute of Physics, Academia Sinica,\\ Nangang, Taipei 11529, Taiwan}
\affiliation[c]{Department of Physics and Astronomy,  University of California, Irvine,\\ CA 92697-4575 USA}

\emailAdd{linden@fysik.su.se}
\emailAdd{thong.nguyen@fysik.su.se}
\emailAdd{ttait@uci.edu}

\abstract{
We model and constrain the unique indirect detection signature produced by dark matter particles that annihilate through a $U(1)$ gauge symmetry into dark photons that subsequently decay into three-photon final states. We focus on scenarios where the dark photon is long-lived, and show that $\gamma$-ray probes of celestial objects can set strong constraints on the dark matter/baryon scattering cross section that in many cases surpass the power of current direct detection constraints, and in some cases even peer into the neutrino fog.
}

\begin{document}
\maketitle
\flushbottom

\section{Introduction}
\label{sect:intro}

Searches for weak-scale dark matter have spanned a wide variety of potential interactions accessible to both terrestrial and astrophysical experiments~\cite{Bertone:2004pz, Bertone:2018krk}. However, most indirect detection searches have focused on two-body final states that originate from dark matter annihilation and decay~\cite{Drlica-Wagner:2022lbd, Chou:2022luk, Cooley:2022ufh, Slatyer:2017sev, Slatyer:2021qgc, DelaTorreLuque:2025zjt, Baryakhtar:2022hbu, Nguyen:2024kwy, Nguyen:2025tkl, Boddy:2022knd}. The consideration of three-body final states is less common, primarily due to the non-trivial nature of three-body phase spaces and energy spectra. Such final states can be a generic consequence of dark photon interactions when the dark photon mass lies below twice the electron mass, causing its decay to proceed primarily into three photons.

In many scenarios, the suppression of the three-body phase space implies that these dark photons may be relatively long-lived. This motivates searches that utilize celestial bodies to gravitationally ``focus" dark matter interactions by producing regions of strongly enhanced dark matter densities~\cite{Bramante:2023djs}. Celestial bodies including: neutron stars~\cite{Bertone:2007ae, Goldman:1989nd, Bauswein:2020kor, Kouvaris:2007ay, deLavallaz:2010wp, Kouvaris:2010vv, Bell:2013xk, Guver:2012ba, Baryakhtar:2017dbj, Bell:2018pkk, Garani:2018kkd, Liang:2023nvo, Chen:2018ohx, Bell:2023ysh, Ruter:2023uzc, Hamaguchi:2019oev, Camargo:2019wou, Bell:2019pyc, Garani:2019fpa, Acevedo:2019agu, Joglekar:2019vzy, Vikiaris:2023vau, Joglekar:2020liw, Bell:2020jou, Garani:2020wge, Acuna:2022ouv, Alvarez:2023fjj, Fujiwara:2023hlj, Bose:2021yhz, Bose:2023yll, Bramante:2021dyx, Brdar:2016ifs, Nguyen:2023ugx, Lin:2022dbl, Nguyen:2022zwb}, brown dwarfs~\cite{Leane:2021ihh, Leane:2020wob, Ilie:2023lbi, John:2023knt}, white dwarfs~\cite{Dasgupta:2019juq, Acevedo:2023xnu}, and other nearby objects such as the Sun~\cite{Garani:2017jcj, Bell:2021pyy, Feng:2016ijc, Leane:2017vag, Kang:2023gef, Bell:2011sn, Chauhan:2023zuf, Maity:2023rez, Niblaeus:2019gjk, Bose:2021cou, HAWC:2018szf, Nguyen:2025ygc, Nguyen:2026nhe, Nguyen:2026apa} or Jupiter~\cite{Leane:2021tjj, Chen:2023fgr, Croon:2023bmu, Ansarifard:2024fan, Blanco:2023qgi, Yan:2023kdg, Li:2022wix}, have recently been explored as targets for dark matter searches. While many searches target the gravitational effects of dark matter interactions~\cite{Goldman:1989nd, Bauswein:2020kor, Bell:2013xk, Garani:2018kkd, Liang:2023nvo, Ruter:2023uzc, Vikiaris:2023vau, Bell:2020jou, Bramante:2021dyx}, the thermal heating of celestial bodies~\cite{Kouvaris:2007ay, deLavallaz:2010wp, Kouvaris:2010vv, Baryakhtar:2017dbj, Bell:2018pkk, Chen:2018ohx, Bell:2023ysh, Hamaguchi:2019oev, Camargo:2019wou, Bell:2019pyc, Garani:2019fpa, Acevedo:2019agu, Joglekar:2019vzy, Joglekar:2020liw, Garani:2020wge, Acuna:2022ouv, Alvarez:2023fjj, Fujiwara:2023hlj, Leane:2020wob, Ilie:2023lbi, John:2023knt, Dasgupta:2019juq, Croon:2023bmu}, or weakly-interacting particles such as neutrinos~\cite{Nguyen:2022zwb, Lin:2022dbl, Bell:2011sn, Bose:2021yhz, Bose:2023yll, Chauhan:2023zuf, Maity:2023rez, Ansarifard:2024fan, Nguyen:2025ygc, Nguyen:2026nhe}, models with long-lived mediators are unique because the mediator can decay outside the celestial body, allowing data to directly probe the unperturbed electromagnetic signal from dark matter annihilations~\cite{Leane:2021ihh, Bell:2021pyy, Acevedo:2023xnu, Feng:2016ijc, Leane:2017vag, Niblaeus:2019gjk, Bose:2021cou, HAWC:2018szf, Leane:2021tjj, Chen:2023fgr, Li:2022wix}.

\begin{figure}[btp]
\centering
\includegraphics[width=0.5\columnwidth]{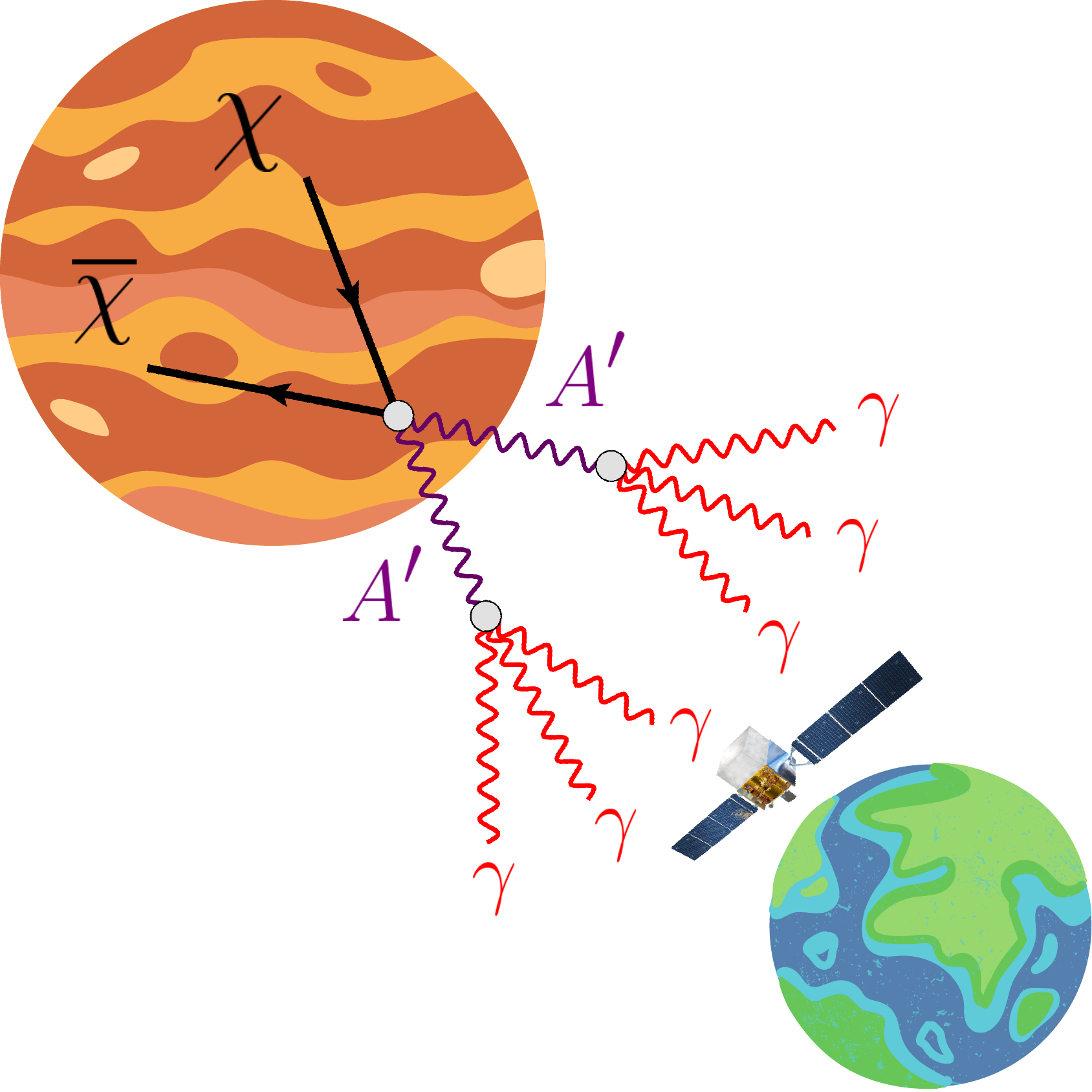}
\caption{Schematic of dark matter ($\chi$) annihilation inside Jupiter into dark photons ($A^{\prime}$). After escaping Jupiter, the mediator decays into three photons that can be measured by the Fermi-LAT.}
\label{fig:DPhotonescape}
\end{figure}

In this study, we perform the first analysis of scenarios where dark matter annihilation produces a long-lived spin-1 dark photon mediator that escapes from celestial objects and subsequently decays into three-photon final states, a process we call the {\em dark photon-photon trident}, as illustrated for the case of Jupiter in Fig.~\ref{fig:DPhotonescape}. Using observational constraints from the Fermi Large Area Telescope (Fermi-LAT), the High Energy Stereoscopic System (H.E.S.S.)~\cite{Malyshev:2015hqa}, and the High-Altitude Water Cherenkov Observatory (HAWC)~\cite{HAWC:2022khj}, we constrain the spin-independent dark matter-nucleon cross section as a function of dark matter mass. We find astrophysical constraints that, in some cases, exceed current constraints from terrestrial dark matter searches. The three-photon spectrum derived in this work has subsequently been applied to indirect detection searches for decaying dark photon dark matter in Ref.~\cite{Linden:2024fby}.

This paper is organized as follows: in Section~\ref{sect:model}, we introduce our dark matter model and provide the spectral formulas for three-body decay. In Section~\ref{sect:captandann}, we describe dark matter capture and annihilation inside a variety of celestial object targets. In Section~\ref{sect:ExpLim}, we assess the observability of our modeled photon signal by comparing benchmark points in our model with limits from Fermi-LAT, H.E.S.S., and HAWC. Finally, in Section~\ref{sect:Xsect}, we perform a parameter space scan to constrain this model. We conclude with a summary and comments on our results, and also outline motivations for future research in Section~\ref{sect:final}.

\section{Dark Matter model and Dark Photon decays}
\label{sect:model}

Assume the dark matter in the Galaxy is composed of equal numbers of particles and anti-particles, we consider a simplified model where the dark matter particle is a pseudo-Dirac fermion that is a singlet under the Standard Model (SM) gauge group where we consider the limit of zero mass splitting. The dark matter interacts with SM particles through a new massive dark photon mediator that stems from an extended $U(1)_{X}$ gauge symmetry. We also consider the kinetic mixing between the massive dark photon $A^{\prime}$ and SM gauge bosons through the Lagrangian:
\begin{equation}
\begin{split}
\mathcal{L}\supset& -\frac{1}{4}F^{\prime}_{\mu\nu}F^{\prime\mu\nu}-\frac{\epsilon}{2}F^{\prime}_{\mu\nu}B^{\mu\nu}-\frac{1}{2}m_{A^{\prime}}^{2}A^{\prime}_{\mu}A^{\prime \mu}+\bar{\chi}(i\slashed{D}_{U(1)_{X}}-m_{\chi})\chi,
\end{split}
\label{eq:lagrangianDP}
\end{equation}
where $\epsilon$ is the kinetic mixing coupling. In general, the SM fermions transform under the new gauge symmetry as well, and thus their interactions typically receive a combination of contributions from kinetic mixing, mass mixing, and direct coupling with the dark photon mediator.  We refer to the mediator generically as a `dark photon', although strictly speaking this label only applies in the limit where the coupling to the SM fermions is dominated by kinetic mixing.

\begin{figure}[t]
\centering
\includegraphics[width=0.7\columnwidth]{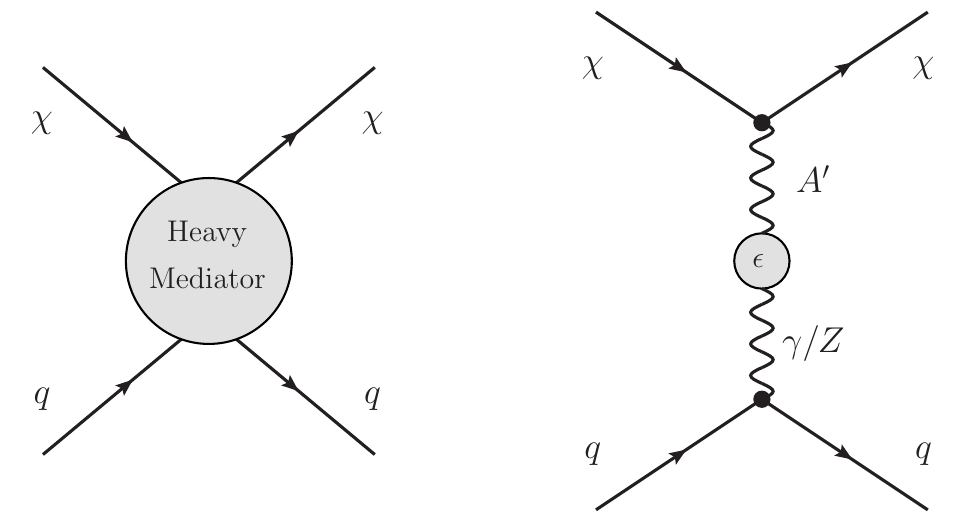}
~
\includegraphics[width=0.7\columnwidth]{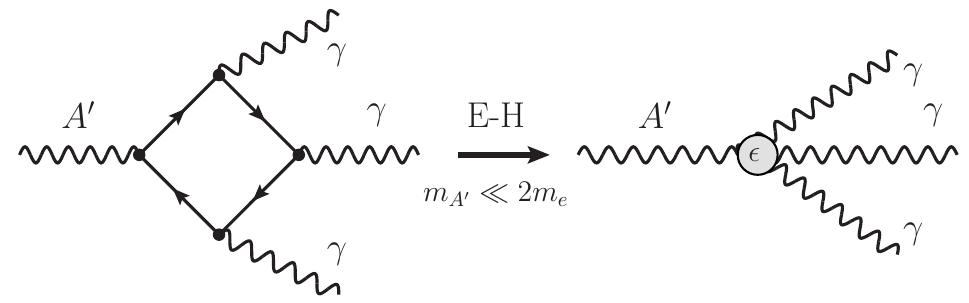}
\caption{{\bf Top}: Feynman diagrams for the interaction between fermionic dark matter with (anti-)quarks inside nucleons. Dark matter interacts with SM particles through heavy hidden sector, and through the mixing between the dark photon $A^{\prime}$ and neutral SM-gauge bosons (right). {\bf Bottom}: Dark photon decays to three photons through fermionic loops. For dark photon masses lighter than twice the mass of the lightest SM fermion, the interaction reduces to the Euler-Heisenberg effective coupling.}
\label{fig:DMSM}
\end{figure}

Figure~\ref{fig:DMSM} (top) illustrates dark matter-SM (nucleon) interactions, which always receives contributions from dark photon kinetic mixing with other SM neutral bosons. 
In addition, there are likely to be additional heavy hidden sector particles which could dominate the DM-nucleon scattering~\cite{Feng:2016ijc} but are too heavy to be kinematically accessible to dark matter annihilation. This heavy mediator plays a dual role in our setup: it decouples $\sigma_{\rm SI}$ from the $A^{\prime}$ lifetime, which would otherwise both be controlled by $\epsilon^{2}$ in the minimal model; and it ensures that the DM-nucleon scattering is isotropic, as assumed in our capture rate calculation. A scattering process mediated by the light $A^{\prime}$ alone would, for $m_{A^{\prime}}$ smaller than the typical momentum transfer, be forward-peaked rather than isotropic.

In addition to final states containing two fermions, the dark photon can decay into three photons through fermion loops, which are illustrated in the bottom left of Fig.~\ref{fig:DMSM} (The two-photon final state is forbidden, according to the Landau-Yang theorem~\cite{Landau:1948kw, Yang:1950rg}). In the limit in which the couplings are dominated by kinetic mixing and for dark photon masses below the lightest charged fermion ($e^{+}e^{-}$), neutrino pair decays are suppressed by $(m_{A^{\prime}}/m_{W^{\pm}})^{4}\simeq 10^{-21}(m_{A^{\prime}}/m_{e})^{4}$ compared to the three-photon decay.

In this low energy regime in the context of Effective Field Theory, we integrate out all heavy SM fields, causing electrons to dominate as the lightest SM field that is integrated out. We employ a dark Euler-Heisenberg Lagrangian that has ``non-linear" dark photon-photon interactions. The Lagrangian contains dimension-8 operators given by
\begin{equation}
\begin{split}
\mathcal{L}^{\rm EH}_{A^{\prime}}=\frac{\epsilon \alpha^{2}}{45 m_{e}^{4}}\Big{(}14F^{\prime}_{\mu\nu}&F^{\nu\lambda}F_{\lambda\rho}F^{\rho\mu}-5F^{\prime}_{\mu\nu}F^{\mu\nu}F_{\alpha\beta}F^{\alpha\beta}\Big{)},
\end{split}
\label{eq:lagrangianEH}
\end{equation}
with $\alpha\equiv \alpha_{\rm EM}$. In this limit, the approximate decay width aligns with Refs.~\cite{McDermott:2017qcg, Pospelov:2008jk}, and is given by:
\begin{equation}
\begin{split}
\Gamma_{\rm EH}&=\frac{17\epsilon^{2}\alpha_{\rm EM}^{4}}{2^{7}3^{6}5^{3} \pi^{3}}\times \frac{m_{A^{\prime}}^{9}}{m_{e}^{8}}\simeq 1{\rm s}^{-1}\times\Big{(}\frac{\epsilon}{0.003}\Big{)}^{2}\times \Big{(}\frac{m_{A^{\prime}}}{m_{e}}\Big{)}^{9}.
\end{split}
\label{eq:GammaEH}
\end{equation}

The full decay width of the dark photon for three massless particle final states can be calculated as
\begin{equation}
\begin{split}
\Gamma_{A^{\prime}\to\gamma\gamma\gamma}&=\frac{1}{(2\pi)^{3}}\frac{1}{32m_{A^{\prime}}^{3}}\frac{1}{6}\times\int\limits_{0}^{m_{A^{\prime}}^{2}}dm_{12}^{2}\int\limits_{0}^{m_{A^{\prime}}^{2}-m_{12}^{2}}dm_{13}^{2}\overline{|\mathcal{M}|}^{2},
\end{split}
\end{equation}
where $\mathcal{M}$ is the amplitude of the decay, which depends on the Dalitz variable $m_{ij}^{2}=(k_{i}+k_{j})^{2}$ for the photon momenta $k_{i}$ and $k_{j}$ in the final state. The factor $1/6$ accounts for the three identical bosons in the final state. In the regime where the dark photon mass is smaller than twice the electron mass, we obtain the full width from the series expansion as
\begin{equation}
\Gamma_{A^{\prime}\to \gamma\gamma\gamma}=\Gamma_{\rm EH}\Big{[}1+\sum\limits_{k=1}^{\infty}c_{k}\Big{(}\frac{m_{A^{\prime}}^{2}}{m_{e}^{2}}\Big{)}^{k}\Big{]}.
\label{eq:GammaApprox}
\end{equation}

\begin{table}[tbp]
\centering
\begin{tabular}{ccc}
\hline
 & $c_{k}$ & $c_{k}\times 4^{k}$\\
\hline
$c_{1}$ & \quad 335 / 714 \quad & 1.88 \\ 
$c_{2}$ & \quad 128,941 / 839,664 \quad & 2.46 \\
$c_{3}$ & \quad 44,787 / 1,026,256 \quad & 2.79 \\
$c_{4}$ & \quad 1,249,649,333 / 108,064,756,800 \quad & 5.92 \\
$c_{5}$ & \quad 36,494,147 / 12,382,420,050 \quad & 3.02 \\
$c_{6}$ & \quad 867,635,449 / 1,614,300,688,000 \quad & 2.20 \\
\hline
\end{tabular}
\caption{Expansion coefficients for the decay $A^{\prime}\to \gamma\gamma\gamma$ when the dark photon mass is smaller than twice the electron mass, taken from Ref.~\cite{McDermott:2017qcg}. For the full dark photon decay width in Eq.~(\ref{eq:GammaApprox}) below the Euler-Heisenberg limit, we take the expansion up to order 6.}
\label{tab:ck}
\end{table}

Here, the expansion contains the series of coefficients $c_{k}$, with values that are shown in Table~\ref{tab:ck} up to order 6. Since we focus on the phase space where the three-photon decay dominates, this approximation closely agrees with the full-width calculation~\cite{McDermott:2017qcg}.
In the context of indirect detection, we need the energy spectrum $dN/dE_{\gamma}$ of photons produced by dark photon decays. This spectrum is related to the distribution of the dark photon decay width by the energy of the final-state photons $E$ as
\begin{equation}
\frac{1}{N}\frac{dN}{dE}=\frac{1}{\Gamma}\frac{d\Gamma}{dE}.
\label{eq:dNNdE}
\end{equation}

\begin{figure}[tbp]
\centering
\includegraphics[width=0.47\columnwidth]{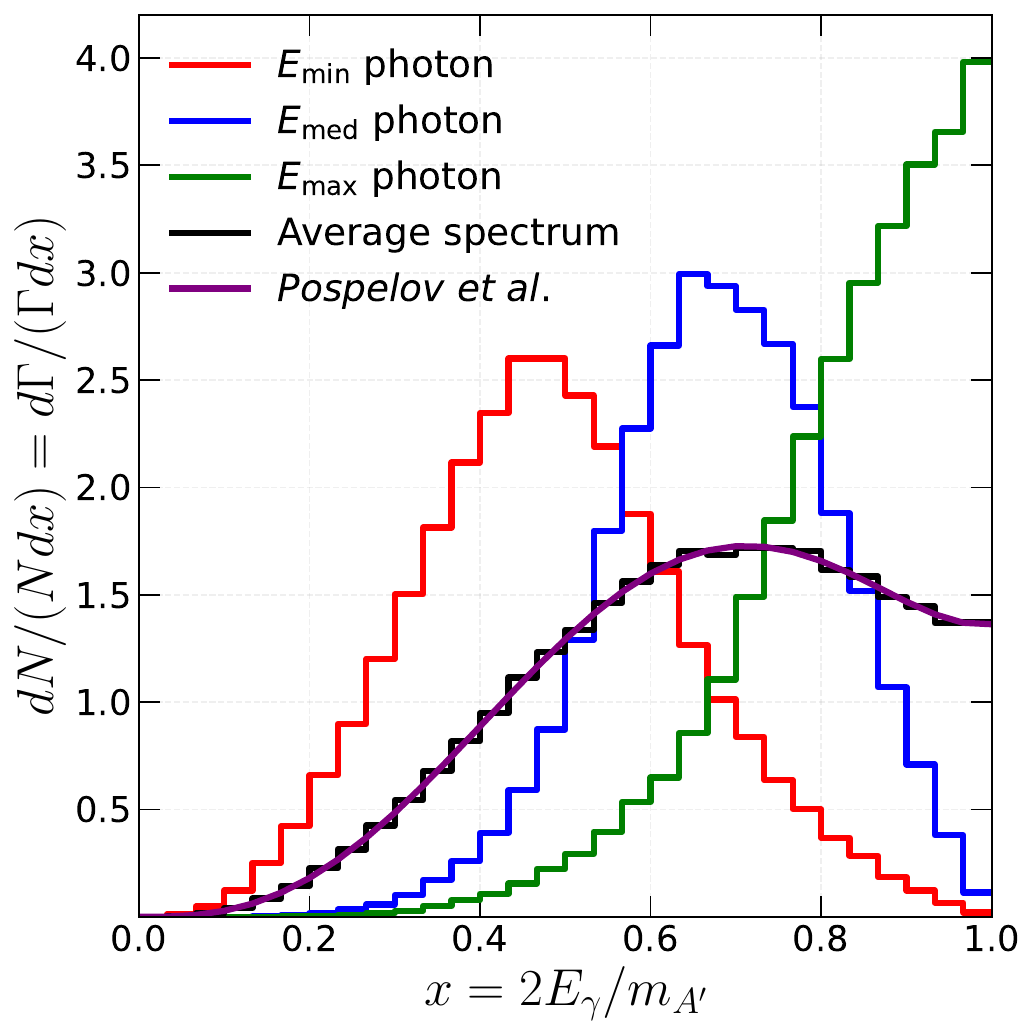}
\hfill
\includegraphics[width=0.47\columnwidth]{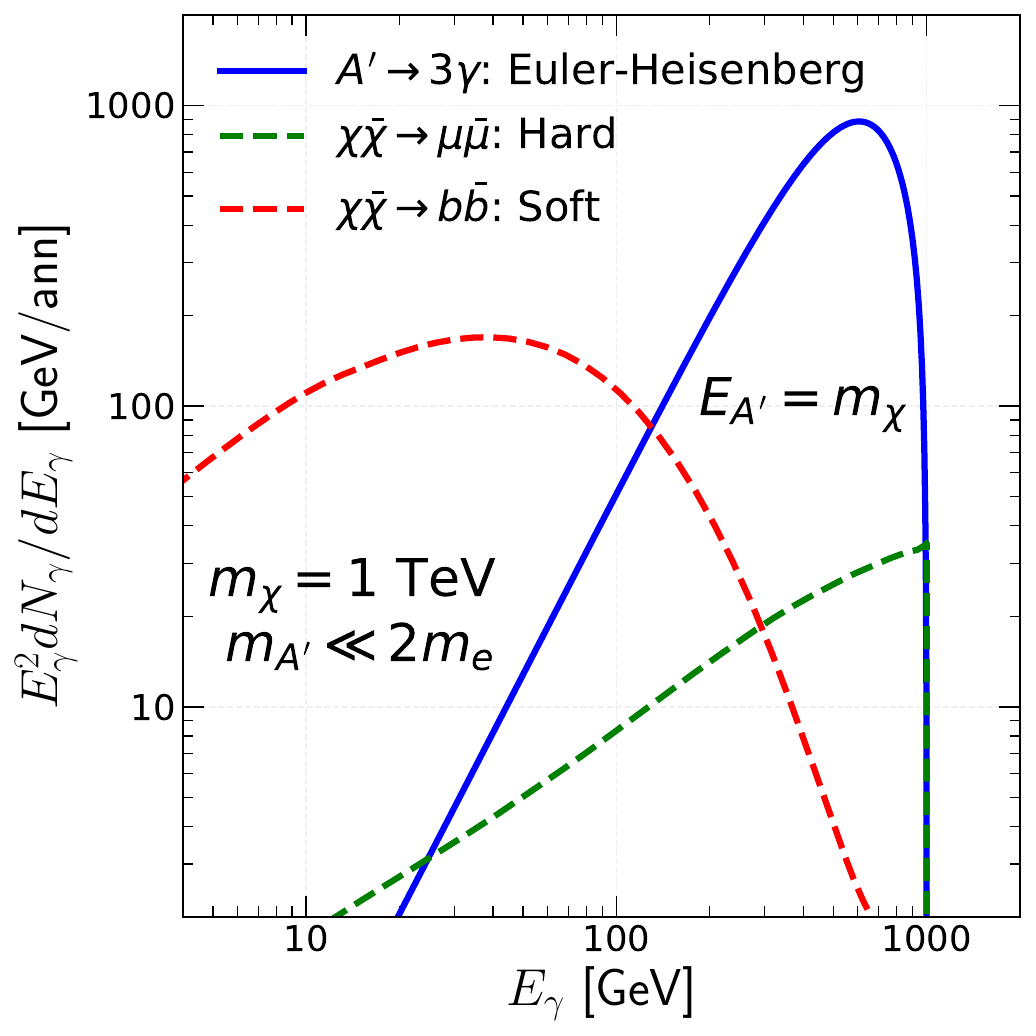}
\caption{{\bf Left}: Normalized $\gamma$-ray spectrum for $x=2E_{\gamma}/m_{A^{\prime}}$, generated by MadGraph ~\cite{Alwall:2011uj, Alwall:2014hca} and MadAnalysis~\cite{Conte:2012fm}.  These results are independent of the dark photon mass. The red, blue, and green histograms represent the spectra of the softest, medium, and hardest photon from each decay. The average photon spectrum (black) is compared with the result  from~Ref.~\cite{Pospelov:2008jk} (violet). {\bf Right}: Boosted differential energy spectrum for dark photon masses below the $e^{+}e^{-}$ threshold. The spectrum for a dark matter mass $m_{\chi}$ = 1~TeV (boost factor $\eta=m_{\chi}/m_{A^{\prime}}$) is compared with the spectra of standard channels such as $\mu\bar{\mu}$ and $b\bar{b}$, which are generated by PPPC4DMID~\cite{Cirelli:2010xx}.}
\label{fig:spectrum}
\end{figure}

From this spectrum, we calculate the modified normalized spectrum $dN/(Ndx)$, with $x=2E_{\gamma}/m_{A^{\prime}}$, which is independent of the dark photon mass. Based on Eq.~(\ref{eq:lagrangianEH}), we generate a UFO model file using FeynRules~\cite{Alloul:2013bka} as an input for MadGraph ~\cite{Alwall:2011uj, Alwall:2014hca} to calculate the dark photon decay width. Since MadGraph only distinguishes identical particles in the final state by their energy, we take the average energy spectrum (black histogram) that is shown in the left part of Fig.~\ref{fig:spectrum}. Our result agrees with the theoretical formula of Ref.~\cite{Pospelov:2008jk} in the rest frame of the dark photon as
\begin{equation}
\frac{d\Gamma}{\Gamma dx}\Big{|}_{\rm rest}=\frac{1}{51}x^{3}\Big{(}1715-3105x+\frac{2919}{2}x^{2}\Big{)}.
\label{eq:spectrumtheory}
\end{equation}

To find the boosted energy spectrum from the decay of an arbitrary dark photon with energy $E_{A^{\prime}}\approx m_{\chi}$, we use the same boost matrix along the $z$-direction as in the 2-body decay, which transforms any momentum from the rest frame to lab-frame as
\begin{equation}
P^{A^{\prime}}_{\rm lab}=\Lambda (m_{\chi}, m_{A^{\prime}})P^{A^{\prime}}_{\rm rest},
\end{equation}
where the boost matrix can be written in terms of the Lorentz boost factor $\eta=m_{\chi}/m_{A^{\prime}}$ as
\begin{equation}
\Lambda(\eta)=\begin{pmatrix}
\eta & 0 & 0 & \sqrt{\eta^{2} -1}\\
0 & 1 & 0 & 0\\
0 & 0 & 1 & 0\\
\sqrt{\eta^{2} - 1} & 0 & 0 & \eta
\end{pmatrix}.
\label{eq:boostmatrix}
\end{equation}

This boost matrix can be applied to the 2-body decay spectrum in the rest frame, which is $dN/(NdE)=\delta(E-m_{\phi}/2)$ as a Dirac-Delta function ($m_{\phi}$ is the mediator mass). In our 3-body decay, the spectrum is not a delta-function, but a continuous distribution, which is shown as a violet line in the left part of Fig.~\ref{fig:spectrum}. This distribution can be discretized into $N_{\rm bin}$ bins, as shown by the black histogram, with $N_{\rm bin}\to \infty$ being equal to the analytic formula in Eq.~(\ref{eq:spectrumtheory}).

To calculate the trident energy spectrum, we apply the boost matrix to our three-body decay process and discretize the results. We note that this process, in the case of a 2-body decay, leads to the well-known box-like spectrum~\cite{Abdullah:2014lla, Ibarra:2012dw, Cirelli:2010xx}. On the other hand, the semi-analytical form of the 3-body photon spectrum from a massive dark photon decay depends on the dark matter mass $m_{\chi}$, and the dark photon mass $m_{A^{\prime}}$, and is given by: 

\begin{equation}
\frac{1}{N}\frac{dN}{dE}(E, m_{\chi}, m_{A^{\prime}})=\lim\limits_{N_{\rm bin}\to \infty}\sum\limits_{n=1}^{N_{\rm bin}}\frac{1}{N}\frac{dN}{dx}\Big{|}_{\rm rest}\big{(}x_{n}\big{)}  \frac{1}{n\sqrt{m_{\chi}^{2}-m_{A^{\prime}}^{2}}}\Theta(E^{\rm max}_{n}-E)\Theta(E-E^{\rm min}_{n}),
\label{eq:spectrum3body_boost}
\end{equation}

where we have the sum of box-shaped spectra from bin $n$ to bin $N_{\rm bin}$. The new variable,  $x_{n}=n/N_{\rm bin}$, is independent of the dark photon mass, and depends on the bin $n$ in $N_{\rm bin}$ that we consider. For each box-shaped distribution from bin $n$, the edges are defined as
\begin{align}
E^{\rm max}_{n}&=\frac{n}{2 N_{\rm bin}}(m_{\chi} + \sqrt{m_{\chi}^{2}-m_{A^{\prime}}^{2}}),\\
E^{\rm min}_{n}&=\frac{n}{2 N_{\rm bin}}(m_{\chi} - \sqrt{m_{\chi}^{2}-m_{A^{\prime}}^{2}}).
\end{align}

In Fig.~\ref{fig:spectrum} (right), we show the differential energy spectrum from dark matter annihilation to two on-shell dark photons, which each decay into three photons. We set $N=6$ because we have 6 photons in the final state, and set $N_{\rm bin}=1000$. We compare our result with common dark matter spectra: $\mu\bar{\mu}$ and $b\bar{b}$, for a benchmark $m_{\chi}=1$~TeV dark matter mass~\cite{Cirelli:2010xx}. The spectrum from three-photon decays produces a hard $\gamma$-ray spectrum.

In addition to the dark matter capture and annihilation scenario presented in this paper, this final state can also be applied to searches for hidden on-shell mediators in standard indirect detection searches~\cite{Abdullah:2014lla}, and could be used to constrain the thermally averaged cross section in vector mediator models. With Eqs.~ (\ref{eq:spectrumtheory}) and (\ref{eq:spectrum3body_boost}), we can generate the spectrum for non-resonant 3-body decay processes without using Pythia~\cite{Sjostrand:2014zea}. 

\section{Dark Matter Capture and Annihilation}
\label{sect:captandann}

\vspace{-0.3cm}

In this section, we explore dark matter capture by celestial objects. We then delve into the annihilation of trapped dark matter and establish a relationship between the capture and annihilation processes. These treatments are applicable across most models that contain dark matter-nucleon interactions.

\subsection{Dark Matter Capture}
\label{ssect:capt}

Many studies have explored interactions between dark matter and celestial baryons~\cite{Ilie:2020vec, Bose:2022ola, Dasgupta:2020dik, Cappiello:2023hza}. The simplest scenario involves dark matter scattering with nucleons in the rest frame. Each scattering event removes some of the dark matter kinetic energy until it is captured when its velocity falls below the escape velocity of the object. 

The capture rate, representing the number of dark matter particles captured per second, is calculated by summing over the capture rate for dark matter particles that scatter with $N$ different nuclei in the celestial object:

\begin{align}
\label{eq:capturerate}
C=\sum\limits_{N=1}^{\infty}C_{N}(\tau)&=f_{\rm cap}\frac{\pi R^{2}p_{N}(\tau)}{(1-2G_{N}M/R)}\frac{\sqrt{6}n_{\chi}}{3\pi \bar{v}}\\
&\times\Big{[}(2\bar{v}^{2}+3v_{\rm esc}^{2})-(2\bar{v}^{2}+3v_{N}^{2})\exp\Big{(}\frac{3(v_{\rm esc}^{2}-v_{N}^{2})}{2\bar{v}^{2}}\Big{)}\Big{]}\nonumber,
\end{align}
where $R$, $M$, and $v_{\rm esc}$ denote the radius, mass, and escape velocity of the celestial body, respectively. The probability of dark matter scattering with $N$ nucleons, denoted as $p_{N}(\tau)$, is calculated by integrating the Poisson distribution over the cosine of the scattering angle, $y$, as
\begin{equation}
p_{N}(\tau)=2\int\limits_{0}^{1}dy\frac{y e^{y\tau}(y\tau)^{N}}{N!},
\label{eq:pN}
\end{equation}
which depends on the optical depth
\begin{equation}
\tau = \frac{3}{2}\frac{\sigma_{\chi n}}{\sigma_{\rm sat}},
\label{eq:tau}
\end{equation}
where $\sigma_{\rm sat}=\pi R^{2}/N_{n}$ is the saturation cross section of the object. Following this picture of multiscattering, we obtain the dark matter velocity after $N$ interactions as
\begin{equation}
v_{N}=v_{\rm esc}(1-\Braket{z}\beta)^{-N/2},
\label{eq:vN}
\end{equation}
where $\beta=4m_{\chi}m_{\rm SM}/(m_{\chi}+m_{\rm SM})^{2}$ $\Braket{z}$ is the cosine of the scattering angle. The $f_{\rm cap}$ function in Eq.~\ref{eq:capturerate} depends on the kinetic regime of DM-nucleon scattering. The treatment of $\Braket{z}$ and $f_{\rm cap}$ is encoded in the \texttt{Asteria} package in Ref.~\cite{Leane:2023woh}, which we use for our calculations of the capture rate.

\begin{table}[tbp]
\centering
\begin{tabular}{ccccc}
\hline
Object & Mass [$M_{\odot}$] & $R$ [km] & Component & $\sigma_{\rm sat}$ [cm$^{2}$]\\
\hline
Neutron Star & 1.4 & 10.0 & neutron & $2.0\times 10^{-45}$\\
Brown Dwarf & 0.0378 & 69,911.0 & hydrogen & $3.63\times 10^{-36}$\\
Jupiter &  1/1047 & 69,911.0 & hydrogen & $1.44\times 10^{-34}$\\
Sun & 1.0 & 696,340.0 & hydrogen & $1.33\times 10^{-35}$\\
\hline
\end{tabular}
\caption{Benchmark celestial object parameters in the capture rate calculation. The saturation cross section is estimated as $\sigma_{\rm sat}=\pi R^{2}/N_{n}$.}
\label{tab:objects}
\end{table}

Table~\ref{tab:objects} presents the benchmark parameters for the celestial objects under consideration: neutron stars, brown dwarfs, Jupiter, and the Sun. These parameters serve as inputs to the \texttt{Asteria} package. It is noteworthy that in Eq.~(\ref{eq:capturerate}), we account for the incoming dark matter blue-shift effect, introducing an enhancement of $1/(1-2G_{N}M/R)$ in the capture rate. This effect, though negligible for brown dwarfs and Jupiter, doubles the capture rate in neutron stars, a consideration not incorporated in {\tt Asteria}.

In models with light dark matter, dark matter evaporation can erode the number density of captured dark matter particles. In our case, we note that the light-dark matter can couple to baryons through the long-lived mediator, which implies that the interaction between DM and the nucleon is likely to proceed through a long-range force, and hence the evaporation effect could be important~\cite{Acevedo:2023owd}. However, this dark photon parameter phase space is already excluded, so we consider DM down to the 10~MeV mass range in these celestial objects, but keep in mind the evaporation limit.

Another crucial input parameter for calculating the capture rate is the initial dark matter velocity dispersion, denoted as $\bar{v}$, and the local dark matter density surrounding the object. For objects in our solar system, we use the value $\bar{v} \approx 220$~km/s, and $\rho_{\chi}=0.3$~GeV/cm$^{3}$.
For celestial objects in the Galactic center region, we use the simplified Milky Way Galaxy mass model in Ref.~\cite{Sofue:2013kja}, and the generalized Navarro-Frenk-White (NFW)~\cite{Navarro:1995iw}:
\begin{equation}
\rho_{\chi}(r)=\frac{\rho_{0}}{(r/r_{s})^{\gamma}(1+(r/r_{s}))^{3-\gamma}},
\label{eq:gNFW}
    \end{equation}
with $\gamma$ as the inner slope of the profile, for which we choose benchmark models $\gamma=1.0$ and 1.5 and assume $r_s$~=~12~kpc. The $\rho_{0}$ parameter is normalized to the local dark matter density in our solar neighborhood, for each value of $\gamma$. To calculate the total dark matter capture rate for celestial bodies near the Galactic center, we convolve this dark matter density distribution with the celestial object morphology as 

\begin{equation}
C_{\rm tot}=4\pi\int\limits_{0.1 {\rm pc}}^{100 {\rm pc}}n_{\star}(r)r^{2}C(r)dr,
\end{equation}
where $n_{\star}(r)$ is the celestial object density distribution. For neutron stars in the Galactic center, we assume an average mass $M_{\rm NS}=1.4M_{\odot}$, and use the number density
\begin{align}
n_{\rm NS}(r)&=5.98\times 10^{3}\Big{(}\frac{r}{1{\rm pc}}\Big{)}^{1.7}{\rm pc}^{-3};\quad 0.1{\rm pc}<r<2{\rm pc},\nonumber\\
&=2.08\times 10^{4}\Big{(}\frac{r}{1{\rm pc}}\Big{)}^{-3.5}{\rm pc}^{-3};\quad r> 2{\rm pc},
\label{eq:nNS}
\end{align}
which is extrapolated from the radial distribution of the `Fiducial $\times$ 10' model in Ref.~\cite{Generozov:2018niv}. In the case of brown dwarfs in the Galactic center, we choose the average mass $M_{\rm BD}=0.0378$~$M_{\odot}$~\cite{Leane:2021ihh}, which is in agreement with the Kroupa Initial Mass function (IMF)~\cite{Kroupa:2011aa}. We adopt the brown dwarf number density in Refs.~\cite{Kroupa:2011aa, Amaro-Seoane:2019umn} for the mass range 0.01--0.07~$M_{\odot}$ as
\begin{equation}
n_{\rm BD}(r)=7.5\times 10^{4}\times\Big{(}\frac{r}{1 {\rm pc}}\Big{)}^{-1.5}{\rm pc}^{-3}.
\end{equation}

We note that neutron stars in the Galactic center region have thermalization timescales that are approximately 1~kyr~\cite{Bell:2023ysh}, allowing us to assume that the dark matter density in all neutron stars is in equilibrium. The same statement can be applied to brown dwarfs, according to Ref.~\cite{Generozov:2018niv}. This equilibrium condition is used later in Sub-section~\ref{ssect:DMann} for the SM flux resulting from dark matter annihilation.

\subsection{Dark Matter Annihilation}
\label{ssect:DMann}
Operating under the assumption that dark matter can interact with the SM, it becomes necessary to consider dark matter annihilation into SM particles. If these SM particles are created within the volume of the celestial object, they will quickly interact with other SM matter in the celestial object, eventually being absorbed by the celestial object and contributing to thermal energy. These processes are collectively referred to as annihilation heating~\cite{Bramante:2023djs}.
However, our study considers a special scenario where dark matter annihilates into a pair of on-shell dark mediators. If the mediator lifetime is sufficiently long to allow it to escape the physical object, it can then decay into stable SM particles such as neutrinos and photons, that would be detectable on Earth. While long-lived particles are investigated in some far-forward detectors such as FASER~\cite{FASER:2018eoc}, celestial objects can probe phase spaces where the lifetime and travel distance of the mediator extends to parsec scales~\cite{Nguyen:2022zwb, Leane:2021ihh, Niblaeus:2019gjk, Leane:2017vag, HAWC:2018szf, Leane:2021tjj}.

\begin{table}[btp]
\centering
\begin{tabular}{ccc}
\hline
\multirow{2}{*}{Target} & Stellar radius ($R$) & Distance to Earth ($D$) \\
                  & (Minimum length) & (Maximum length) \\ \hline
    Sun & 696,340 km & 1 AU \\
    Jupiter & 69,911 km & $7.78\times 10^{8}$ km \\
\multirow{2}{*}{Galactic center} & NS: 10 km & \multirow{2}{*}{8 kpc} \\
                  & BD: 69,911 km &                   \\ \hline
\end{tabular}
\label{tab:distance}
\caption{The celestial body radius and distance to Earth for each of our celestial targets. These values combine to specify the minimum and maximum decay lengths for the dark photon.}
\end{table}

The minimum and maximum decay lengths of the mediator are set by the size of the celestial object and the distance of the celestial object to Earth, respectively,
\begin{equation}
R\lesssim L=\eta \beta \tau_{A^{\prime}}\simeq \frac{m_{\chi}}{m_{A^{\prime}}}c\tau_{A^{\prime}}\lesssim D,
\label{eq:limL}
\end{equation}
where the Lorentz boost factor $\eta=E_{A^{\prime}}/m_{A^{\prime}}\simeq m_{\chi}/m_{A^{\prime}}$, assuming the non-relativistic dark matter annihilation. The distances $R$ and $D$ are the stellar radius and the source distance to our detectors. Since the travel distance depends on the mediator lifetime $\tau_{A^{\prime}}$, for every dark matter mass, we calculate upper and lower limits for the mediator $A^{\prime}$ couplings and mass in this scenario.

\begin{figure}[btp]
\centering
\includegraphics[width=0.99\columnwidth]{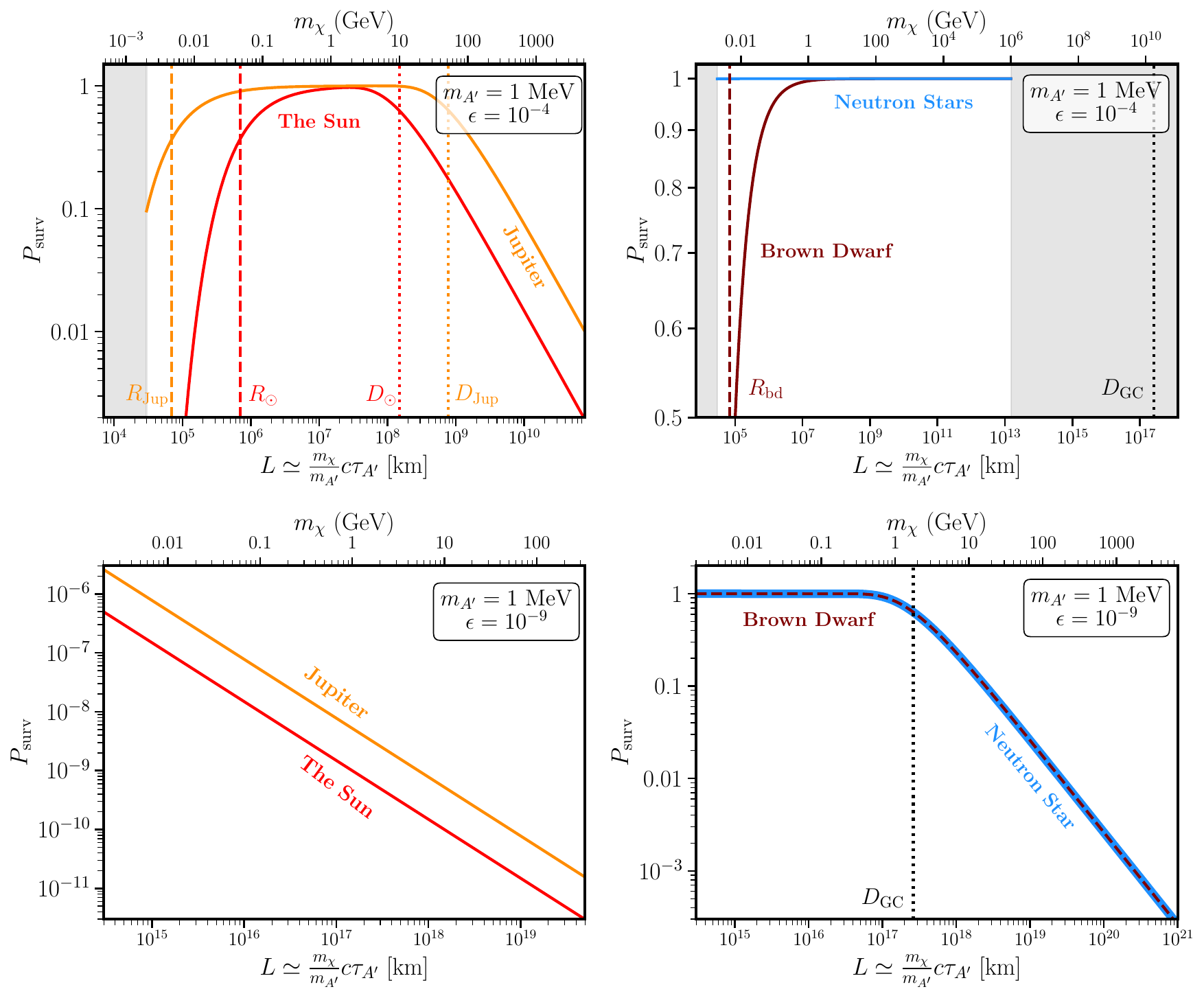}
\caption{Survival probability as a function of the dark photon decay length (bottom x-axes), with the equivalent DM mass shown on the upper $x$-axes, for a dark photon mass of 1~MeV. Upper panels are for $\epsilon=10^{-4}$, lower panels for $\epsilon=10^{-9}$. Left panels show results for the Sun (red) and Jupiter (orange); right panels show brown dwarfs (maroon) and neutron stars (light blue) in the Galactic Center. The equivalent radii of these objects are shown as dashed lines, the distances to Earth as dotted lines, and the Galactic Center distance as a black dotted line. The gray shaded regions mark decay lengths outside the DM mass scan range, from 20~MeV to 1~PeV.}
\label{fig:PSurv}
\end{figure}

\begin{figure}[btp]
\centering
\includegraphics[width=0.99\columnwidth]{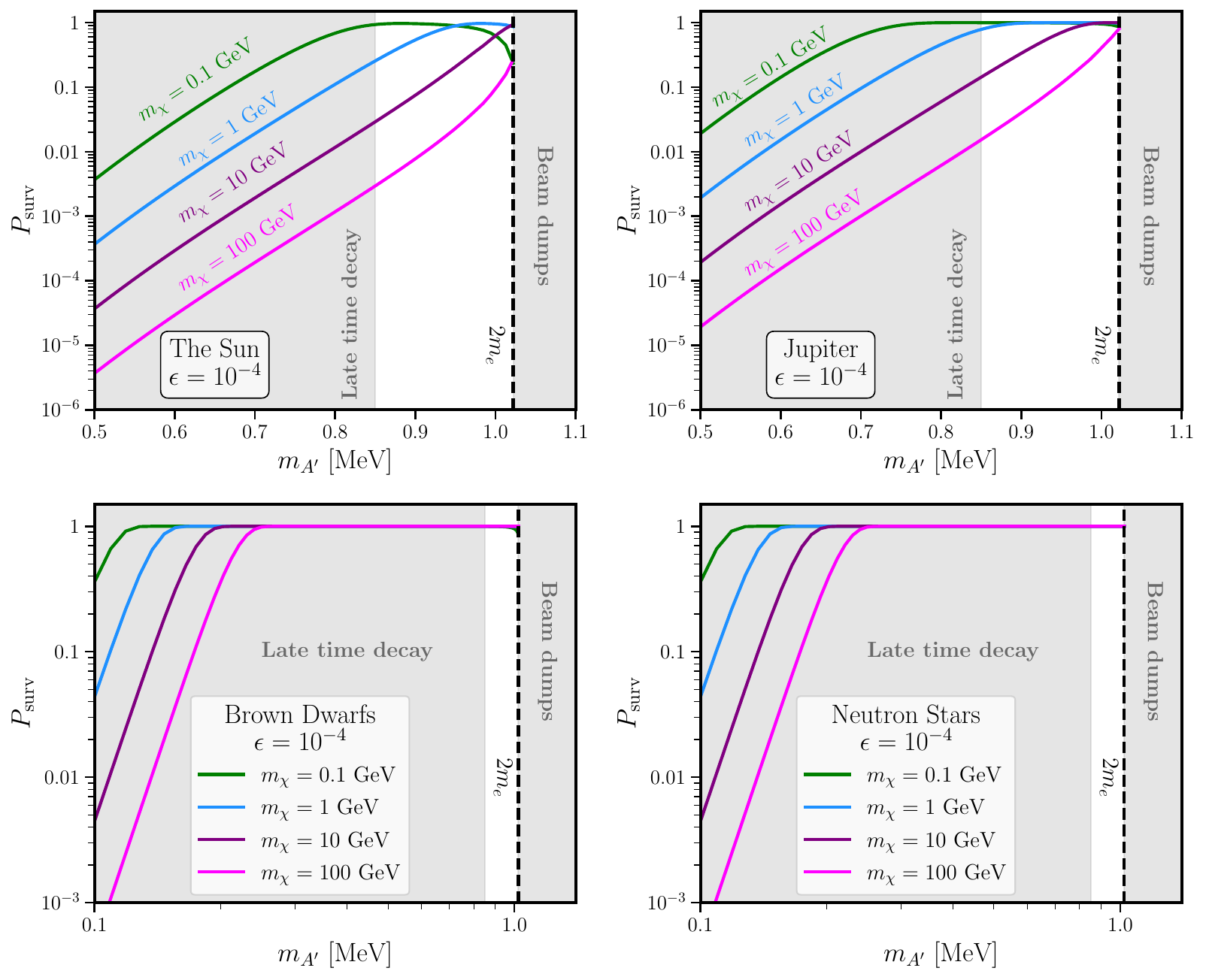}
\caption{Survival probability as a function of the dark photon mass for a benchmark kinetic mixing $\epsilon=10^{-4}$, shown for four benchmark DM masses: $0.1$~GeV (green), $1$~GeV (blue), $10$~GeV (purple), and $100$~GeV (magenta). The four panels correspond to different celestial objects: the Sun (top left), Jupiter (top right), brown dwarfs (bottom left), and neutron stars (bottom right). The gray shaded regions are excluded by the requirement that the dark photon decay lifetime be shorter than $1$~second, before the epoch of ionization~\cite{McDermott:2017qcg} (left) and by beam-dump experiments~\cite{Essig:2013lka} (right). The dashed lines mark the $2m_e$ threshold, where the $A^{\prime}\to 3\gamma$ decay starts to be suppressed.}
\label{fig:PSurv_mA}
\end{figure}

In the case that the dark matter density in a celestial object is in equilibrium, the rate of dark matter annihilation is related to the capture rate as
\begin{equation}
\Gamma_{\rm ann}=\frac{C}{2},
\label{eq:annrate}
\end{equation}
which relates the annihilation rate to the dark matter-nucleon scattering cross section. The factor 1/2 arises because each annihilation removes two dark matter particles.

Once the mediators escape the gravitational bounds of the stellar object, they decay into SM particles. Given that the mediators are on-shell particles with a lifetime $\tau_{A^{\prime}}$, the differential energy flux of the SM signal reaching our detectors from a distance $D$ takes the form:
\begin{equation}
E^{2}\frac{d\Phi}{dE}=\frac{\Gamma_{\rm ann}}{4\pi D^{2}}\times E^{2}\frac{dN}{dE}\times{\rm BR}(A^{\prime}\to {\rm SM})\times P_{\rm surv}.
\label{eq:E2dNdE}
\end{equation}
Here, $E\equiv E_{\rm SM}$ and $\Phi\equiv \Phi_{\rm SM}$ represent the energy and flux of the SM signal we aim to detect. Given our focus on the on-shell decays of our mediator $A^{\prime}$, $\Gamma_{\rm ann}$ denotes the annihilation rate from Eq.~(\ref{eq:annrate}), while ${\rm BR}(A^{\prime}\to {\rm SM})$ represents the branching ratio of $A^{\prime}$ decay into these SM signals. The survival probability is determined as
\begin{equation}
P_{\rm surv}=e^{-R/\eta c\tau_{A^{\prime}}}-e^{-D/\eta c\tau_{A^{\prime}}},
\end{equation}
which depends on the mediator lifetime $\tau_{A^{\prime}}$.

Fig.~\ref{fig:PSurv} shows the survival probability as a function of the decay length, and equivalently the dark matter mass, for the benchmark dark photon mass of 1~MeV. This benchmark point sits in a regime where the $(m_{A^{\prime}}/m_{e})^{9}$ suppression is not prohibitively severe while ensuring that the $e^{+}e^{-}$ decay channel remains kinematically forbidden. We choose two benchmark kinetic mixing values: $\epsilon=10^{-4}$, which avoids dark photon decay after the epoch of ionization~\cite{McDermott:2017qcg}, as well as constraints from SN1987A~\cite{Chang:2016ntp}, the electron anomalous magnetic moment~\cite{Pospelov:2008zw}, and beam dump experiments~\cite{Essig:2013lka}. We also show results for the lower value $\epsilon=10^{-9}$, which avoids the constraints from $\Delta N_{\rm eff}$~\cite{Bennett:2020zkv, Caputo:2025avc}.

Similarly, we show the survival probability as a function of the dark photon mass $m_{A^{\prime}}$ in Fig.~\ref{fig:PSurv_mA}, fixing $\epsilon = 10^{-4}$ and showing four benchmark dark matter masses for each celestial target. Because the decay width scales as $\Gamma_{A^{\prime}} \propto m_{A^{\prime}}^{9}$, the lab-frame decay length grows as $L \propto m_{A^{\prime}}^{-10}$, so even a modest decrease in $m_{A^{\prime}}$ rapidly increases the decay length and suppresses the survival probability. The gray shaded regions are excluded by the requirement that the dark photon not decay after the epoch of ionization~\cite{McDermott:2017qcg} (left) and by beam dump experiments~\cite{Essig:2013lka} (right), while the dashed line marks the $2m_e$ threshold above which the $A^{\prime} \to 3\gamma$ channel becomes suppressed. Combined with these constraints from late-time decay and beam dump experiments, the available parameter space in which this decay is significant is restricted to a narrow band of dark photon mass, from roughly $850$~keV to $1022$~keV.

Our results show that celestial objects in both the solar system and the Galactic Center favor larger values of kinetic mixing, for which the survival probability approaches unity. Objects near the Galactic Center also unlock larger mass ranges in for the dark matter parameter space. For lower kinetic mixing probabilities, the Sun and Jupiter quickly lose sensitivity due to the fact that the tridents do not decay before passing the Earth, while brown dwarfs and neutron stars in the Galactic Center can still reach the maximum probability and only begin to be suppressed for dark matter masses above $\sim 2$~GeV. These results demonstrate that celestial bodies at different distances from Earth are sensitive to different regions of parameter space in this class of long-lived mediator models.

Alongside this probability and the branching ratio, the energy spectrum $dN/dE$ is also contingent on the final state of the SM signals produced by the decays. In this study, we apply the three-photon final state spectrum in Eq.~(\ref{eq:spectrum3body_boost}).

\section{Gamma-Ray Spectra and Limits}
\label{sect:ExpLim}

\begin{figure}[btp]
\centering
\includegraphics[width=0.47\columnwidth]{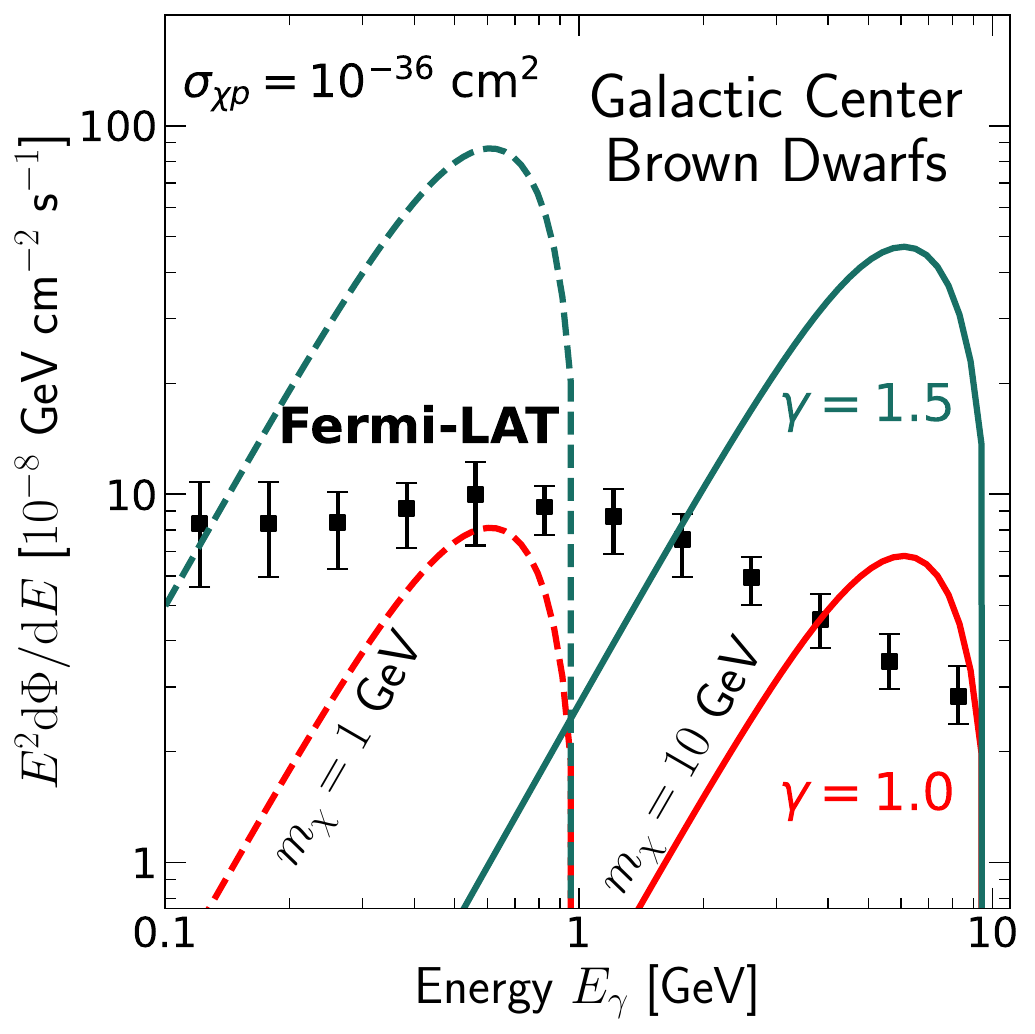}
\hfill
\includegraphics[width=0.47\columnwidth]{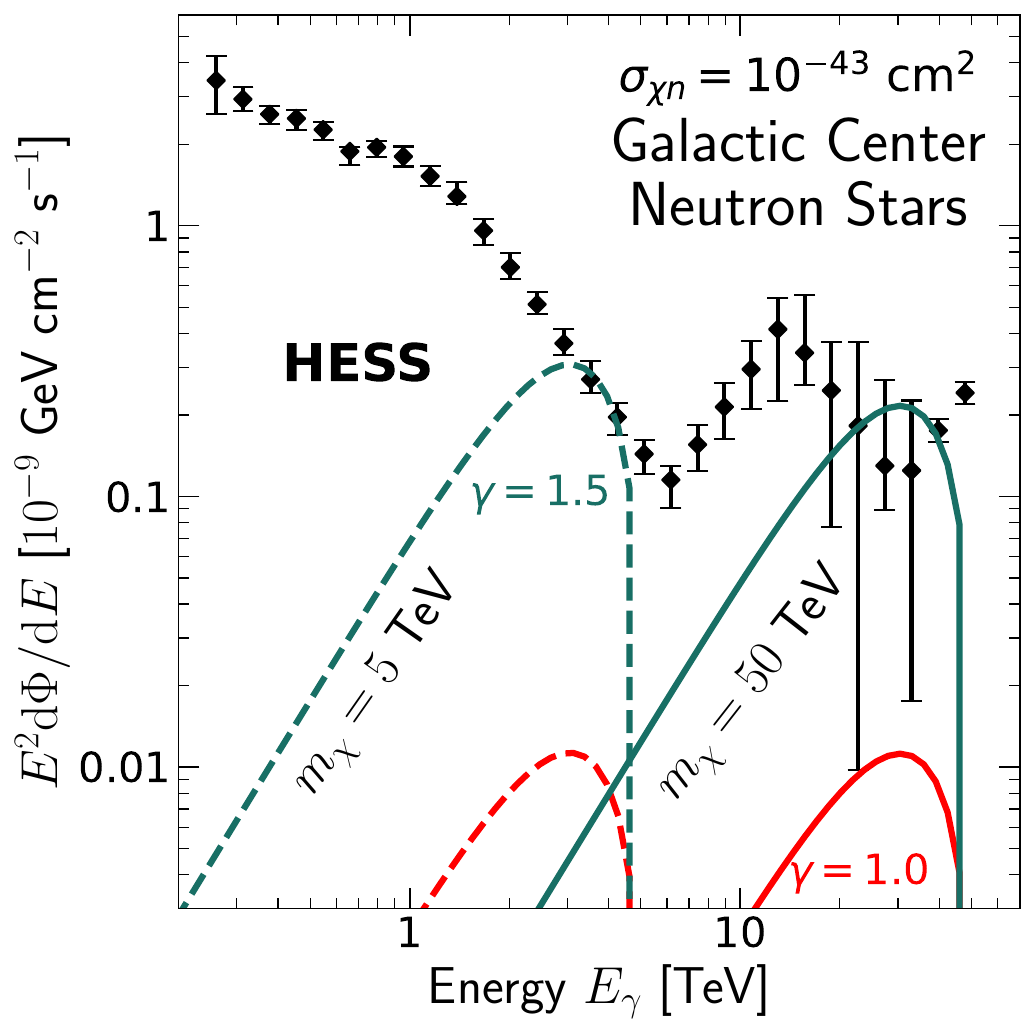}
\caption{Gamma-ray spectra from the Galactic center region. {\bf Left}: Fermi-LAT spectrum from 0.1--10~GeV. The solid lines are for $m_{\chi}=10$~GeV and the dashed lines are for $m_{\chi}=1$~GeV. We consider dark matter scattering with protons inside brown dwarfs in the Galactic center with the cross section $\sigma_{\chi p}=10^{-36}$~cm$^{2}$. We calculate the spectrum coming from the Galactic center for two different gNFW profiles: $\gamma=1.0$ (red lines) and $\gamma=1.5$ (teal lines). {\bf Right}: Same as left, but showing H.E.S.S. data for 0.2--50~TeV $\gamma$-rays. The dark matter flux comes from neutron stars, with a dark matter-neutron cross section of $\sigma_{\chi n}=10^{-43}$~cm$^{2}$. We show benchmark dark matter masses at 5~TeV and 50~TeV.}
\label{fig:GCFlux}
\end{figure}

In this section, we determine whether or not current $\gamma$-ray observations can probe the on-shell 3-body decay from dark matter annihilation inside celestial objects. We consider two cases: the ensemble of celestial objects near the Galactic center such as neutron stars or brown dwarfs, and single solar system objects such as the Sun and Jupiter. We utilize $\gamma$-ray data from Fermi-LAT and H.E.S.S. for Galactic center observations, Fermi-LAT data for Jupiter and HAWC observations of the Sun.

\begin{figure}[btp]
\centering
\includegraphics[width=0.5\columnwidth]{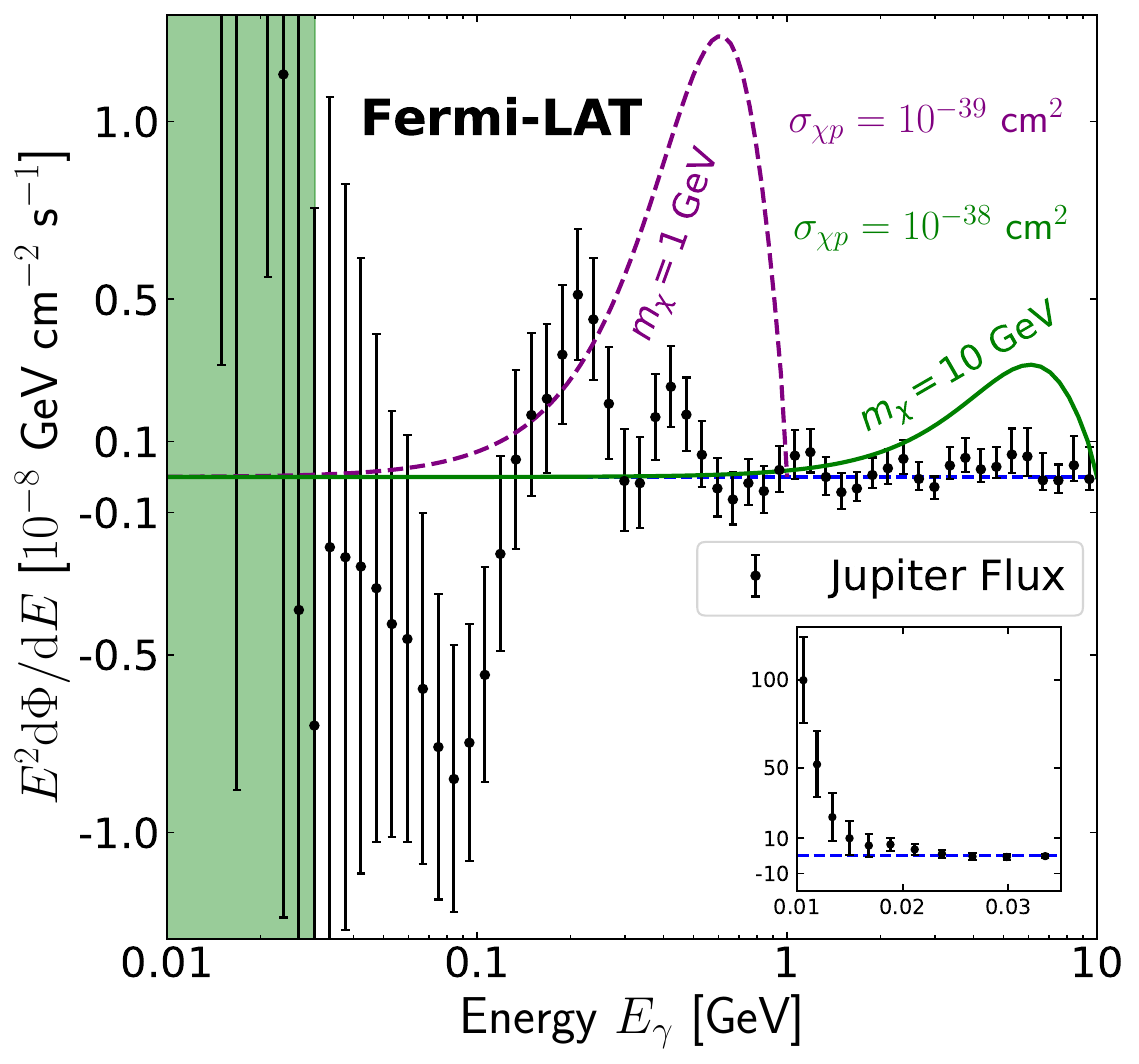}
\caption{$\gamma$-ray upper limits from Jupiter by the Fermi-LAT. The blue horizontal line depicts zero flux. The green shaded region for energies below 30~MeV is zoomed in the sub-figure. We show two benchmark points: The violet dashed line is for $m_{\chi}=$ 1~GeV, with a dark matter-proton cross section of $10^{-39}$~cm$^{2}$. The green solid line is for $m_{\chi}=10$~GeV, which has the benchmark cross section that is larger by 1 order of magnitude.}
\label{fig:JupLim}
\end{figure}

\begin{figure}[btp]
\centering
\includegraphics[width=0.5\columnwidth]{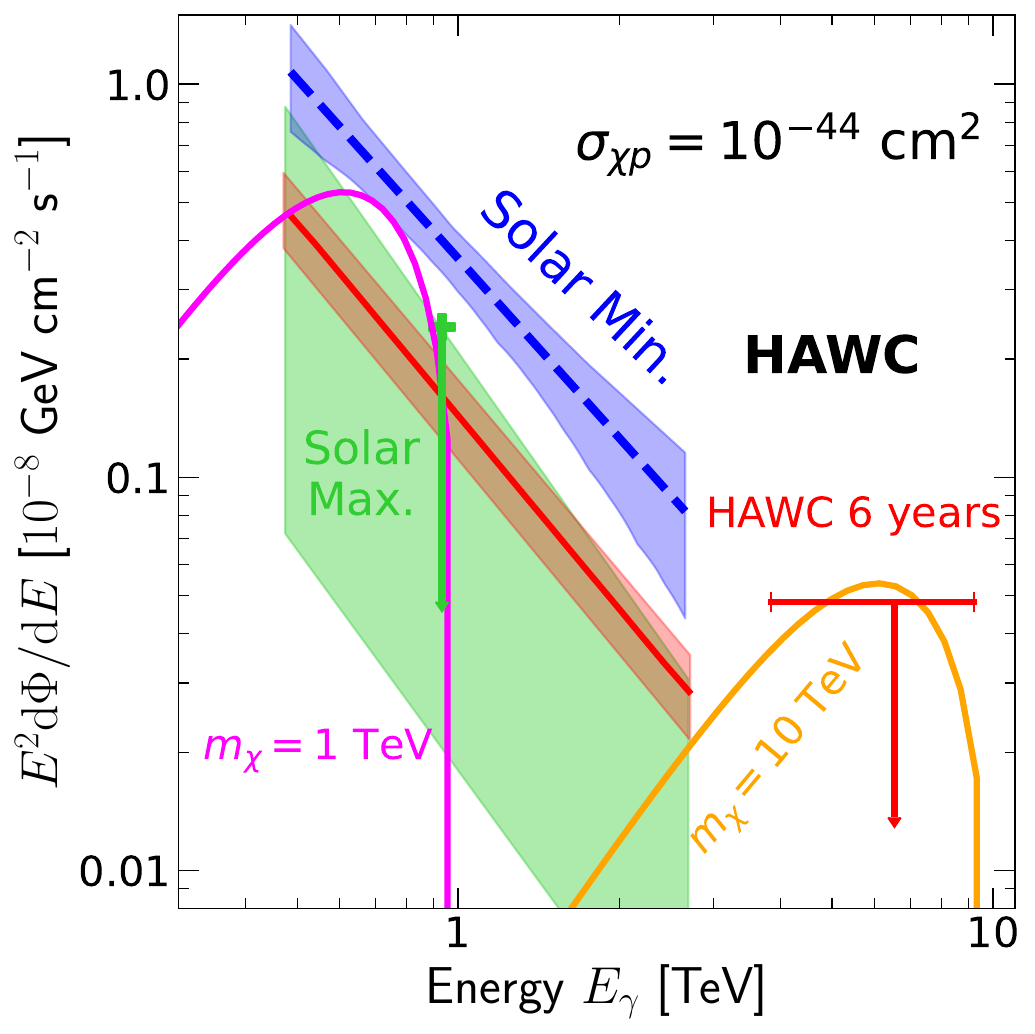}
\caption{Gamma-ray spectra and upper limits from the Sun. The 6.1 year-spectrum and the 90\% CL upper limit at 7~TeV from HAWC are in the red line and red bin respectively. The solar minimum spectrum is shown in the blue dashed line. The 1$\sigma$ upper limit at 1~TeV photon energy is presented as a green point. The shaded area for each limit includes systematic uncertainties. The two spectra from dark photon decay for dark matter-proton cross sections of $\sigma_{\chi p}=10^{-44}$~cm$^{2}$ are shown with the  magenta line for $m_{\chi}= 1$~TeV and the orange line for $m_{\chi}=10$~TeV.}
\label{fig:SunLim}
\end{figure}

We note that for the Galactic center, both Fermi-LAT and H.E.S.S. data include statistically significant detections of $\gamma$-ray emission, which are shown in Fig.~\ref{fig:GCFlux}. These $\gamma$-ray fluxes are assumed to originate from astrophysical sources in the Galactic center, and not from dark matter. However, we choose to conservatively set upper limits on the dark matter interaction cross-section by assuming that dark matter does not overproduce the entirety of the $\gamma$-ray signal. For observations of Jupiter and the Sun, we employ upper limits from Fermi-LAT and HAWC in Fig.s~\ref{fig:JupLim} and \ref{fig:SunLim}, respectively. We note that the Sun is detected by HAWC during solar minimum, but owing to the fact that the dark matter signal should not vary with solar cycle, we strongly constrain dark matter by ensuring that the annihilation signal does not overproduce the $\gamma$-ray upper limits obtained during solar maximum. 

In these figures, we also sketch the photon spectra from dark matter annihilation through the dark photon-photon trident for several benchmark dark matter masses and scattering cross sections with nucleons. We assume that all dark photons escape the stellar volume and decay within the length scales in Table~\ref{tab:distance}, such that the branching ratio into photons can be assumed to be 100\%. Since the couplings between the dark photon and other baryonic particles are weak, we also neglect the attenuation of the photon signal and assume a 100\% survival probability. The cross-section benchmarks for these spectra are chosen based on the saturation cross section in Table~\ref{tab:objects} depending on the target objects.

Fig.~\ref{fig:GCFlux} (left) shows the Fermi-LAT spectrum from Ref.~\cite{Malyshev:2015hqa}, between 0.1--10~GeV. We indicate two dark matter benchmarks at 1~GeV (dashed lines) and 10~GeV (solid lines). We consider two gNFW dark matter density profiles, with $\gamma=1.0$ equivalent to the classical NFW (red lines), and $\gamma=1.5$ that assumes more dark matter inside the Galactic central region (teal lines). The dark matter-proton cross section is chosen to be $10^{-36}$ cm$^{2}$, which is near the saturation cross section of brown dwarfs. For dark matter masses below 10~GeV, Fermi-LAT data can probe the dark matter-nucleon cross-section near the brown dwarf saturation cross-section for both classical and generalized NFW profiles.

Similarly, we show the H.E.S.S. spectrum for high energy $\gamma$-rays from the Galactic center in the right panel of Fig.~\ref{fig:GCFlux}. Because H.E.S.S. observations probe photon energies between 0.2--50~TeV, we choose two dark matter masses at 5~TeV (dashed lines) and 50~TeV (solid lines). In this high-energy regime, our analysis is more sensitive to signals from neutron stars than at Fermi-LAT energies~\cite{Nguyen:2022zwb, Leane:2021ihh}. The dark matter-neutron cross-section is chosen to be 10$^{-43}$~cm$^{2}$, which is bigger than the NS saturation cross section by about 2 orders of magnitude. With this cross section, the capture rate achieves its maximum value, as discussed in Ref.~\cite{Nguyen:2022zwb}. Based on these constraints, we see that neutron stars only provide sensitivity to generalized NFW profiles with $\gamma \sim 1.5$.

For Jupiter and the Sun, we show the $\gamma$-ray upper limits and spectra in Figs.~\ref{fig:JupLim} and \ref{fig:SunLim}, along with our model dark matter spectra. The benchmark scattering cross sections are chosen to be much smaller than the saturation cross sections in Table~\ref{tab:objects}. These choices emphasize the sensitivities of nearby objects as dark matter detectors.

Fig.~\ref{fig:JupLim} shows Fermi-LAT upper limits on the Jupiter $\gamma$-ray flux in 60~energy bins between 10~MeV and 10~GeV.  These upper limits are analyzed by using the fully {\it data-driven} background model of Ref.~\cite{Leane:2021tjj}. The limits come with the full likelihood profiles in every bin. Unlike the Galactic center, with Jupiter, we can evaluate the $\gamma$-ray flux all the way down to 10~MeV. We also sketch differential energy spectra for two dark matter masses at 1 and 10~GeV, at the dark matter-proton cross section of $10^{-39}$~cm$^{2}$ (violet dashed line) and $10^{-38}$~cm$^{2}$(green solid line), respectively. These cross sections are smaller than the cross-section where Jupiter captures dark matter particles with an efficiency of unity by 8--9 orders of magnitude. We note that the cross-section bounds become weaker for heavier dark matter masses, demonstrating the difficulty in capturing dark matter particles with masses heavier than the nucleon targets.

Finally, we considered a celestial object that is very familiar to us in Fig.~\ref{fig:SunLim}: the Sun. We use the latest spectra and upper limits from HAWC between 200~GeV and 10~TeV~\cite{HAWC:2022khj}. The HAWC's 6.1-year spectrum (which constitutes a detection of solar $\gamma$-ray emission) is shown in the red solid lines at energies between 500~GeV to 2.7~TeV. This data is divided into two time periods, which demonstrates the variation of the emission over the solar cycle. The bright solar $\gamma$-ray emission observed during solar minimum is shown with the  blue dashed line. The strong upper limit on solar $\gamma$-ray emission during solar minimum is presented by the green scatter point. The shaded regions demonstrate the statistical uncertainty in each analysis. At higher energies, we show the 90\% CL upper limit at 7 TeV as the red bin, which spans an energy range from 3.8--10~TeV.

Similar to Jupiter in Fig.~\ref{fig:JupLim}, we sketch two dark matter spectra for dark matter-proton cross sections at $\sigma_{\chi p}=10^{-44}$~cm$^{2}$,  which is smaller than the Sun's efficient capture cross section by 6 orders of magnitude. The magenta line shows a dark matter mass of 1~TeV, while the orange line is for $m_{\chi}=10$~TeV.

Based on these spectra and upper limits, we can estimate the sensitivities of these celestial objects in different energy ranges. Below 10~GeV, brown dwarfs in the Galactic center are sensitive to dark matter-nucleon cross section near their saturation cross section, and for very large decay lengths with both NFW and gNFW profiles, while Jupiter can probe cross sections that are 4--5 orders of magnitude smaller than its geometric cross section, but only for smaller decay length. Similarly, for the energy range above 100~GeV, neutron stars can probe cross sections near their saturation cross section for a large decay length, but only for the gNFW density profile. On the other hand, the Sun can probe the cross sections that are 9 orders of magnitude smaller than its saturation cross section for short decay lengths, with an upper limit that is independent of the dark matter density profile. Notably, nearby celestial objects can be used to study dark matter capture in the optically thin limit.

\section{Constraints on the spin-independent dark matter-nucleon cross section}
\label{sect:Xsect}


We scan the dark matter-nucleon cross section for each dark matter mass. As discussed in Section~\ref{sect:model}, the spin-independent cross section is controlled by the heavy mediator and is therefore independent of the dark photon parameters $(\epsilon, m_{A^{\prime}})$. The dark photon mass and kinetic mixing instead govern only the survival probability of the mediator before it decays. We scan the dark photon mass over the 850--1022~keV range, and the kinetic mixing coupling $\epsilon$ over the range from $10^{-5}$--$10^{-4}$, which satisfy existing constraints from Refs.~\cite{McDermott:2017qcg, Caputo:2021eaa, Carenza:2025uwx, Frerick:2023xnf, Essig:2013lka}, and for each target there exists a choice within this band that yields a survival probability of order unity, as shown in Figs.~\ref{fig:PSurv} and~\ref{fig:PSurv_mA}. The cross-section limits we present therefore correspond to this maximal-signal benchmark. Because the $\gamma$-ray energy lies a factor of $\sim$2 below the dark matter mass, we scan dark matter masses above 10~MeV to remain within the energy range of our telescopes. This implies that the Lorentz boost factor is larger than 10 and increases rapidly for heavier dark matter particles. Because the dark photon mass is negligible compared to its energy, the energy spectrum we show in Eq.~(\ref{eq:dNNdE}) has an identical spectral shape when renormalized to the dark matter mass, which simplifies the analysis.

In the case of the Galactic center and solar data, observations only provide either best-fit fluxes (with 1$\sigma$ uncertainties), or $\gamma$-ray upper limits. Thus, for these targets, we perform our scan by conservatively assuming that the dark matter annihilation signal does not overproduce the total $\gamma$-ray flux (or upper limit) in any energy bin. For Jupiter, however, the data is produced along with a likelihood profile for the flux in every energy bin, allowing us to perform a more powerful analysis. Assuming the null hypothesis that there is no $\gamma$-ray flux from Jupiter, for every dark matter-nucleon cross section with a specific dark matter mass, we calculate the $\gamma$-ray spectrum, and the resulting $TS$ as
\begin{equation}
TS(\sigma_{\chi p}, m_{\chi})=-2\log\frac{\mathcal{L}_{0}}{\mathcal{L}(\sigma_{\chi p}, m_{\chi})},
\end{equation}
where $\mathcal{L}_{0}$ is the likelihood of the null hypothesis that describes no dark matter interaction with the nucleon, and $\mathcal{L}(\sigma_{\chi p}, m_{\chi})$ is likelihood of a dark matter signal. To find the bounds using this method, we scanned the parameter space to find the highest cross section that can give the $TS\simeq \chi^{2}=3.841$, which is equivalent to the 95\% CL upper limit. With this method, we can provide strong bounds on the cross section with dark matter masses down to 20~MeV.

\begin{figure}[btp]
\centering
\includegraphics[width=0.47\columnwidth]{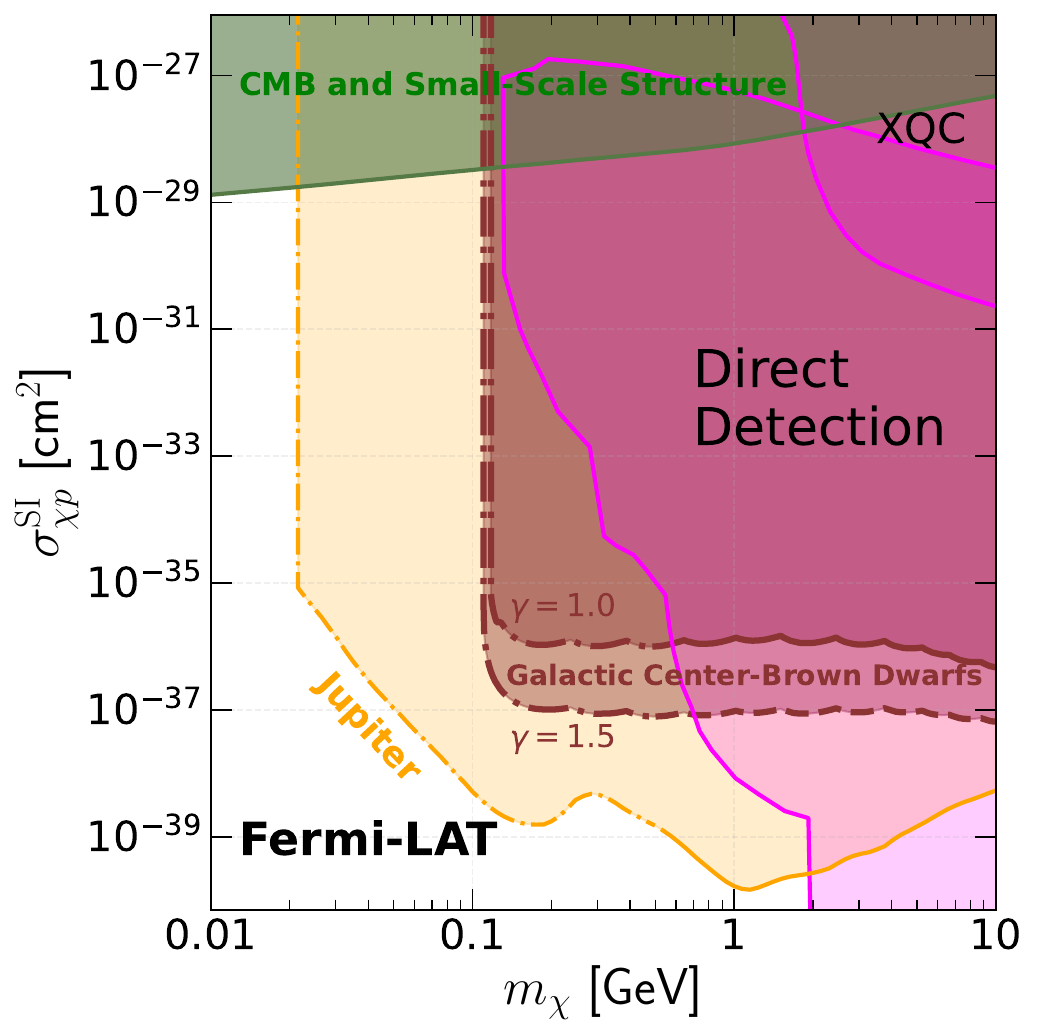}
\hfill
\includegraphics[width=0.47\columnwidth]{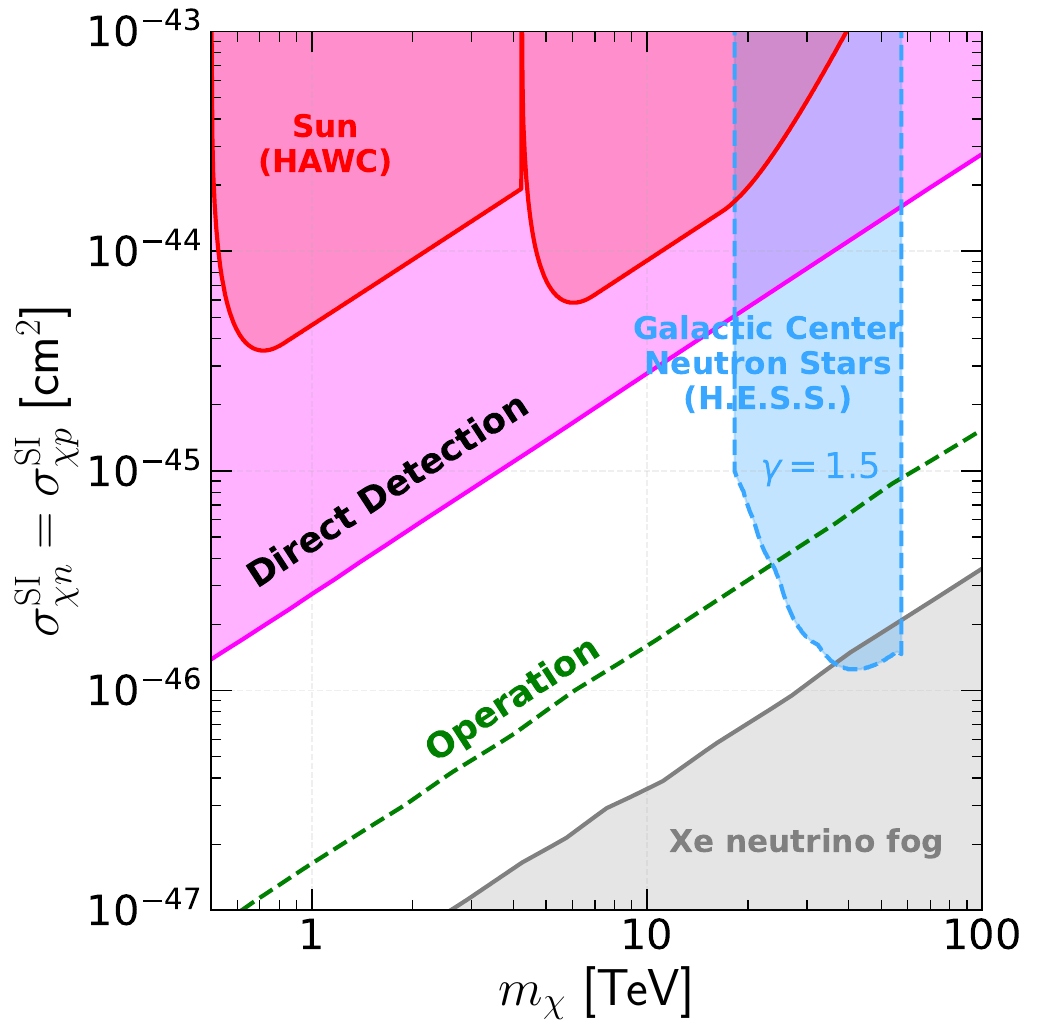}
\caption{Dark matter-nucleon cross section bounds from celestial body searches, compared to other current bounds from Ref.~\cite{Chou:2022luk}. {\bf Left}: Bounds in the 10~MeV to 10~GeV dark matter mass range using Fermi-LAT data. The red-violet solid and dashed lines are for brown dwarfs near the Galactic center, with gNFW profiles that have $\gamma=1.0$ and 1.5 respectively. The orange line is the bound calculated from observations of Jupiter. Below 500~MeV these bounds are shown as dashed-dotted-lines, since these objects may have significant dark matter evaporation~\cite{Garani:2021feo, Ilie:2023lbi}. We include the bound coming from CMB and Small Scale Structure in green, as well as direct detection and XQC limits in magenta shaded. {\bf Right}: Bounds for 500~GeV to 100~TeV dark matter. The cross-section upper limits using HAWC solar observations are the red solid lines. The H.E.S.S. bound for all neutron stars near the Galactic center, assuming the gNFW profile with $\gamma=1.5$, is the light-blue dashed line. The Xenon neutrino fog region is the gray area, along with both the current bound from direct detection magenta shaded), and the projection from currently operating direct detection experiments (green dashed line) (LZ, XENONnT, PandaX-4T, SuperCDMS SNOLAB, SBC)~\cite{Akerib:2022ort}.}
\label{fig:bound}
\end{figure}

We show the resulting cross-section upper limits in Fig.~\ref{fig:bound}. For comparison, we include the current bounds from direct detection~\cite{DEAP:2019yzn} as magenta lines from a combination of the results from XENON~\cite{XENON:2020gfr}, COSINE-100~\cite{Adhikari:2018ljm}, CRESST-III~\cite{CRESST:2019jnq}, DAMA/LIBRA~\cite{Bernabei:2018yyw}, and DarkSide~\cite{DarkSide:2018bpj}. We also include bounds from the X-ray Quantum Calorimeter (XQC)~\cite{Erickcek:2007jv}, as well as constraints from CMB and Small Scale Structure in green solid line~\cite{Gluscevic:2017ywp, Boddy:2018wzy, DES:2020fxi, Rogers:2021byl}. For TeV dark matter (right), we show the projection from the operating experiments such as LZ~\cite{LZ:2022lsv}, XENONnT~\cite{XENON:2020kmp}, PandaX-4T~\cite{PandaX-4T:2021bab}, SuperCDMS SNOLAB~\cite{SuperCDMS:2018gro}, SBC~\cite{Alfonso-Pita:2022akn}. The Xenon neutrino fog region, taken from Ref. \cite{Akerib:2022ort}, is shown in the gray area.

For low-mass dark matter (left), we show constraints based on all brown dwarfs near the Galactic center red-violet lines, in the case of a classical NFW profile ($\gamma=1.0$, solid), and for the gNFW with $\gamma=1.5$ (dashed).
Due to the large uncertainty in the dark matter profile, the actual bound is likely to lie in between these two curves, from $10^{-35}$ down to $10^{-37}$~cm$^{2}$. This shows that brown dwarfs can probe cross sections smaller than the bounds from direct detection by 4--6 orders of magnitude for dark matter masses near 100~MeV, and 6--9 orders of magnitude smaller than the CMB and Small Scale Structure bounds from 10~MeV to 10~GeV. On the other hand, Jupiter (orange line) can provide cross-section bounds that are 2--5 orders of magnitude stronger than brown dwarf limits and extend down to nearly 10~MeV. Note that since brown dwarfs and Jupiter have an evaporation limit around 500~MeV~\cite{Garani:2021feo, Ilie:2023lbi}, below this limit we present the bounds as dashed-dotted-lines.  The bounds above the evaporation limit are still 1 order of magnitude stronger than the direct detection constraints.

Comparing the results, we see that Jupiter provides cross-section bounds that are stronger and with a wider dark matter mass range than brown dwarfs near the Galactic center. Bounds from Jupiter also do not suffer from the high uncertainty coming from the dark matter density profile like brown dwarfs. On the other hand, since we require the dark photon to decay before reaching Earth from the annihilation source, brown dwarfs can probe much larger decay lengths, opening more parameter space for the kinetic mixing coupling and the dark photon mass, compared to Jupiter.

Moving to the heavier dark matter models in Fig.~\ref{fig:bound} (right), we show constraints from the Sun (red line) with HAWC, along with neutron stars near the Galactic center (light-blue dashed line) using H.E.S.S. Bounds from the Sun are already excluded by direct detection since spin-independent cross-section constraints in this regime are very stringent. For neutron stars, as shown in Fig.~\ref{fig:GCFlux} (right), the classical NFW profile cannot produce strong enough photon signals to be probed by the H.E.S.S. data. Thus, we only show the constraint for the gNFW profile with $\gamma=1.5$, On the other hand, the large optical depth of neutron stars can push the cross-section upper limit down to $10^{-46}$~cm$^{2}$ up to 70~TeV dark matter masses, surpassing the projection from current direct detection experiments and even touching the neutrino fog. This result showcases the advantage of using astrophysical objects and signals to investigate dark matter-nucleon interactions in the high-mass regime, which requires new technology and strategies in direct detection experiments.

These results also highlight the importance of $\gamma$-ray observations to study the dark matter-nucleon interaction in the dark photon mediator model. Previous studies of this model in Ref.~\cite{Nguyen:2022zwb} studied the neutrino signal, which can only probe the heavy dark matter regime where the dark matter mass is heavier than 100~GeV. Since $\gamma$-ray observations can probe signals down to nearly 10~MeV, studying the dark photon-photon trident decay can probe this model in the sub-GeV regime. Moreover, more accurate background models can help push down the cross-section bounds.

\section{Conclusions and Outlook}
\label{sect:final}

In this paper, we have studied the indirect detection signal from {\it dark photon-photon tridents}, which is the three-photon final state of a dark photon decay. We derived the semi-analytic formula for the photon energy spectrum from a boosted dark photon in Eq.~(\ref{eq:spectrum3body_boost}). We applied this formula to the case where this dark photon can be long-lived and thus be produced by the annihilation of accumulated dark matter inside celestial objects, such as neutron stars, brown dwarfs, Jupiter, and the Sun. Using current $\gamma$-ray observations from the Fermi-LAT, H.E.S.S., and HAWC, we set strong bounds on the spin-independent cross section of dark matter with nucleons for various dark matter mass ranges in Fig.~\ref{fig:bound}.

These upper limits, which exceed those of direct detection experiments in several regimes, motivate future $\gamma$-ray observations using CTA~\cite{Keith:2022xbd}, APT~\cite{Xu:2023zyz}, AMEGO~\cite{Kierans_2020}, and e-ASTROGAM~\cite{e-ASTROGAM:2017pxr} to improve the dark matter sensitivity and extend the dark matter mass range. The astrophysical uncertainty in these results can be reduced using more complicated Milky Way Halo mass models~\cite{Foster:2021ngm}, as well as more accurate stellar population models based on future observations from JWST~\cite{Schoedel:2023ott} and Euclid~\cite{Euclid:2023jih}.

Finally, while this study focuses on scenarios where the dark photon is long-lived, the dark photon-photon trident can also be considered in generic indirect detection searches~\cite{Redondo:2008ec, Krnjaic:2022wor}. Because the final state only contains photons, the three-photon spectrum can be strongly constrained by indirect searches for dark matter annihilation and dark matter decay, which has been used to further study the properties of dark photon dark matter~\cite{Linden:2024fby}.

\acknowledgments

We would like to thank Chris Cappiello, Ciaran O'Hare, Kenny Ng, Martin Spinrath, Sebastian Trojanowski and Yi-Ming Zhong, for fruitful discussions during the production of this paper. We would also like to thank Chris Cappiello, Rebecca Leane, and Gordan Krnjaic for comments on the draft. Special thanks to Felix Kahlhoefer for pointing out the Euler-Heisenberg dark photon Lagrangian, Juri Smirnov for helping with the {\tt Asteria} package, and Le Minh Ngoc for helping with Fig.~\ref{fig:DPhotonescape}. TTQN and TL are supported in part by the Swedish Research Council under contract 2022-04283. TTQN is also supported as a research assistant by Prof. Tzu-Chiang Yuan's NSTC grant No. 111-2112-M-001-035. TL is also supported by the Swedish National Space Agency under contract 117/19. The work of TMPT is supported in part by the U.S.\ National Science Foundation under Grant PHY-2210283.  TTQN would like to thank NCTS-Hsinchu Hub and National Tsing Hua University, Taiwan for their hospitality during the progress of this work.

This work made use of {\tt Numpy}~\cite{Harris_2020}, {\tt SciPy}~\cite{Virtanen:2019joe}, {\tt matplotlib}~\cite{HunterMatplotlib}, {\tt Jupyter}~\cite{2016ppap.book...87K}, {\tt Jaxodraw}~\cite{Binosi:2008ig}, as well as Webplotdigitizer~\cite{Rohatgi2022}.




\bibliography{biblio.bib}

\begin{thebibliography}{143}%
\makeatletter
\providecommand \@ifxundefined [1]{%
 \@ifx{#1\undefined}
}%
\providecommand \@ifnum [1]{%
 \ifnum #1\expandafter \@firstoftwo
 \else \expandafter \@secondoftwo
 \fi
}%
\providecommand \@ifx [1]{%
 \ifx #1\expandafter \@firstoftwo
 \else \expandafter \@secondoftwo
 \fi
}%
\providecommand \natexlab [1]{#1}%
\providecommand \enquote  [1]{``#1''}%
\providecommand \bibnamefont  [1]{#1}%
\providecommand \bibfnamefont [1]{#1}%
\providecommand \citenamefont [1]{#1}%
\providecommand \href@noop [0]{\@secondoftwo}%
\providecommand \href [0]{\begingroup \@sanitize@url \@href}%
\providecommand \@href[1]{\@@startlink{#1}\@@href}%
\providecommand \@@href[1]{\endgroup#1\@@endlink}%
\providecommand \@sanitize@url [0]{\catcode `\\12\catcode `\$12\catcode `\&12\catcode `\#12\catcode `\^12\catcode `\_12\catcode `\%12\relax}%
\providecommand \@@startlink[1]{}%
\providecommand \@@endlink[0]{}%
\providecommand \url  [0]{\begingroup\@sanitize@url \@url }%
\providecommand \@url [1]{\endgroup\@href {#1}{\urlprefix }}%
\providecommand \urlprefix  [0]{URL }%
\providecommand \Eprint [0]{\href }%
\providecommand \doibase [0]{http://dx.doi.org/}%
\providecommand \selectlanguage [0]{\@gobble}%
\providecommand \bibinfo  [0]{\@secondoftwo}%
\providecommand \bibfield  [0]{\@secondoftwo}%
\providecommand \translation [1]{[#1]}%
\providecommand \BibitemOpen [0]{}%
\providecommand \bibitemStop [0]{}%
\providecommand \bibitemNoStop [0]{.\EOS\space}%
\providecommand \EOS [0]{\spacefactor3000\relax}%
\providecommand \BibitemShut  [1]{\csname bibitem#1\endcsname}%
\let\auto@bib@innerbib\@empty
\bibitem [{\citenamefont {Bertone}\ \emph {et~al.}(2005)\citenamefont {Bertone}, \citenamefont {Hooper},\ and\ \citenamefont {Silk}}]{Bertone:2004pz}%
  \BibitemOpen
  \bibfield  {author} {\bibinfo {author} {\bibfnamefont {G.}~\bibnamefont {Bertone}}, \bibinfo {author} {\bibfnamefont {D.}~\bibnamefont {Hooper}}, \ and\ \bibinfo {author} {\bibfnamefont {J.}~\bibnamefont {Silk}},\ }\href {\doibase 10.1016/j.physrep.2004.08.031} {\bibfield  {journal} {\bibinfo  {journal} {Phys. Rept.}\ }\textbf {\bibinfo {volume} {405}},\ \bibinfo {pages} {279} (\bibinfo {year} {2005})},\ \Eprint {http://arxiv.org/abs/hep-ph/0404175} {arXiv:hep-ph/0404175} \BibitemShut {NoStop}%
\bibitem [{\citenamefont {Bertone}\ and\ \citenamefont {Tait}(2018)}]{Bertone:2018krk}%
  \BibitemOpen
  \bibfield  {author} {\bibinfo {author} {\bibfnamefont {G.}~\bibnamefont {Bertone}}\ and\ \bibinfo {author} {\bibfnamefont {T.}~\bibnamefont {Tait}, \bibfnamefont {M.~P.}},\ }\href {\doibase 10.1038/s41586-018-0542-z} {\bibfield  {journal} {\bibinfo  {journal} {Nature}\ }\textbf {\bibinfo {volume} {562}},\ \bibinfo {pages} {51} (\bibinfo {year} {2018})},\ \Eprint {http://arxiv.org/abs/1810.01668} {arXiv:1810.01668 [astro-ph.CO]} \BibitemShut {NoStop}%
\bibitem [{\citenamefont {Drlica-Wagner}\ \emph {et~al.}(2022)\citenamefont {Drlica-Wagner} \emph {et~al.}}]{Drlica-Wagner:2022lbd}%
  \BibitemOpen
  \bibfield  {author} {\bibinfo {author} {\bibfnamefont {A.}~\bibnamefont {Drlica-Wagner}} \emph {et~al.},\ }\href@noop {} {\  (\bibinfo {year} {2022})},\ \Eprint {http://arxiv.org/abs/2209.08215} {arXiv:2209.08215 [hep-ph]} \BibitemShut {NoStop}%
\bibitem [{\citenamefont {Chou}\ \emph {et~al.}(2022)\citenamefont {Chou} \emph {et~al.}}]{Chou:2022luk}%
  \BibitemOpen
  \bibfield  {author} {\bibinfo {author} {\bibfnamefont {A.~S.}\ \bibnamefont {Chou}} \emph {et~al.},\ }in\ \href@noop {} {\emph {\bibinfo {booktitle} {{Snowmass 2021}}}}\ (\bibinfo {year} {2022})\ \Eprint {http://arxiv.org/abs/2211.09978} {arXiv:2211.09978 [hep-ex]} \BibitemShut {NoStop}%
\bibitem [{\citenamefont {Cooley}\ \emph {et~al.}(2022)\citenamefont {Cooley} \emph {et~al.}}]{Cooley:2022ufh}%
  \BibitemOpen
  \bibfield  {author} {\bibinfo {author} {\bibfnamefont {J.}~\bibnamefont {Cooley}} \emph {et~al.},\ }\href@noop {} {\  (\bibinfo {year} {2022})},\ \Eprint {http://arxiv.org/abs/2209.07426} {arXiv:2209.07426 [hep-ph]} \BibitemShut {NoStop}%
\bibitem [{\citenamefont {Slatyer}(2018)}]{Slatyer:2017sev}%
  \BibitemOpen
  \bibfield  {author} {\bibinfo {author} {\bibfnamefont {T.~R.}\ \bibnamefont {Slatyer}},\ }in\ \href {\doibase 10.1142/9789813233348_0005} {\emph {\bibinfo {booktitle} {{Theoretical Advanced Study Institute in Elementary Particle Physics}: {Anticipating the Next Discoveries in Particle Physics}}}}\ (\bibinfo {year} {2018})\ pp.\ \bibinfo {pages} {297--353},\ \Eprint {http://arxiv.org/abs/1710.05137} {arXiv:1710.05137 [hep-ph]} \BibitemShut {NoStop}%
\bibitem [{\citenamefont {Slatyer}(2022)}]{Slatyer:2021qgc}%
  \BibitemOpen
  \bibfield  {author} {\bibinfo {author} {\bibfnamefont {T.~R.}\ \bibnamefont {Slatyer}},\ }\href {\doibase 10.21468/SciPostPhysLectNotes.53} {\bibfield  {journal} {\bibinfo  {journal} {SciPost Phys. Lect. Notes}\ }\textbf {\bibinfo {volume} {53}},\ \bibinfo {pages} {1} (\bibinfo {year} {2022})},\ \Eprint {http://arxiv.org/abs/2109.02696} {arXiv:2109.02696 [hep-ph]} \BibitemShut {NoStop}%
\bibitem [{\citenamefont {De~la Torre~Luque}\ \emph {et~al.}(2026)\citenamefont {De~la Torre~Luque}, \citenamefont {Carenza},\ and\ \citenamefont {Nguyen}}]{DelaTorreLuque:2025zjt}%
  \BibitemOpen
  \bibfield  {author} {\bibinfo {author} {\bibfnamefont {P.}~\bibnamefont {De~la Torre~Luque}}, \bibinfo {author} {\bibfnamefont {P.}~\bibnamefont {Carenza}}, \ and\ \bibinfo {author} {\bibfnamefont {T.~T.~Q.}\ \bibnamefont {Nguyen}},\ }\href {\doibase 10.1103/qmw9-1k4k} {\bibfield  {journal} {\bibinfo  {journal} {Phys. Rev. D}\ }\textbf {\bibinfo {volume} {113}},\ \bibinfo {pages} {063035} (\bibinfo {year} {2026})},\ \Eprint {http://arxiv.org/abs/2507.01962} {arXiv:2507.01962 [hep-ph]} \BibitemShut {NoStop}%
\bibitem [{\citenamefont {Baryakhtar}\ \emph {et~al.}(2022)\citenamefont {Baryakhtar} \emph {et~al.}}]{Baryakhtar:2022hbu}%
  \BibitemOpen
  \bibfield  {author} {\bibinfo {author} {\bibfnamefont {M.}~\bibnamefont {Baryakhtar}} \emph {et~al.},\ }in\ \href@noop {} {\emph {\bibinfo {booktitle} {{Snowmass 2021}}}}\ (\bibinfo {year} {2022})\ \Eprint {http://arxiv.org/abs/2203.07984} {arXiv:2203.07984 [hep-ph]} \BibitemShut {NoStop}%
\bibitem [{\citenamefont {Nguyen}\ \emph {et~al.}(2026{\natexlab{a}})\citenamefont {Nguyen}, \citenamefont {John}, \citenamefont {Linden},\ and\ \citenamefont {Tait}}]{Nguyen:2024kwy}%
  \BibitemOpen
  \bibfield  {author} {\bibinfo {author} {\bibfnamefont {T.~T.~Q.}\ \bibnamefont {Nguyen}}, \bibinfo {author} {\bibfnamefont {I.}~\bibnamefont {John}}, \bibinfo {author} {\bibfnamefont {T.}~\bibnamefont {Linden}}, \ and\ \bibinfo {author} {\bibfnamefont {T.~M.~P.}\ \bibnamefont {Tait}},\ }\href {\doibase 10.1103/5wrw-7dz8} {\bibfield  {journal} {\bibinfo  {journal} {Phys. Rev. D}\ }\textbf {\bibinfo {volume} {113}},\ \bibinfo {pages} {103051} (\bibinfo {year} {2026}{\natexlab{a}})},\ \Eprint {http://arxiv.org/abs/2412.00180} {arXiv:2412.00180 [hep-ph]} \BibitemShut {NoStop}%
\bibitem [{\citenamefont {Nguyen}\ \emph {et~al.}(2026{\natexlab{b}})\citenamefont {Nguyen}, \citenamefont {De~la Torre~Luque}, \citenamefont {John}, \citenamefont {Balaji}, \citenamefont {Carenza},\ and\ \citenamefont {Linden}}]{Nguyen:2025tkl}%
  \BibitemOpen
  \bibfield  {author} {\bibinfo {author} {\bibfnamefont {T.~T.~Q.}\ \bibnamefont {Nguyen}}, \bibinfo {author} {\bibfnamefont {P.}~\bibnamefont {De~la Torre~Luque}}, \bibinfo {author} {\bibfnamefont {I.}~\bibnamefont {John}}, \bibinfo {author} {\bibfnamefont {S.}~\bibnamefont {Balaji}}, \bibinfo {author} {\bibfnamefont {P.}~\bibnamefont {Carenza}}, \ and\ \bibinfo {author} {\bibfnamefont {T.}~\bibnamefont {Linden}},\ }\href {\doibase 10.1103/fn6k-1nlc} {\bibfield  {journal} {\bibinfo  {journal} {Phys. Rev. D}\ }\textbf {\bibinfo {volume} {113}},\ \bibinfo {pages} {103010} (\bibinfo {year} {2026}{\natexlab{b}})},\ \Eprint {http://arxiv.org/abs/2507.13432} {arXiv:2507.13432 [hep-ph]} \BibitemShut {NoStop}%
\bibitem [{\citenamefont {Boddy}\ \emph {et~al.}(2022)\citenamefont {Boddy} \emph {et~al.}}]{Boddy:2022knd}%
  \BibitemOpen
  \bibfield  {author} {\bibinfo {author} {\bibfnamefont {K.~K.}\ \bibnamefont {Boddy}} \emph {et~al.},\ }\href {\doibase 10.1016/j.jheap.2022.06.005} {\bibfield  {journal} {\bibinfo  {journal} {JHEAp}\ }\textbf {\bibinfo {volume} {35}},\ \bibinfo {pages} {112} (\bibinfo {year} {2022})},\ \Eprint {http://arxiv.org/abs/2203.06380} {arXiv:2203.06380 [hep-ph]} \BibitemShut {NoStop}%
\bibitem [{\citenamefont {Bramante}\ and\ \citenamefont {Raj}(2024)}]{Bramante:2023djs}%
  \BibitemOpen
  \bibfield  {author} {\bibinfo {author} {\bibfnamefont {J.}~\bibnamefont {Bramante}}\ and\ \bibinfo {author} {\bibfnamefont {N.}~\bibnamefont {Raj}},\ }\href {\doibase 10.1016/j.physrep.2023.12.001} {\bibfield  {journal} {\bibinfo  {journal} {Phys. Rept.}\ }\textbf {\bibinfo {volume} {1052}},\ \bibinfo {pages} {1} (\bibinfo {year} {2024})},\ \Eprint {http://arxiv.org/abs/2307.14435} {arXiv:2307.14435 [hep-ph]} \BibitemShut {NoStop}%
\bibitem [{\citenamefont {Bertone}\ and\ \citenamefont {Fairbairn}(2008)}]{Bertone:2007ae}%
  \BibitemOpen
  \bibfield  {author} {\bibinfo {author} {\bibfnamefont {G.}~\bibnamefont {Bertone}}\ and\ \bibinfo {author} {\bibfnamefont {M.}~\bibnamefont {Fairbairn}},\ }\href {\doibase 10.1103/PhysRevD.77.043515} {\bibfield  {journal} {\bibinfo  {journal} {Phys. Rev. D}\ }\textbf {\bibinfo {volume} {77}},\ \bibinfo {pages} {043515} (\bibinfo {year} {2008})},\ \Eprint {http://arxiv.org/abs/0709.1485} {arXiv:0709.1485 [astro-ph]} \BibitemShut {NoStop}%
\bibitem [{\citenamefont {Goldman}\ and\ \citenamefont {Nussinov}(1989)}]{Goldman:1989nd}%
  \BibitemOpen
  \bibfield  {author} {\bibinfo {author} {\bibfnamefont {I.}~\bibnamefont {Goldman}}\ and\ \bibinfo {author} {\bibfnamefont {S.}~\bibnamefont {Nussinov}},\ }\href {\doibase 10.1103/PhysRevD.40.3221} {\bibfield  {journal} {\bibinfo  {journal} {Phys. Rev. D}\ }\textbf {\bibinfo {volume} {40}},\ \bibinfo {pages} {3221} (\bibinfo {year} {1989})}\BibitemShut {NoStop}%
\bibitem [{\citenamefont {Bauswein}\ \emph {et~al.}(2023)\citenamefont {Bauswein}, \citenamefont {Guo}, \citenamefont {Lien}, \citenamefont {Lin},\ and\ \citenamefont {Wu}}]{Bauswein:2020kor}%
  \BibitemOpen
  \bibfield  {author} {\bibinfo {author} {\bibfnamefont {A.}~\bibnamefont {Bauswein}}, \bibinfo {author} {\bibfnamefont {G.}~\bibnamefont {Guo}}, \bibinfo {author} {\bibfnamefont {J.-H.}\ \bibnamefont {Lien}}, \bibinfo {author} {\bibfnamefont {Y.-H.}\ \bibnamefont {Lin}}, \ and\ \bibinfo {author} {\bibfnamefont {M.-R.}\ \bibnamefont {Wu}},\ }\href {\doibase 10.1103/PhysRevD.107.083002} {\bibfield  {journal} {\bibinfo  {journal} {Phys. Rev. D}\ }\textbf {\bibinfo {volume} {107}},\ \bibinfo {pages} {083002} (\bibinfo {year} {2023})},\ \Eprint {http://arxiv.org/abs/2012.11908} {arXiv:2012.11908 [astro-ph.HE]} \BibitemShut {NoStop}%
\bibitem [{\citenamefont {Kouvaris}(2008)}]{Kouvaris:2007ay}%
  \BibitemOpen
  \bibfield  {author} {\bibinfo {author} {\bibfnamefont {C.}~\bibnamefont {Kouvaris}},\ }\href {\doibase 10.1103/PhysRevD.77.023006} {\bibfield  {journal} {\bibinfo  {journal} {Phys. Rev. D}\ }\textbf {\bibinfo {volume} {77}},\ \bibinfo {pages} {023006} (\bibinfo {year} {2008})},\ \Eprint {http://arxiv.org/abs/0708.2362} {arXiv:0708.2362 [astro-ph]} \BibitemShut {NoStop}%
\bibitem [{\citenamefont {de~Lavallaz}\ and\ \citenamefont {Fairbairn}(2010)}]{deLavallaz:2010wp}%
  \BibitemOpen
  \bibfield  {author} {\bibinfo {author} {\bibfnamefont {A.}~\bibnamefont {de~Lavallaz}}\ and\ \bibinfo {author} {\bibfnamefont {M.}~\bibnamefont {Fairbairn}},\ }\href {\doibase 10.1103/PhysRevD.81.123521} {\bibfield  {journal} {\bibinfo  {journal} {Phys. Rev. D}\ }\textbf {\bibinfo {volume} {81}},\ \bibinfo {pages} {123521} (\bibinfo {year} {2010})},\ \Eprint {http://arxiv.org/abs/1004.0629} {arXiv:1004.0629 [astro-ph.GA]} \BibitemShut {NoStop}%
\bibitem [{\citenamefont {Kouvaris}\ and\ \citenamefont {Tinyakov}(2010)}]{Kouvaris:2010vv}%
  \BibitemOpen
  \bibfield  {author} {\bibinfo {author} {\bibfnamefont {C.}~\bibnamefont {Kouvaris}}\ and\ \bibinfo {author} {\bibfnamefont {P.}~\bibnamefont {Tinyakov}},\ }\href {\doibase 10.1103/PhysRevD.82.063531} {\bibfield  {journal} {\bibinfo  {journal} {Phys. Rev. D}\ }\textbf {\bibinfo {volume} {82}},\ \bibinfo {pages} {063531} (\bibinfo {year} {2010})},\ \Eprint {http://arxiv.org/abs/1004.0586} {arXiv:1004.0586 [astro-ph.GA]} \BibitemShut {NoStop}%
\bibitem [{\citenamefont {Bell}\ \emph {et~al.}(2013)\citenamefont {Bell}, \citenamefont {Melatos},\ and\ \citenamefont {Petraki}}]{Bell:2013xk}%
  \BibitemOpen
  \bibfield  {author} {\bibinfo {author} {\bibfnamefont {N.~F.}\ \bibnamefont {Bell}}, \bibinfo {author} {\bibfnamefont {A.}~\bibnamefont {Melatos}}, \ and\ \bibinfo {author} {\bibfnamefont {K.}~\bibnamefont {Petraki}},\ }\href {\doibase 10.1103/PhysRevD.87.123507} {\bibfield  {journal} {\bibinfo  {journal} {Phys. Rev. D}\ }\textbf {\bibinfo {volume} {87}},\ \bibinfo {pages} {123507} (\bibinfo {year} {2013})},\ \Eprint {http://arxiv.org/abs/1301.6811} {arXiv:1301.6811 [hep-ph]} \BibitemShut {NoStop}%
\bibitem [{\citenamefont {G\"uver}\ \emph {et~al.}(2014)\citenamefont {G\"uver}, \citenamefont {Erkoca}, \citenamefont {Hall~Reno},\ and\ \citenamefont {Sarcevic}}]{Guver:2012ba}%
  \BibitemOpen
  \bibfield  {author} {\bibinfo {author} {\bibfnamefont {T.}~\bibnamefont {G\"uver}}, \bibinfo {author} {\bibfnamefont {A.~E.}\ \bibnamefont {Erkoca}}, \bibinfo {author} {\bibfnamefont {M.}~\bibnamefont {Hall~Reno}}, \ and\ \bibinfo {author} {\bibfnamefont {I.}~\bibnamefont {Sarcevic}},\ }\href {\doibase 10.1088/1475-7516/2014/05/013} {\bibfield  {journal} {\bibinfo  {journal} {JCAP}\ }\textbf {\bibinfo {volume} {05}},\ \bibinfo {pages} {013} (\bibinfo {year} {2014})},\ \Eprint {http://arxiv.org/abs/1201.2400} {arXiv:1201.2400 [hep-ph]} \BibitemShut {NoStop}%
\bibitem [{\citenamefont {Baryakhtar}\ \emph {et~al.}(2017)\citenamefont {Baryakhtar}, \citenamefont {Bramante}, \citenamefont {Li}, \citenamefont {Linden},\ and\ \citenamefont {Raj}}]{Baryakhtar:2017dbj}%
  \BibitemOpen
  \bibfield  {author} {\bibinfo {author} {\bibfnamefont {M.}~\bibnamefont {Baryakhtar}}, \bibinfo {author} {\bibfnamefont {J.}~\bibnamefont {Bramante}}, \bibinfo {author} {\bibfnamefont {S.~W.}\ \bibnamefont {Li}}, \bibinfo {author} {\bibfnamefont {T.}~\bibnamefont {Linden}}, \ and\ \bibinfo {author} {\bibfnamefont {N.}~\bibnamefont {Raj}},\ }\href {\doibase 10.1103/PhysRevLett.119.131801} {\bibfield  {journal} {\bibinfo  {journal} {Phys. Rev. Lett.}\ }\textbf {\bibinfo {volume} {119}},\ \bibinfo {pages} {131801} (\bibinfo {year} {2017})},\ \Eprint {http://arxiv.org/abs/1704.01577} {arXiv:1704.01577 [hep-ph]} \BibitemShut {NoStop}%
\bibitem [{\citenamefont {Bell}\ \emph {et~al.}(2018)\citenamefont {Bell}, \citenamefont {Busoni},\ and\ \citenamefont {Robles}}]{Bell:2018pkk}%
  \BibitemOpen
  \bibfield  {author} {\bibinfo {author} {\bibfnamefont {N.~F.}\ \bibnamefont {Bell}}, \bibinfo {author} {\bibfnamefont {G.}~\bibnamefont {Busoni}}, \ and\ \bibinfo {author} {\bibfnamefont {S.}~\bibnamefont {Robles}},\ }\href {\doibase 10.1088/1475-7516/2018/09/018} {\bibfield  {journal} {\bibinfo  {journal} {JCAP}\ }\textbf {\bibinfo {volume} {09}},\ \bibinfo {pages} {018} (\bibinfo {year} {2018})},\ \Eprint {http://arxiv.org/abs/1807.02840} {arXiv:1807.02840 [hep-ph]} \BibitemShut {NoStop}%
\bibitem [{\citenamefont {Garani}\ \emph {et~al.}(2019)\citenamefont {Garani}, \citenamefont {Genolini},\ and\ \citenamefont {Hambye}}]{Garani:2018kkd}%
  \BibitemOpen
  \bibfield  {author} {\bibinfo {author} {\bibfnamefont {R.}~\bibnamefont {Garani}}, \bibinfo {author} {\bibfnamefont {Y.}~\bibnamefont {Genolini}}, \ and\ \bibinfo {author} {\bibfnamefont {T.}~\bibnamefont {Hambye}},\ }\href {\doibase 10.1088/1475-7516/2019/05/035} {\bibfield  {journal} {\bibinfo  {journal} {JCAP}\ }\textbf {\bibinfo {volume} {05}},\ \bibinfo {pages} {035} (\bibinfo {year} {2019})},\ \Eprint {http://arxiv.org/abs/1812.08773} {arXiv:1812.08773 [hep-ph]} \BibitemShut {NoStop}%
\bibitem [{\citenamefont {Liang}\ and\ \citenamefont {Shao}(2023)}]{Liang:2023nvo}%
  \BibitemOpen
  \bibfield  {author} {\bibinfo {author} {\bibfnamefont {D.}~\bibnamefont {Liang}}\ and\ \bibinfo {author} {\bibfnamefont {L.}~\bibnamefont {Shao}},\ }\href {\doibase 10.1088/1475-7516/2023/08/016} {\bibfield  {journal} {\bibinfo  {journal} {JCAP}\ }\textbf {\bibinfo {volume} {08}},\ \bibinfo {pages} {016} (\bibinfo {year} {2023})},\ \Eprint {http://arxiv.org/abs/2303.05107} {arXiv:2303.05107 [astro-ph.HE]} \BibitemShut {NoStop}%
\bibitem [{\citenamefont {Chen}\ and\ \citenamefont {Lin}(2018)}]{Chen:2018ohx}%
  \BibitemOpen
  \bibfield  {author} {\bibinfo {author} {\bibfnamefont {C.-S.}\ \bibnamefont {Chen}}\ and\ \bibinfo {author} {\bibfnamefont {Y.-H.}\ \bibnamefont {Lin}},\ }\href {\doibase 10.1007/JHEP08(2018)069} {\bibfield  {journal} {\bibinfo  {journal} {JHEP}\ }\textbf {\bibinfo {volume} {08}},\ \bibinfo {pages} {069} (\bibinfo {year} {2018})},\ \Eprint {http://arxiv.org/abs/1804.03409} {arXiv:1804.03409 [hep-ph]} \BibitemShut {NoStop}%
\bibitem [{\citenamefont {Bell}\ \emph {et~al.}(2023)\citenamefont {Bell}, \citenamefont {Busoni}, \citenamefont {Robles},\ and\ \citenamefont {Virgato}}]{Bell:2023ysh}%
  \BibitemOpen
  \bibfield  {author} {\bibinfo {author} {\bibfnamefont {N.~F.}\ \bibnamefont {Bell}}, \bibinfo {author} {\bibfnamefont {G.}~\bibnamefont {Busoni}}, \bibinfo {author} {\bibfnamefont {S.}~\bibnamefont {Robles}}, \ and\ \bibinfo {author} {\bibfnamefont {M.}~\bibnamefont {Virgato}},\ }\href@noop {} {\  (\bibinfo {year} {2023})},\ \Eprint {http://arxiv.org/abs/2312.11892} {arXiv:2312.11892 [hep-ph]} \BibitemShut {NoStop}%
\bibitem [{\citenamefont {R\"uter}\ \emph {et~al.}(2023)\citenamefont {R\"uter}, \citenamefont {Sagun}, \citenamefont {Tichy},\ and\ \citenamefont {Dietrich}}]{Ruter:2023uzc}%
  \BibitemOpen
  \bibfield  {author} {\bibinfo {author} {\bibfnamefont {H.~R.}\ \bibnamefont {R\"uter}}, \bibinfo {author} {\bibfnamefont {V.}~\bibnamefont {Sagun}}, \bibinfo {author} {\bibfnamefont {W.}~\bibnamefont {Tichy}}, \ and\ \bibinfo {author} {\bibfnamefont {T.}~\bibnamefont {Dietrich}},\ }\href {\doibase 10.1103/PhysRevD.108.124080} {\bibfield  {journal} {\bibinfo  {journal} {Phys. Rev. D}\ }\textbf {\bibinfo {volume} {108}},\ \bibinfo {pages} {124080} (\bibinfo {year} {2023})},\ \Eprint {http://arxiv.org/abs/2301.03568} {arXiv:2301.03568 [gr-qc]} \BibitemShut {NoStop}%
\bibitem [{\citenamefont {Hamaguchi}\ \emph {et~al.}(2019)\citenamefont {Hamaguchi}, \citenamefont {Nagata},\ and\ \citenamefont {Yanagi}}]{Hamaguchi:2019oev}%
  \BibitemOpen
  \bibfield  {author} {\bibinfo {author} {\bibfnamefont {K.}~\bibnamefont {Hamaguchi}}, \bibinfo {author} {\bibfnamefont {N.}~\bibnamefont {Nagata}}, \ and\ \bibinfo {author} {\bibfnamefont {K.}~\bibnamefont {Yanagi}},\ }\href {\doibase 10.1016/j.physletb.2019.06.060} {\bibfield  {journal} {\bibinfo  {journal} {Phys. Lett. B}\ }\textbf {\bibinfo {volume} {795}},\ \bibinfo {pages} {484} (\bibinfo {year} {2019})},\ \Eprint {http://arxiv.org/abs/1905.02991} {arXiv:1905.02991 [hep-ph]} \BibitemShut {NoStop}%
\bibitem [{\citenamefont {Camargo}\ \emph {et~al.}(2019)\citenamefont {Camargo}, \citenamefont {Queiroz},\ and\ \citenamefont {Sturani}}]{Camargo:2019wou}%
  \BibitemOpen
  \bibfield  {author} {\bibinfo {author} {\bibfnamefont {D.~A.}\ \bibnamefont {Camargo}}, \bibinfo {author} {\bibfnamefont {F.~S.}\ \bibnamefont {Queiroz}}, \ and\ \bibinfo {author} {\bibfnamefont {R.}~\bibnamefont {Sturani}},\ }\href {\doibase 10.1088/1475-7516/2019/09/051} {\bibfield  {journal} {\bibinfo  {journal} {JCAP}\ }\textbf {\bibinfo {volume} {09}},\ \bibinfo {pages} {051} (\bibinfo {year} {2019})},\ \Eprint {http://arxiv.org/abs/1901.05474} {arXiv:1901.05474 [hep-ph]} \BibitemShut {NoStop}%
\bibitem [{\citenamefont {Bell}\ \emph {et~al.}(2019)\citenamefont {Bell}, \citenamefont {Busoni},\ and\ \citenamefont {Robles}}]{Bell:2019pyc}%
  \BibitemOpen
  \bibfield  {author} {\bibinfo {author} {\bibfnamefont {N.~F.}\ \bibnamefont {Bell}}, \bibinfo {author} {\bibfnamefont {G.}~\bibnamefont {Busoni}}, \ and\ \bibinfo {author} {\bibfnamefont {S.}~\bibnamefont {Robles}},\ }\href {\doibase 10.1088/1475-7516/2019/06/054} {\bibfield  {journal} {\bibinfo  {journal} {JCAP}\ }\textbf {\bibinfo {volume} {06}},\ \bibinfo {pages} {054} (\bibinfo {year} {2019})},\ \Eprint {http://arxiv.org/abs/1904.09803} {arXiv:1904.09803 [hep-ph]} \BibitemShut {NoStop}%
\bibitem [{\citenamefont {Garani}\ and\ \citenamefont {Heeck}(2019)}]{Garani:2019fpa}%
  \BibitemOpen
  \bibfield  {author} {\bibinfo {author} {\bibfnamefont {R.}~\bibnamefont {Garani}}\ and\ \bibinfo {author} {\bibfnamefont {J.}~\bibnamefont {Heeck}},\ }\href {\doibase 10.1103/PhysRevD.100.035039} {\bibfield  {journal} {\bibinfo  {journal} {Phys. Rev. D}\ }\textbf {\bibinfo {volume} {100}},\ \bibinfo {pages} {035039} (\bibinfo {year} {2019})},\ \Eprint {http://arxiv.org/abs/1906.10145} {arXiv:1906.10145 [hep-ph]} \BibitemShut {NoStop}%
\bibitem [{\citenamefont {Acevedo}\ \emph {et~al.}(2020)\citenamefont {Acevedo}, \citenamefont {Bramante}, \citenamefont {Leane},\ and\ \citenamefont {Raj}}]{Acevedo:2019agu}%
  \BibitemOpen
  \bibfield  {author} {\bibinfo {author} {\bibfnamefont {J.~F.}\ \bibnamefont {Acevedo}}, \bibinfo {author} {\bibfnamefont {J.}~\bibnamefont {Bramante}}, \bibinfo {author} {\bibfnamefont {R.~K.}\ \bibnamefont {Leane}}, \ and\ \bibinfo {author} {\bibfnamefont {N.}~\bibnamefont {Raj}},\ }\href {\doibase 10.1088/1475-7516/2020/03/038} {\bibfield  {journal} {\bibinfo  {journal} {JCAP}\ }\textbf {\bibinfo {volume} {03}},\ \bibinfo {pages} {038} (\bibinfo {year} {2020})},\ \Eprint {http://arxiv.org/abs/1911.06334} {arXiv:1911.06334 [hep-ph]} \BibitemShut {NoStop}%
\bibitem [{\citenamefont {Joglekar}\ \emph {et~al.}(2020{\natexlab{a}})\citenamefont {Joglekar}, \citenamefont {Raj}, \citenamefont {Tanedo},\ and\ \citenamefont {Yu}}]{Joglekar:2019vzy}%
  \BibitemOpen
  \bibfield  {author} {\bibinfo {author} {\bibfnamefont {A.}~\bibnamefont {Joglekar}}, \bibinfo {author} {\bibfnamefont {N.}~\bibnamefont {Raj}}, \bibinfo {author} {\bibfnamefont {P.}~\bibnamefont {Tanedo}}, \ and\ \bibinfo {author} {\bibfnamefont {H.-B.}\ \bibnamefont {Yu}},\ }\href {\doibase 10.1016/j.physletb.2020.135767} {\bibfield  {journal} {\bibinfo  {journal} {Phys. Lett.}\ }\textbf {\bibinfo {volume} {B}},\ \bibinfo {pages} {135767} (\bibinfo {year} {2020}{\natexlab{a}})},\ \Eprint {http://arxiv.org/abs/1911.13293} {arXiv:1911.13293 [hep-ph]} \BibitemShut {NoStop}%
\bibitem [{\citenamefont {Vikiaris}\ \emph {et~al.}(2023)\citenamefont {Vikiaris}, \citenamefont {Petousis}, \citenamefont {Veselsky},\ and\ \citenamefont {Moustakidis}}]{Vikiaris:2023vau}%
  \BibitemOpen
  \bibfield  {author} {\bibinfo {author} {\bibfnamefont {M.}~\bibnamefont {Vikiaris}}, \bibinfo {author} {\bibfnamefont {V.}~\bibnamefont {Petousis}}, \bibinfo {author} {\bibfnamefont {M.}~\bibnamefont {Veselsky}}, \ and\ \bibinfo {author} {\bibfnamefont {C.~C.}\ \bibnamefont {Moustakidis}},\ }\href@noop {} {\  (\bibinfo {year} {2023})},\ \Eprint {http://arxiv.org/abs/2312.07412} {arXiv:2312.07412 [astro-ph.HE]} \BibitemShut {NoStop}%
\bibitem [{\citenamefont {Joglekar}\ \emph {et~al.}(2020{\natexlab{b}})\citenamefont {Joglekar}, \citenamefont {Raj}, \citenamefont {Tanedo},\ and\ \citenamefont {Yu}}]{Joglekar:2020liw}%
  \BibitemOpen
  \bibfield  {author} {\bibinfo {author} {\bibfnamefont {A.}~\bibnamefont {Joglekar}}, \bibinfo {author} {\bibfnamefont {N.}~\bibnamefont {Raj}}, \bibinfo {author} {\bibfnamefont {P.}~\bibnamefont {Tanedo}}, \ and\ \bibinfo {author} {\bibfnamefont {H.-B.}\ \bibnamefont {Yu}},\ }\href {\doibase 10.1103/PhysRevD.102.123002} {\bibfield  {journal} {\bibinfo  {journal} {Phys. Rev. D}\ }\textbf {\bibinfo {volume} {102}},\ \bibinfo {pages} {123002} (\bibinfo {year} {2020}{\natexlab{b}})},\ \Eprint {http://arxiv.org/abs/2004.09539} {arXiv:2004.09539 [hep-ph]} \BibitemShut {NoStop}%
\bibitem [{\citenamefont {Bell}\ \emph {et~al.}(2020)\citenamefont {Bell}, \citenamefont {Busoni}, \citenamefont {Robles},\ and\ \citenamefont {Virgato}}]{Bell:2020jou}%
  \BibitemOpen
  \bibfield  {author} {\bibinfo {author} {\bibfnamefont {N.~F.}\ \bibnamefont {Bell}}, \bibinfo {author} {\bibfnamefont {G.}~\bibnamefont {Busoni}}, \bibinfo {author} {\bibfnamefont {S.}~\bibnamefont {Robles}}, \ and\ \bibinfo {author} {\bibfnamefont {M.}~\bibnamefont {Virgato}},\ }\href {\doibase 10.1088/1475-7516/2020/09/028} {\bibfield  {journal} {\bibinfo  {journal} {JCAP}\ }\textbf {\bibinfo {volume} {09}},\ \bibinfo {pages} {028} (\bibinfo {year} {2020})},\ \Eprint {http://arxiv.org/abs/2004.14888} {arXiv:2004.14888 [hep-ph]} \BibitemShut {NoStop}%
\bibitem [{\citenamefont {Garani}\ \emph {et~al.}(2021)\citenamefont {Garani}, \citenamefont {Gupta},\ and\ \citenamefont {Raj}}]{Garani:2020wge}%
  \BibitemOpen
  \bibfield  {author} {\bibinfo {author} {\bibfnamefont {R.}~\bibnamefont {Garani}}, \bibinfo {author} {\bibfnamefont {A.}~\bibnamefont {Gupta}}, \ and\ \bibinfo {author} {\bibfnamefont {N.}~\bibnamefont {Raj}},\ }\href {\doibase 10.1103/PhysRevD.103.043019} {\bibfield  {journal} {\bibinfo  {journal} {Phys. Rev. D}\ }\textbf {\bibinfo {volume} {103}},\ \bibinfo {pages} {043019} (\bibinfo {year} {2021})},\ \Eprint {http://arxiv.org/abs/2009.10728} {arXiv:2009.10728 [hep-ph]} \BibitemShut {NoStop}%
\bibitem [{\citenamefont {Acu\~na}\ \emph {et~al.}(2022)\citenamefont {Acu\~na}, \citenamefont {Stengel},\ and\ \citenamefont {Ullio}}]{Acuna:2022ouv}%
  \BibitemOpen
  \bibfield  {author} {\bibinfo {author} {\bibfnamefont {J.~T.}\ \bibnamefont {Acu\~na}}, \bibinfo {author} {\bibfnamefont {P.}~\bibnamefont {Stengel}}, \ and\ \bibinfo {author} {\bibfnamefont {P.}~\bibnamefont {Ullio}},\ }\href@noop {} {\  (\bibinfo {year} {2022})},\ \Eprint {http://arxiv.org/abs/2209.12552} {arXiv:2209.12552 [hep-ph]} \BibitemShut {NoStop}%
\bibitem [{\citenamefont {Alvarez}\ \emph {et~al.}(2023)\citenamefont {Alvarez}, \citenamefont {Joglekar}, \citenamefont {Phoroutan-Mehr},\ and\ \citenamefont {Yu}}]{Alvarez:2023fjj}%
  \BibitemOpen
  \bibfield  {author} {\bibinfo {author} {\bibfnamefont {G.}~\bibnamefont {Alvarez}}, \bibinfo {author} {\bibfnamefont {A.}~\bibnamefont {Joglekar}}, \bibinfo {author} {\bibfnamefont {M.}~\bibnamefont {Phoroutan-Mehr}}, \ and\ \bibinfo {author} {\bibfnamefont {H.-B.}\ \bibnamefont {Yu}},\ }\href {\doibase 10.1103/PhysRevD.107.103024} {\bibfield  {journal} {\bibinfo  {journal} {Phys. Rev. D}\ }\textbf {\bibinfo {volume} {107}},\ \bibinfo {pages} {103024} (\bibinfo {year} {2023})},\ \Eprint {http://arxiv.org/abs/2301.08767} {arXiv:2301.08767 [hep-ph]} \BibitemShut {NoStop}%
\bibitem [{\citenamefont {Fujiwara}\ \emph {et~al.}(2023)\citenamefont {Fujiwara}, \citenamefont {Hamaguchi}, \citenamefont {Nagata},\ and\ \citenamefont {Ramirez-Quezada}}]{Fujiwara:2023hlj}%
  \BibitemOpen
  \bibfield  {author} {\bibinfo {author} {\bibfnamefont {M.}~\bibnamefont {Fujiwara}}, \bibinfo {author} {\bibfnamefont {K.}~\bibnamefont {Hamaguchi}}, \bibinfo {author} {\bibfnamefont {N.}~\bibnamefont {Nagata}}, \ and\ \bibinfo {author} {\bibfnamefont {M.~E.}\ \bibnamefont {Ramirez-Quezada}},\ }\href@noop {} {\  (\bibinfo {year} {2023})},\ \Eprint {http://arxiv.org/abs/2309.02633} {arXiv:2309.02633 [hep-ph]} \BibitemShut {NoStop}%
\bibitem [{\citenamefont {Bose}\ \emph {et~al.}(2022{\natexlab{a}})\citenamefont {Bose}, \citenamefont {Maity},\ and\ \citenamefont {Ray}}]{Bose:2021yhz}%
  \BibitemOpen
  \bibfield  {author} {\bibinfo {author} {\bibfnamefont {D.}~\bibnamefont {Bose}}, \bibinfo {author} {\bibfnamefont {T.~N.}\ \bibnamefont {Maity}}, \ and\ \bibinfo {author} {\bibfnamefont {T.~S.}\ \bibnamefont {Ray}},\ }\href {\doibase 10.1088/1475-7516/2022/05/001} {\bibfield  {journal} {\bibinfo  {journal} {JCAP}\ }\textbf {\bibinfo {volume} {05}},\ \bibinfo {pages} {001} (\bibinfo {year} {2022}{\natexlab{a}})},\ \Eprint {http://arxiv.org/abs/2108.12420} {arXiv:2108.12420 [hep-ph]} \BibitemShut {NoStop}%
\bibitem [{\citenamefont {Bose}\ \emph {et~al.}(2023)\citenamefont {Bose}, \citenamefont {Chowdhury}, \citenamefont {Mondal},\ and\ \citenamefont {Ray}}]{Bose:2023yll}%
  \BibitemOpen
  \bibfield  {author} {\bibinfo {author} {\bibfnamefont {D.}~\bibnamefont {Bose}}, \bibinfo {author} {\bibfnamefont {D.}~\bibnamefont {Chowdhury}}, \bibinfo {author} {\bibfnamefont {P.}~\bibnamefont {Mondal}}, \ and\ \bibinfo {author} {\bibfnamefont {T.~S.}\ \bibnamefont {Ray}},\ }\href@noop {} {\  (\bibinfo {year} {2023})},\ \Eprint {http://arxiv.org/abs/2312.05131} {arXiv:2312.05131 [hep-ph]} \BibitemShut {NoStop}%
\bibitem [{\citenamefont {Bramante}\ \emph {et~al.}(2022)\citenamefont {Bramante}, \citenamefont {Kavanagh},\ and\ \citenamefont {Raj}}]{Bramante:2021dyx}%
  \BibitemOpen
  \bibfield  {author} {\bibinfo {author} {\bibfnamefont {J.}~\bibnamefont {Bramante}}, \bibinfo {author} {\bibfnamefont {B.~J.}\ \bibnamefont {Kavanagh}}, \ and\ \bibinfo {author} {\bibfnamefont {N.}~\bibnamefont {Raj}},\ }\href {\doibase 10.1103/PhysRevLett.128.231801} {\bibfield  {journal} {\bibinfo  {journal} {Phys. Rev. Lett.}\ }\textbf {\bibinfo {volume} {128}},\ \bibinfo {pages} {231801} (\bibinfo {year} {2022})},\ \Eprint {http://arxiv.org/abs/2109.04582} {arXiv:2109.04582 [hep-ph]} \BibitemShut {NoStop}%
\bibitem [{\citenamefont {Brdar}\ \emph {et~al.}(2017)\citenamefont {Brdar}, \citenamefont {Kopp},\ and\ \citenamefont {Liu}}]{Brdar:2016ifs}%
  \BibitemOpen
  \bibfield  {author} {\bibinfo {author} {\bibfnamefont {V.}~\bibnamefont {Brdar}}, \bibinfo {author} {\bibfnamefont {J.}~\bibnamefont {Kopp}}, \ and\ \bibinfo {author} {\bibfnamefont {J.}~\bibnamefont {Liu}},\ }\href {\doibase 10.1103/PhysRevD.95.055031} {\bibfield  {journal} {\bibinfo  {journal} {Phys. Rev. D}\ }\textbf {\bibinfo {volume} {95}},\ \bibinfo {pages} {055031} (\bibinfo {year} {2017})},\ \Eprint {http://arxiv.org/abs/1607.04278} {arXiv:1607.04278 [hep-ph]} \BibitemShut {NoStop}%
\bibitem [{\citenamefont {Nguyen}(2023)}]{Nguyen:2023ugx}%
  \BibitemOpen
  \bibfield  {author} {\bibinfo {author} {\bibfnamefont {T.~T.~Q.}\ \bibnamefont {Nguyen}}\ }(\bibinfo {year} {2023})\ \Eprint {http://arxiv.org/abs/2312.12292} {arXiv:2312.12292 [hep-ph]} \BibitemShut {NoStop}%
\bibitem [{\citenamefont {Lin}\ \emph {et~al.}(2023)\citenamefont {Lin}, \citenamefont {Wu}, \citenamefont {Wu},\ and\ \citenamefont {Wong}}]{Lin:2022dbl}%
  \BibitemOpen
  \bibfield  {author} {\bibinfo {author} {\bibfnamefont {Y.-H.}\ \bibnamefont {Lin}}, \bibinfo {author} {\bibfnamefont {W.-H.}\ \bibnamefont {Wu}}, \bibinfo {author} {\bibfnamefont {M.-R.}\ \bibnamefont {Wu}}, \ and\ \bibinfo {author} {\bibfnamefont {H.~T.-K.}\ \bibnamefont {Wong}},\ }\href {\doibase 10.1103/PhysRevLett.130.111002} {\bibfield  {journal} {\bibinfo  {journal} {Phys. Rev. Lett.}\ }\textbf {\bibinfo {volume} {130}},\ \bibinfo {pages} {111002} (\bibinfo {year} {2023})},\ \Eprint {http://arxiv.org/abs/2206.06864} {arXiv:2206.06864 [hep-ph]} \BibitemShut {NoStop}%
\bibitem [{\citenamefont {Nguyen}\ and\ \citenamefont {Tait}(2023)}]{Nguyen:2022zwb}%
  \BibitemOpen
  \bibfield  {author} {\bibinfo {author} {\bibfnamefont {T.~T.~Q.}\ \bibnamefont {Nguyen}}\ and\ \bibinfo {author} {\bibfnamefont {T.~M.~P.}\ \bibnamefont {Tait}},\ }\href {\doibase 10.1103/PhysRevD.107.115016} {\bibfield  {journal} {\bibinfo  {journal} {Phys. Rev. D}\ }\textbf {\bibinfo {volume} {107}},\ \bibinfo {pages} {115016} (\bibinfo {year} {2023})},\ \Eprint {http://arxiv.org/abs/2212.12547} {arXiv:2212.12547 [hep-ph]} \BibitemShut {NoStop}%
\bibitem [{\citenamefont {Leane}\ \emph {et~al.}(2021)\citenamefont {Leane}, \citenamefont {Linden}, \citenamefont {Mukhopadhyay},\ and\ \citenamefont {Toro}}]{Leane:2021ihh}%
  \BibitemOpen
  \bibfield  {author} {\bibinfo {author} {\bibfnamefont {R.~K.}\ \bibnamefont {Leane}}, \bibinfo {author} {\bibfnamefont {T.}~\bibnamefont {Linden}}, \bibinfo {author} {\bibfnamefont {P.}~\bibnamefont {Mukhopadhyay}}, \ and\ \bibinfo {author} {\bibfnamefont {N.}~\bibnamefont {Toro}},\ }\href {\doibase 10.1103/PhysRevD.103.075030} {\bibfield  {journal} {\bibinfo  {journal} {Phys. Rev. D}\ }\textbf {\bibinfo {volume} {103}},\ \bibinfo {pages} {075030} (\bibinfo {year} {2021})},\ \Eprint {http://arxiv.org/abs/2101.12213} {arXiv:2101.12213 [astro-ph.HE]} \BibitemShut {NoStop}%
\bibitem [{\citenamefont {Leane}\ and\ \citenamefont {Smirnov}(2021)}]{Leane:2020wob}%
  \BibitemOpen
  \bibfield  {author} {\bibinfo {author} {\bibfnamefont {R.~K.}\ \bibnamefont {Leane}}\ and\ \bibinfo {author} {\bibfnamefont {J.}~\bibnamefont {Smirnov}},\ }\href {\doibase 10.1103/PhysRevLett.126.161101} {\bibfield  {journal} {\bibinfo  {journal} {Phys. Rev. Lett.}\ }\textbf {\bibinfo {volume} {126}},\ \bibinfo {pages} {161101} (\bibinfo {year} {2021})},\ \Eprint {http://arxiv.org/abs/2010.00015} {arXiv:2010.00015 [hep-ph]} \BibitemShut {NoStop}%
\bibitem [{\citenamefont {Ilie}\ \emph {et~al.}(2023)\citenamefont {Ilie}, \citenamefont {Levy},\ and\ \citenamefont {Diks}}]{Ilie:2023lbi}%
  \BibitemOpen
  \bibfield  {author} {\bibinfo {author} {\bibfnamefont {C.}~\bibnamefont {Ilie}}, \bibinfo {author} {\bibfnamefont {C.}~\bibnamefont {Levy}}, \ and\ \bibinfo {author} {\bibfnamefont {J.}~\bibnamefont {Diks}},\ }\href@noop {} {\  (\bibinfo {year} {2023})},\ \Eprint {http://arxiv.org/abs/2312.13979} {arXiv:2312.13979 [astro-ph.CO]} \BibitemShut {NoStop}%
\bibitem [{\citenamefont {John}\ \emph {et~al.}(2023)\citenamefont {John}, \citenamefont {Leane},\ and\ \citenamefont {Linden}}]{John:2023knt}%
  \BibitemOpen
  \bibfield  {author} {\bibinfo {author} {\bibfnamefont {I.}~\bibnamefont {John}}, \bibinfo {author} {\bibfnamefont {R.~K.}\ \bibnamefont {Leane}}, \ and\ \bibinfo {author} {\bibfnamefont {T.}~\bibnamefont {Linden}},\ }\href@noop {} {\  (\bibinfo {year} {2023})},\ \Eprint {http://arxiv.org/abs/2311.16228} {arXiv:2311.16228 [astro-ph.HE]} \BibitemShut {NoStop}%
\bibitem [{\citenamefont {Dasgupta}\ \emph {et~al.}(2019)\citenamefont {Dasgupta}, \citenamefont {Gupta},\ and\ \citenamefont {Ray}}]{Dasgupta:2019juq}%
  \BibitemOpen
  \bibfield  {author} {\bibinfo {author} {\bibfnamefont {B.}~\bibnamefont {Dasgupta}}, \bibinfo {author} {\bibfnamefont {A.}~\bibnamefont {Gupta}}, \ and\ \bibinfo {author} {\bibfnamefont {A.}~\bibnamefont {Ray}},\ }\href {\doibase 10.1088/1475-7516/2019/08/018} {\bibfield  {journal} {\bibinfo  {journal} {JCAP}\ }\textbf {\bibinfo {volume} {08}},\ \bibinfo {pages} {018} (\bibinfo {year} {2019})},\ \Eprint {http://arxiv.org/abs/1906.04204} {arXiv:1906.04204 [hep-ph]} \BibitemShut {NoStop}%
\bibitem [{\citenamefont {Acevedo}\ \emph {et~al.}(2023{\natexlab{a}})\citenamefont {Acevedo}, \citenamefont {Leane},\ and\ \citenamefont {Santos-Olmsted}}]{Acevedo:2023xnu}%
  \BibitemOpen
  \bibfield  {author} {\bibinfo {author} {\bibfnamefont {J.~F.}\ \bibnamefont {Acevedo}}, \bibinfo {author} {\bibfnamefont {R.~K.}\ \bibnamefont {Leane}}, \ and\ \bibinfo {author} {\bibfnamefont {L.}~\bibnamefont {Santos-Olmsted}},\ }\href@noop {} {\  (\bibinfo {year} {2023}{\natexlab{a}})},\ \Eprint {http://arxiv.org/abs/2309.10843} {arXiv:2309.10843 [hep-ph]} \BibitemShut {NoStop}%
\bibitem [{\citenamefont {Garani}\ and\ \citenamefont {Palomares-Ruiz}(2017)}]{Garani:2017jcj}%
  \BibitemOpen
  \bibfield  {author} {\bibinfo {author} {\bibfnamefont {R.}~\bibnamefont {Garani}}\ and\ \bibinfo {author} {\bibfnamefont {S.}~\bibnamefont {Palomares-Ruiz}},\ }\href {\doibase 10.1088/1475-7516/2017/05/007} {\bibfield  {journal} {\bibinfo  {journal} {JCAP}\ }\textbf {\bibinfo {volume} {05}},\ \bibinfo {pages} {007} (\bibinfo {year} {2017})},\ \Eprint {http://arxiv.org/abs/1702.02768} {arXiv:1702.02768 [hep-ph]} \BibitemShut {NoStop}%
\bibitem [{\citenamefont {Bell}\ \emph {et~al.}(2021)\citenamefont {Bell}, \citenamefont {Dent},\ and\ \citenamefont {Sanderson}}]{Bell:2021pyy}%
  \BibitemOpen
  \bibfield  {author} {\bibinfo {author} {\bibfnamefont {N.~F.}\ \bibnamefont {Bell}}, \bibinfo {author} {\bibfnamefont {J.~B.}\ \bibnamefont {Dent}}, \ and\ \bibinfo {author} {\bibfnamefont {I.~W.}\ \bibnamefont {Sanderson}},\ }\href {\doibase 10.1103/PhysRevD.104.023024} {\bibfield  {journal} {\bibinfo  {journal} {Phys. Rev. D}\ }\textbf {\bibinfo {volume} {104}},\ \bibinfo {pages} {023024} (\bibinfo {year} {2021})},\ \Eprint {http://arxiv.org/abs/2103.16794} {arXiv:2103.16794 [hep-ph]} \BibitemShut {NoStop}%
\bibitem [{\citenamefont {Feng}\ \emph {et~al.}(2016)\citenamefont {Feng}, \citenamefont {Smolinsky},\ and\ \citenamefont {Tanedo}}]{Feng:2016ijc}%
  \BibitemOpen
  \bibfield  {author} {\bibinfo {author} {\bibfnamefont {J.~L.}\ \bibnamefont {Feng}}, \bibinfo {author} {\bibfnamefont {J.}~\bibnamefont {Smolinsky}}, \ and\ \bibinfo {author} {\bibfnamefont {P.}~\bibnamefont {Tanedo}},\ }\href {\doibase 10.1103/PhysRevD.93.115036} {\bibfield  {journal} {\bibinfo  {journal} {Phys. Rev. D}\ }\textbf {\bibinfo {volume} {93}},\ \bibinfo {pages} {115036} (\bibinfo {year} {2016})},\ \bibinfo {note} {[Erratum: Phys.Rev.D 96, 099903 (2017)]},\ \Eprint {http://arxiv.org/abs/1602.01465} {arXiv:1602.01465 [hep-ph]} \BibitemShut {NoStop}%
\bibitem [{\citenamefont {Leane}\ \emph {et~al.}(2017)\citenamefont {Leane}, \citenamefont {Ng},\ and\ \citenamefont {Beacom}}]{Leane:2017vag}%
  \BibitemOpen
  \bibfield  {author} {\bibinfo {author} {\bibfnamefont {R.~K.}\ \bibnamefont {Leane}}, \bibinfo {author} {\bibfnamefont {K.~C.~Y.}\ \bibnamefont {Ng}}, \ and\ \bibinfo {author} {\bibfnamefont {J.~F.}\ \bibnamefont {Beacom}},\ }\href {\doibase 10.1103/PhysRevD.95.123016} {\bibfield  {journal} {\bibinfo  {journal} {Phys. Rev. D}\ }\textbf {\bibinfo {volume} {95}},\ \bibinfo {pages} {123016} (\bibinfo {year} {2017})},\ \Eprint {http://arxiv.org/abs/1703.04629} {arXiv:1703.04629 [astro-ph.HE]} \BibitemShut {NoStop}%
\bibitem [{\citenamefont {Kang}\ \emph {et~al.}(2023)\citenamefont {Kang}, \citenamefont {Kar},\ and\ \citenamefont {Scopel}}]{Kang:2023gef}%
  \BibitemOpen
  \bibfield  {author} {\bibinfo {author} {\bibfnamefont {S.}~\bibnamefont {Kang}}, \bibinfo {author} {\bibfnamefont {A.}~\bibnamefont {Kar}}, \ and\ \bibinfo {author} {\bibfnamefont {S.}~\bibnamefont {Scopel}},\ }\href {\doibase 10.1088/1475-7516/2023/11/077} {\bibfield  {journal} {\bibinfo  {journal} {JCAP}\ }\textbf {\bibinfo {volume} {11}},\ \bibinfo {pages} {077} (\bibinfo {year} {2023})},\ \Eprint {http://arxiv.org/abs/2308.13203} {arXiv:2308.13203 [hep-ph]} \BibitemShut {NoStop}%
\bibitem [{\citenamefont {Bell}\ and\ \citenamefont {Petraki}(2011)}]{Bell:2011sn}%
  \BibitemOpen
  \bibfield  {author} {\bibinfo {author} {\bibfnamefont {N.~F.}\ \bibnamefont {Bell}}\ and\ \bibinfo {author} {\bibfnamefont {K.}~\bibnamefont {Petraki}},\ }\href {\doibase 10.1088/1475-7516/2011/04/003} {\bibfield  {journal} {\bibinfo  {journal} {JCAP}\ }\textbf {\bibinfo {volume} {04}},\ \bibinfo {pages} {003} (\bibinfo {year} {2011})},\ \Eprint {http://arxiv.org/abs/1102.2958} {arXiv:1102.2958 [hep-ph]} \BibitemShut {NoStop}%
\bibitem [{\citenamefont {Chauhan}\ \emph {et~al.}(2024)\citenamefont {Chauhan}, \citenamefont {Reno}, \citenamefont {Rott},\ and\ \citenamefont {Sarcevic}}]{Chauhan:2023zuf}%
  \BibitemOpen
  \bibfield  {author} {\bibinfo {author} {\bibfnamefont {B.}~\bibnamefont {Chauhan}}, \bibinfo {author} {\bibfnamefont {M.~H.}\ \bibnamefont {Reno}}, \bibinfo {author} {\bibfnamefont {C.}~\bibnamefont {Rott}}, \ and\ \bibinfo {author} {\bibfnamefont {I.}~\bibnamefont {Sarcevic}},\ }\href {\doibase 10.1088/1475-7516/2024/01/030} {\bibfield  {journal} {\bibinfo  {journal} {JCAP}\ }\textbf {\bibinfo {volume} {01}},\ \bibinfo {pages} {030} (\bibinfo {year} {2024})},\ \Eprint {http://arxiv.org/abs/2308.16134} {arXiv:2308.16134 [hep-ph]} \BibitemShut {NoStop}%
\bibitem [{\citenamefont {Maity}\ \emph {et~al.}(2023)\citenamefont {Maity}, \citenamefont {Saha}, \citenamefont {Mondal},\ and\ \citenamefont {Laha}}]{Maity:2023rez}%
  \BibitemOpen
  \bibfield  {author} {\bibinfo {author} {\bibfnamefont {T.~N.}\ \bibnamefont {Maity}}, \bibinfo {author} {\bibfnamefont {A.~K.}\ \bibnamefont {Saha}}, \bibinfo {author} {\bibfnamefont {S.}~\bibnamefont {Mondal}}, \ and\ \bibinfo {author} {\bibfnamefont {R.}~\bibnamefont {Laha}},\ }\href@noop {} {\  (\bibinfo {year} {2023})},\ \Eprint {http://arxiv.org/abs/2308.12336} {arXiv:2308.12336 [hep-ph]} \BibitemShut {NoStop}%
\bibitem [{\citenamefont {Niblaeus}\ \emph {et~al.}(2019)\citenamefont {Niblaeus}, \citenamefont {Beniwal},\ and\ \citenamefont {Edsjo}}]{Niblaeus:2019gjk}%
  \BibitemOpen
  \bibfield  {author} {\bibinfo {author} {\bibfnamefont {C.}~\bibnamefont {Niblaeus}}, \bibinfo {author} {\bibfnamefont {A.}~\bibnamefont {Beniwal}}, \ and\ \bibinfo {author} {\bibfnamefont {J.}~\bibnamefont {Edsjo}},\ }\href {\doibase 10.1088/1475-7516/2019/11/011} {\bibfield  {journal} {\bibinfo  {journal} {JCAP}\ }\textbf {\bibinfo {volume} {11}},\ \bibinfo {pages} {011} (\bibinfo {year} {2019})},\ \Eprint {http://arxiv.org/abs/1903.11363} {arXiv:1903.11363 [astro-ph.HE]} \BibitemShut {NoStop}%
\bibitem [{\citenamefont {Bose}\ \emph {et~al.}(2022{\natexlab{b}})\citenamefont {Bose}, \citenamefont {Maity},\ and\ \citenamefont {Ray}}]{Bose:2021cou}%
  \BibitemOpen
  \bibfield  {author} {\bibinfo {author} {\bibfnamefont {D.}~\bibnamefont {Bose}}, \bibinfo {author} {\bibfnamefont {T.~N.}\ \bibnamefont {Maity}}, \ and\ \bibinfo {author} {\bibfnamefont {T.~S.}\ \bibnamefont {Ray}},\ }\href {\doibase 10.1103/PhysRevD.105.123013} {\bibfield  {journal} {\bibinfo  {journal} {Phys. Rev. D}\ }\textbf {\bibinfo {volume} {105}},\ \bibinfo {pages} {123013} (\bibinfo {year} {2022}{\natexlab{b}})},\ \Eprint {http://arxiv.org/abs/2112.08286} {arXiv:2112.08286 [hep-ph]} \BibitemShut {NoStop}%
\bibitem [{\citenamefont {Albert}\ \emph {et~al.}(2018)\citenamefont {Albert} \emph {et~al.}}]{HAWC:2018szf}%
  \BibitemOpen
  \bibfield  {author} {\bibinfo {author} {\bibfnamefont {A.}~\bibnamefont {Albert}} \emph {et~al.} (\bibinfo {collaboration} {HAWC}),\ }\href {\doibase 10.1103/PhysRevD.98.123012} {\bibfield  {journal} {\bibinfo  {journal} {Phys. Rev. D}\ }\textbf {\bibinfo {volume} {98}},\ \bibinfo {pages} {123012} (\bibinfo {year} {2018})},\ \Eprint {http://arxiv.org/abs/1808.05624} {arXiv:1808.05624 [hep-ph]} \BibitemShut {NoStop}%
\bibitem [{\citenamefont {Nguyen}\ \emph {et~al.}(2026{\natexlab{c}})\citenamefont {Nguyen}, \citenamefont {Linden}, \citenamefont {Carenza},\ and\ \citenamefont {Widmark}}]{Nguyen:2025ygc}%
  \BibitemOpen
  \bibfield  {author} {\bibinfo {author} {\bibfnamefont {T.~T.~Q.}\ \bibnamefont {Nguyen}}, \bibinfo {author} {\bibfnamefont {T.}~\bibnamefont {Linden}}, \bibinfo {author} {\bibfnamefont {P.}~\bibnamefont {Carenza}}, \ and\ \bibinfo {author} {\bibfnamefont {A.}~\bibnamefont {Widmark}},\ }\href {\doibase 10.1103/txsw-2kdf} {\bibfield  {journal} {\bibinfo  {journal} {Phys. Rev. D}\ }\textbf {\bibinfo {volume} {113}},\ \bibinfo {pages} {L101305} (\bibinfo {year} {2026}{\natexlab{c}})},\ \Eprint {http://arxiv.org/abs/2501.14864} {arXiv:2501.14864 [astro-ph.HE]} \BibitemShut {NoStop}%
\bibitem [{\citenamefont {Nguyen}\ and\ \citenamefont {Linden}(2026)}]{Nguyen:2026nhe}%
  \BibitemOpen
  \bibfield  {author} {\bibinfo {author} {\bibfnamefont {T.~T.~Q.}\ \bibnamefont {Nguyen}}\ and\ \bibinfo {author} {\bibfnamefont {T.}~\bibnamefont {Linden}},\ }\href@noop {} {\  (\bibinfo {year} {2026})},\ \Eprint {http://arxiv.org/abs/2602.15113} {arXiv:2602.15113 [hep-ph]} \BibitemShut {NoStop}%
\bibitem [{\citenamefont {Nguyen}(2026)}]{Nguyen:2026apa}%
  \BibitemOpen
  \bibfield  {author} {\bibinfo {author} {\bibfnamefont {T.~T.~Q.}\ \bibnamefont {Nguyen}},\ }\href@noop {} {\  (\bibinfo {year} {2026})},\ \Eprint {http://arxiv.org/abs/2606.09998} {arXiv:2606.09998 [hep-ph]} \BibitemShut {NoStop}%
\bibitem [{\citenamefont {Leane}\ and\ \citenamefont {Linden}(2023)}]{Leane:2021tjj}%
  \BibitemOpen
  \bibfield  {author} {\bibinfo {author} {\bibfnamefont {R.~K.}\ \bibnamefont {Leane}}\ and\ \bibinfo {author} {\bibfnamefont {T.}~\bibnamefont {Linden}},\ }\href {\doibase 10.1103/PhysRevLett.131.071001} {\bibfield  {journal} {\bibinfo  {journal} {Phys. Rev. Lett.}\ }\textbf {\bibinfo {volume} {131}},\ \bibinfo {pages} {071001} (\bibinfo {year} {2023})},\ \Eprint {http://arxiv.org/abs/2104.02068} {arXiv:2104.02068 [astro-ph.HE]} \BibitemShut {NoStop}%
\bibitem [{\citenamefont {Chen}\ \emph {et~al.}(2023)\citenamefont {Chen}, \citenamefont {Zu}, \citenamefont {Xia}, \citenamefont {Tsai},\ and\ \citenamefont {Fan}}]{Chen:2023fgr}%
  \BibitemOpen
  \bibfield  {author} {\bibinfo {author} {\bibfnamefont {Y.-X.}\ \bibnamefont {Chen}}, \bibinfo {author} {\bibfnamefont {L.}~\bibnamefont {Zu}}, \bibinfo {author} {\bibfnamefont {Z.-Q.}\ \bibnamefont {Xia}}, \bibinfo {author} {\bibfnamefont {Y.-L.~S.}\ \bibnamefont {Tsai}}, \ and\ \bibinfo {author} {\bibfnamefont {Y.-Z.}\ \bibnamefont {Fan}},\ }\href {\doibase 10.1088/1475-7516/2023/08/036} {\bibfield  {journal} {\bibinfo  {journal} {JCAP}\ }\textbf {\bibinfo {volume} {08}},\ \bibinfo {pages} {036} (\bibinfo {year} {2023})},\ \Eprint {http://arxiv.org/abs/2302.09951} {arXiv:2302.09951 [astro-ph.HE]} \BibitemShut {NoStop}%
\bibitem [{\citenamefont {Croon}\ and\ \citenamefont {Smirnov}(2023)}]{Croon:2023bmu}%
  \BibitemOpen
  \bibfield  {author} {\bibinfo {author} {\bibfnamefont {D.}~\bibnamefont {Croon}}\ and\ \bibinfo {author} {\bibfnamefont {J.}~\bibnamefont {Smirnov}},\ }\href@noop {} {\  (\bibinfo {year} {2023})},\ \Eprint {http://arxiv.org/abs/2309.02495} {arXiv:2309.02495 [hep-ph]} \BibitemShut {NoStop}%
\bibitem [{\citenamefont {Ansarifard}\ and\ \citenamefont {Farzan}(2024)}]{Ansarifard:2024fan}%
  \BibitemOpen
  \bibfield  {author} {\bibinfo {author} {\bibfnamefont {S.}~\bibnamefont {Ansarifard}}\ and\ \bibinfo {author} {\bibfnamefont {Y.}~\bibnamefont {Farzan}},\ }\href@noop {} {\  (\bibinfo {year} {2024})},\ \Eprint {http://arxiv.org/abs/2401.13043} {arXiv:2401.13043 [hep-ph]} \BibitemShut {NoStop}%
\bibitem [{\citenamefont {Blanco}\ and\ \citenamefont {Leane}(2023)}]{Blanco:2023qgi}%
  \BibitemOpen
  \bibfield  {author} {\bibinfo {author} {\bibfnamefont {C.}~\bibnamefont {Blanco}}\ and\ \bibinfo {author} {\bibfnamefont {R.~K.}\ \bibnamefont {Leane}},\ }\href@noop {} {\  (\bibinfo {year} {2023})},\ \Eprint {http://arxiv.org/abs/2312.06758} {arXiv:2312.06758 [hep-ph]} \BibitemShut {NoStop}%
\bibitem [{\citenamefont {Yan}\ \emph {et~al.}(2023)\citenamefont {Yan}, \citenamefont {Li},\ and\ \citenamefont {Fan}}]{Yan:2023kdg}%
  \BibitemOpen
  \bibfield  {author} {\bibinfo {author} {\bibfnamefont {S.}~\bibnamefont {Yan}}, \bibinfo {author} {\bibfnamefont {L.}~\bibnamefont {Li}}, \ and\ \bibinfo {author} {\bibfnamefont {J.}~\bibnamefont {Fan}},\ }\href@noop {} {\  (\bibinfo {year} {2023})},\ \Eprint {http://arxiv.org/abs/2312.06746} {arXiv:2312.06746 [hep-ph]} \BibitemShut {NoStop}%
\bibitem [{\citenamefont {Li}\ and\ \citenamefont {Fan}(2022)}]{Li:2022wix}%
  \BibitemOpen
  \bibfield  {author} {\bibinfo {author} {\bibfnamefont {L.}~\bibnamefont {Li}}\ and\ \bibinfo {author} {\bibfnamefont {J.}~\bibnamefont {Fan}},\ }\href {\doibase 10.1007/JHEP10(2022)186} {\bibfield  {journal} {\bibinfo  {journal} {JHEP}\ }\textbf {\bibinfo {volume} {10}},\ \bibinfo {pages} {186} (\bibinfo {year} {2022})},\ \Eprint {http://arxiv.org/abs/2207.13709} {arXiv:2207.13709 [hep-ph]} \BibitemShut {NoStop}%
\bibitem [{\citenamefont {Malyshev}\ \emph {et~al.}(2015)\citenamefont {Malyshev}, \citenamefont {Chernyakova}, \citenamefont {Neronov},\ and\ \citenamefont {Walter}}]{Malyshev:2015hqa}%
  \BibitemOpen
  \bibfield  {author} {\bibinfo {author} {\bibfnamefont {D.}~\bibnamefont {Malyshev}}, \bibinfo {author} {\bibfnamefont {M.}~\bibnamefont {Chernyakova}}, \bibinfo {author} {\bibfnamefont {A.}~\bibnamefont {Neronov}}, \ and\ \bibinfo {author} {\bibfnamefont {R.}~\bibnamefont {Walter}},\ }\href {\doibase 10.1051/0004-6361/201526120} {\bibfield  {journal} {\bibinfo  {journal} {Astron. Astrophys.}\ }\textbf {\bibinfo {volume} {582}},\ \bibinfo {pages} {A11} (\bibinfo {year} {2015})},\ \Eprint {http://arxiv.org/abs/1503.05120} {arXiv:1503.05120 [astro-ph.HE]} \BibitemShut {NoStop}%
\bibitem [{\citenamefont {Albert}\ \emph {et~al.}(2023)\citenamefont {Albert} \emph {et~al.}}]{HAWC:2022khj}%
  \BibitemOpen
  \bibfield  {author} {\bibinfo {author} {\bibfnamefont {A.}~\bibnamefont {Albert}} \emph {et~al.} (\bibinfo {collaboration} {HAWC}),\ }\href {\doibase 10.1103/PhysRevLett.131.051201} {\bibfield  {journal} {\bibinfo  {journal} {Phys. Rev. Lett.}\ }\textbf {\bibinfo {volume} {131}},\ \bibinfo {pages} {051201} (\bibinfo {year} {2023})},\ \Eprint {http://arxiv.org/abs/2212.00815} {arXiv:2212.00815 [astro-ph.HE]} \BibitemShut {NoStop}%
\bibitem [{\citenamefont {Linden}\ \emph {et~al.}(2025)\citenamefont {Linden}, \citenamefont {Nguyen},\ and\ \citenamefont {Tait}}]{Linden:2024fby}%
  \BibitemOpen
  \bibfield  {author} {\bibinfo {author} {\bibfnamefont {T.}~\bibnamefont {Linden}}, \bibinfo {author} {\bibfnamefont {T.~T.~Q.}\ \bibnamefont {Nguyen}}, \ and\ \bibinfo {author} {\bibfnamefont {T.~M.~P.}\ \bibnamefont {Tait}},\ }\href {\doibase 10.1103/37gn-x3y1} {\bibfield  {journal} {\bibinfo  {journal} {Phys. Rev. D}\ }\textbf {\bibinfo {volume} {112}},\ \bibinfo {pages} {023026} (\bibinfo {year} {2025})},\ \Eprint {http://arxiv.org/abs/2406.19445} {arXiv:2406.19445 [hep-ph]} \BibitemShut {NoStop}%
\bibitem [{\citenamefont {Landau}(1948)}]{Landau:1948kw}%
  \BibitemOpen
  \bibfield  {author} {\bibinfo {author} {\bibfnamefont {L.~D.}\ \bibnamefont {Landau}},\ }\href {\doibase 10.1016/B978-0-08-010586-4.50070-5} {\bibfield  {journal} {\bibinfo  {journal} {Dokl. Akad. Nauk SSSR}\ }\textbf {\bibinfo {volume} {60}},\ \bibinfo {pages} {207} (\bibinfo {year} {1948})}\BibitemShut {NoStop}%
\bibitem [{\citenamefont {Yang}(1950)}]{Yang:1950rg}%
  \BibitemOpen
  \bibfield  {author} {\bibinfo {author} {\bibfnamefont {C.-N.}\ \bibnamefont {Yang}},\ }\href {\doibase 10.1103/PhysRev.77.242} {\bibfield  {journal} {\bibinfo  {journal} {Phys. Rev.}\ }\textbf {\bibinfo {volume} {77}},\ \bibinfo {pages} {242} (\bibinfo {year} {1950})}\BibitemShut {NoStop}%
\bibitem [{\citenamefont {McDermott}\ \emph {et~al.}(2018)\citenamefont {McDermott}, \citenamefont {Patel},\ and\ \citenamefont {Ramani}}]{McDermott:2017qcg}%
  \BibitemOpen
  \bibfield  {author} {\bibinfo {author} {\bibfnamefont {S.~D.}\ \bibnamefont {McDermott}}, \bibinfo {author} {\bibfnamefont {H.~H.}\ \bibnamefont {Patel}}, \ and\ \bibinfo {author} {\bibfnamefont {H.}~\bibnamefont {Ramani}},\ }\href {\doibase 10.1103/PhysRevD.97.073005} {\bibfield  {journal} {\bibinfo  {journal} {Phys. Rev. D}\ }\textbf {\bibinfo {volume} {97}},\ \bibinfo {pages} {073005} (\bibinfo {year} {2018})},\ \Eprint {http://arxiv.org/abs/1705.00619} {arXiv:1705.00619 [hep-ph]} \BibitemShut {NoStop}%
\bibitem [{\citenamefont {Pospelov}\ \emph {et~al.}(2008)\citenamefont {Pospelov}, \citenamefont {Ritz},\ and\ \citenamefont {Voloshin}}]{Pospelov:2008jk}%
  \BibitemOpen
  \bibfield  {author} {\bibinfo {author} {\bibfnamefont {M.}~\bibnamefont {Pospelov}}, \bibinfo {author} {\bibfnamefont {A.}~\bibnamefont {Ritz}}, \ and\ \bibinfo {author} {\bibfnamefont {M.~B.}\ \bibnamefont {Voloshin}},\ }\href {\doibase 10.1103/PhysRevD.78.115012} {\bibfield  {journal} {\bibinfo  {journal} {Phys. Rev. D}\ }\textbf {\bibinfo {volume} {78}},\ \bibinfo {pages} {115012} (\bibinfo {year} {2008})},\ \Eprint {http://arxiv.org/abs/0807.3279} {arXiv:0807.3279 [hep-ph]} \BibitemShut {NoStop}%
\bibitem [{\citenamefont {Alloul}\ \emph {et~al.}(2014)\citenamefont {Alloul}, \citenamefont {Christensen}, \citenamefont {Degrande}, \citenamefont {Duhr},\ and\ \citenamefont {Fuks}}]{Alloul:2013bka}%
  \BibitemOpen
  \bibfield  {author} {\bibinfo {author} {\bibfnamefont {A.}~\bibnamefont {Alloul}}, \bibinfo {author} {\bibfnamefont {N.~D.}\ \bibnamefont {Christensen}}, \bibinfo {author} {\bibfnamefont {C.}~\bibnamefont {Degrande}}, \bibinfo {author} {\bibfnamefont {C.}~\bibnamefont {Duhr}}, \ and\ \bibinfo {author} {\bibfnamefont {B.}~\bibnamefont {Fuks}},\ }\href {\doibase 10.1016/j.cpc.2014.04.012} {\bibfield  {journal} {\bibinfo  {journal} {Comput. Phys. Commun.}\ }\textbf {\bibinfo {volume} {185}},\ \bibinfo {pages} {2250} (\bibinfo {year} {2014})},\ \Eprint {http://arxiv.org/abs/1310.1921} {arXiv:1310.1921 [hep-ph]} \BibitemShut {NoStop}%
\bibitem [{\citenamefont {Alwall}\ \emph {et~al.}(2011)\citenamefont {Alwall}, \citenamefont {Herquet}, \citenamefont {Maltoni}, \citenamefont {Mattelaer},\ and\ \citenamefont {Stelzer}}]{Alwall:2011uj}%
  \BibitemOpen
  \bibfield  {author} {\bibinfo {author} {\bibfnamefont {J.}~\bibnamefont {Alwall}}, \bibinfo {author} {\bibfnamefont {M.}~\bibnamefont {Herquet}}, \bibinfo {author} {\bibfnamefont {F.}~\bibnamefont {Maltoni}}, \bibinfo {author} {\bibfnamefont {O.}~\bibnamefont {Mattelaer}}, \ and\ \bibinfo {author} {\bibfnamefont {T.}~\bibnamefont {Stelzer}},\ }\href {\doibase 10.1007/JHEP06(2011)128} {\bibfield  {journal} {\bibinfo  {journal} {JHEP}\ }\textbf {\bibinfo {volume} {06}},\ \bibinfo {pages} {128} (\bibinfo {year} {2011})},\ \Eprint {http://arxiv.org/abs/1106.0522} {arXiv:1106.0522 [hep-ph]} \BibitemShut {NoStop}%
\bibitem [{\citenamefont {Alwall}\ \emph {et~al.}(2014)\citenamefont {Alwall}, \citenamefont {Frederix}, \citenamefont {Frixione}, \citenamefont {Hirschi}, \citenamefont {Maltoni}, \citenamefont {Mattelaer}, \citenamefont {Shao}, \citenamefont {Stelzer}, \citenamefont {Torrielli},\ and\ \citenamefont {Zaro}}]{Alwall:2014hca}%
  \BibitemOpen
  \bibfield  {author} {\bibinfo {author} {\bibfnamefont {J.}~\bibnamefont {Alwall}}, \bibinfo {author} {\bibfnamefont {R.}~\bibnamefont {Frederix}}, \bibinfo {author} {\bibfnamefont {S.}~\bibnamefont {Frixione}}, \bibinfo {author} {\bibfnamefont {V.}~\bibnamefont {Hirschi}}, \bibinfo {author} {\bibfnamefont {F.}~\bibnamefont {Maltoni}}, \bibinfo {author} {\bibfnamefont {O.}~\bibnamefont {Mattelaer}}, \bibinfo {author} {\bibfnamefont {H.~S.}\ \bibnamefont {Shao}}, \bibinfo {author} {\bibfnamefont {T.}~\bibnamefont {Stelzer}}, \bibinfo {author} {\bibfnamefont {P.}~\bibnamefont {Torrielli}}, \ and\ \bibinfo {author} {\bibfnamefont {M.}~\bibnamefont {Zaro}},\ }\href {\doibase 10.1007/JHEP07(2014)079} {\bibfield  {journal} {\bibinfo  {journal} {JHEP}\ }\textbf {\bibinfo {volume} {07}},\ \bibinfo {pages} {079} (\bibinfo {year} {2014})},\ \Eprint {http://arxiv.org/abs/1405.0301} {arXiv:1405.0301 [hep-ph]} \BibitemShut {NoStop}%
\bibitem [{\citenamefont {Conte}\ \emph {et~al.}(2013)\citenamefont {Conte}, \citenamefont {Fuks},\ and\ \citenamefont {Serret}}]{Conte:2012fm}%
  \BibitemOpen
  \bibfield  {author} {\bibinfo {author} {\bibfnamefont {E.}~\bibnamefont {Conte}}, \bibinfo {author} {\bibfnamefont {B.}~\bibnamefont {Fuks}}, \ and\ \bibinfo {author} {\bibfnamefont {G.}~\bibnamefont {Serret}},\ }\href {\doibase 10.1016/j.cpc.2012.09.009} {\bibfield  {journal} {\bibinfo  {journal} {Comput. Phys. Commun.}\ }\textbf {\bibinfo {volume} {184}},\ \bibinfo {pages} {222} (\bibinfo {year} {2013})},\ \Eprint {http://arxiv.org/abs/1206.1599} {arXiv:1206.1599 [hep-ph]} \BibitemShut {NoStop}%
\bibitem [{\citenamefont {Cirelli}\ \emph {et~al.}(2011)\citenamefont {Cirelli}, \citenamefont {Corcella}, \citenamefont {Hektor}, \citenamefont {Hutsi}, \citenamefont {Kadastik}, \citenamefont {Panci}, \citenamefont {Raidal}, \citenamefont {Sala},\ and\ \citenamefont {Strumia}}]{Cirelli:2010xx}%
  \BibitemOpen
  \bibfield  {author} {\bibinfo {author} {\bibfnamefont {M.}~\bibnamefont {Cirelli}}, \bibinfo {author} {\bibfnamefont {G.}~\bibnamefont {Corcella}}, \bibinfo {author} {\bibfnamefont {A.}~\bibnamefont {Hektor}}, \bibinfo {author} {\bibfnamefont {G.}~\bibnamefont {Hutsi}}, \bibinfo {author} {\bibfnamefont {M.}~\bibnamefont {Kadastik}}, \bibinfo {author} {\bibfnamefont {P.}~\bibnamefont {Panci}}, \bibinfo {author} {\bibfnamefont {M.}~\bibnamefont {Raidal}}, \bibinfo {author} {\bibfnamefont {F.}~\bibnamefont {Sala}}, \ and\ \bibinfo {author} {\bibfnamefont {A.}~\bibnamefont {Strumia}},\ }\href {\doibase 10.1088/1475-7516/2012/10/E01} {\bibfield  {journal} {\bibinfo  {journal} {JCAP}\ }\textbf {\bibinfo {volume} {03}},\ \bibinfo {pages} {051} (\bibinfo {year} {2011})},\ \bibinfo {note} {[Erratum: JCAP 10, E01 (2012)]},\ \Eprint {http://arxiv.org/abs/1012.4515} {arXiv:1012.4515 [hep-ph]} \BibitemShut {NoStop}%
\bibitem [{\citenamefont {Abdullah}\ \emph {et~al.}(2014)\citenamefont {Abdullah}, \citenamefont {DiFranzo}, \citenamefont {Rajaraman}, \citenamefont {Tait}, \citenamefont {Tanedo},\ and\ \citenamefont {Wijangco}}]{Abdullah:2014lla}%
  \BibitemOpen
  \bibfield  {author} {\bibinfo {author} {\bibfnamefont {M.}~\bibnamefont {Abdullah}}, \bibinfo {author} {\bibfnamefont {A.}~\bibnamefont {DiFranzo}}, \bibinfo {author} {\bibfnamefont {A.}~\bibnamefont {Rajaraman}}, \bibinfo {author} {\bibfnamefont {T.~M.~P.}\ \bibnamefont {Tait}}, \bibinfo {author} {\bibfnamefont {P.}~\bibnamefont {Tanedo}}, \ and\ \bibinfo {author} {\bibfnamefont {A.~M.}\ \bibnamefont {Wijangco}},\ }\href {\doibase 10.1103/PhysRevD.90.035004} {\bibfield  {journal} {\bibinfo  {journal} {Phys. Rev. D}\ }\textbf {\bibinfo {volume} {90}},\ \bibinfo {pages} {035004} (\bibinfo {year} {2014})},\ \Eprint {http://arxiv.org/abs/1404.6528} {arXiv:1404.6528 [hep-ph]} \BibitemShut {NoStop}%
\bibitem [{\citenamefont {Ibarra}\ \emph {et~al.}(2012)\citenamefont {Ibarra}, \citenamefont {Lopez~Gehler},\ and\ \citenamefont {Pato}}]{Ibarra:2012dw}%
  \BibitemOpen
  \bibfield  {author} {\bibinfo {author} {\bibfnamefont {A.}~\bibnamefont {Ibarra}}, \bibinfo {author} {\bibfnamefont {S.}~\bibnamefont {Lopez~Gehler}}, \ and\ \bibinfo {author} {\bibfnamefont {M.}~\bibnamefont {Pato}},\ }\href {\doibase 10.1088/1475-7516/2012/07/043} {\bibfield  {journal} {\bibinfo  {journal} {JCAP}\ }\textbf {\bibinfo {volume} {07}},\ \bibinfo {pages} {043} (\bibinfo {year} {2012})},\ \Eprint {http://arxiv.org/abs/1205.0007} {arXiv:1205.0007 [hep-ph]} \BibitemShut {NoStop}%
\bibitem [{\citenamefont {Sj\"ostrand}\ \emph {et~al.}(2015)\citenamefont {Sj\"ostrand}, \citenamefont {Ask}, \citenamefont {Christiansen}, \citenamefont {Corke}, \citenamefont {Desai}, \citenamefont {Ilten}, \citenamefont {Mrenna}, \citenamefont {Prestel}, \citenamefont {Rasmussen},\ and\ \citenamefont {Skands}}]{Sjostrand:2014zea}%
  \BibitemOpen
  \bibfield  {author} {\bibinfo {author} {\bibfnamefont {T.}~\bibnamefont {Sj\"ostrand}}, \bibinfo {author} {\bibfnamefont {S.}~\bibnamefont {Ask}}, \bibinfo {author} {\bibfnamefont {J.~R.}\ \bibnamefont {Christiansen}}, \bibinfo {author} {\bibfnamefont {R.}~\bibnamefont {Corke}}, \bibinfo {author} {\bibfnamefont {N.}~\bibnamefont {Desai}}, \bibinfo {author} {\bibfnamefont {P.}~\bibnamefont {Ilten}}, \bibinfo {author} {\bibfnamefont {S.}~\bibnamefont {Mrenna}}, \bibinfo {author} {\bibfnamefont {S.}~\bibnamefont {Prestel}}, \bibinfo {author} {\bibfnamefont {C.~O.}\ \bibnamefont {Rasmussen}}, \ and\ \bibinfo {author} {\bibfnamefont {P.~Z.}\ \bibnamefont {Skands}},\ }\href {\doibase 10.1016/j.cpc.2015.01.024} {\bibfield  {journal} {\bibinfo  {journal} {Comput. Phys. Commun.}\ }\textbf {\bibinfo {volume} {191}},\ \bibinfo {pages} {159} (\bibinfo {year} {2015})},\ \Eprint {http://arxiv.org/abs/1410.3012} {arXiv:1410.3012 [hep-ph]} \BibitemShut {NoStop}%
\bibitem [{\citenamefont {Ilie}\ \emph {et~al.}(2020)\citenamefont {Ilie}, \citenamefont {Pilawa},\ and\ \citenamefont {Zhang}}]{Ilie:2020vec}%
  \BibitemOpen
  \bibfield  {author} {\bibinfo {author} {\bibfnamefont {C.}~\bibnamefont {Ilie}}, \bibinfo {author} {\bibfnamefont {J.}~\bibnamefont {Pilawa}}, \ and\ \bibinfo {author} {\bibfnamefont {S.}~\bibnamefont {Zhang}},\ }\href {\doibase 10.1103/PhysRevD.102.048301} {\bibfield  {journal} {\bibinfo  {journal} {Phys. Rev. D}\ }\textbf {\bibinfo {volume} {102}},\ \bibinfo {pages} {048301} (\bibinfo {year} {2020})},\ \Eprint {http://arxiv.org/abs/2005.05946} {arXiv:2005.05946 [astro-ph.CO]} \BibitemShut {NoStop}%
\bibitem [{\citenamefont {Bose}\ and\ \citenamefont {Sarkar}(2023)}]{Bose:2022ola}%
  \BibitemOpen
  \bibfield  {author} {\bibinfo {author} {\bibfnamefont {D.}~\bibnamefont {Bose}}\ and\ \bibinfo {author} {\bibfnamefont {S.}~\bibnamefont {Sarkar}},\ }\href {\doibase 10.1103/PhysRevD.107.063010} {\bibfield  {journal} {\bibinfo  {journal} {Phys. Rev. D}\ }\textbf {\bibinfo {volume} {107}},\ \bibinfo {pages} {063010} (\bibinfo {year} {2023})},\ \Eprint {http://arxiv.org/abs/2211.16982} {arXiv:2211.16982 [astro-ph.CO]} \BibitemShut {NoStop}%
\bibitem [{\citenamefont {Dasgupta}\ \emph {et~al.}(2020)\citenamefont {Dasgupta}, \citenamefont {Gupta},\ and\ \citenamefont {Ray}}]{Dasgupta:2020dik}%
  \BibitemOpen
  \bibfield  {author} {\bibinfo {author} {\bibfnamefont {B.}~\bibnamefont {Dasgupta}}, \bibinfo {author} {\bibfnamefont {A.}~\bibnamefont {Gupta}}, \ and\ \bibinfo {author} {\bibfnamefont {A.}~\bibnamefont {Ray}},\ }\href {\doibase 10.1088/1475-7516/2020/10/023} {\bibfield  {journal} {\bibinfo  {journal} {JCAP}\ }\textbf {\bibinfo {volume} {10}},\ \bibinfo {pages} {023} (\bibinfo {year} {2020})},\ \Eprint {http://arxiv.org/abs/2006.10773} {arXiv:2006.10773 [hep-ph]} \BibitemShut {NoStop}%
\bibitem [{\citenamefont {Cappiello}(2023)}]{Cappiello:2023hza}%
  \BibitemOpen
  \bibfield  {author} {\bibinfo {author} {\bibfnamefont {C.~V.}\ \bibnamefont {Cappiello}},\ }\href {\doibase 10.1103/PhysRevLett.130.221001} {\bibfield  {journal} {\bibinfo  {journal} {Phys. Rev. Lett.}\ }\textbf {\bibinfo {volume} {130}},\ \bibinfo {pages} {221001} (\bibinfo {year} {2023})},\ \Eprint {http://arxiv.org/abs/2301.07728} {arXiv:2301.07728 [hep-ph]} \BibitemShut {NoStop}%
\bibitem [{\citenamefont {Leane}\ and\ \citenamefont {Smirnov}(2023)}]{Leane:2023woh}%
  \BibitemOpen
  \bibfield  {author} {\bibinfo {author} {\bibfnamefont {R.~K.}\ \bibnamefont {Leane}}\ and\ \bibinfo {author} {\bibfnamefont {J.}~\bibnamefont {Smirnov}},\ }\href {\doibase 10.1088/1475-7516/2023/12/040} {\bibfield  {journal} {\bibinfo  {journal} {JCAP}\ }\textbf {\bibinfo {volume} {12}},\ \bibinfo {pages} {040} (\bibinfo {year} {2023})},\ \Eprint {http://arxiv.org/abs/2309.00669} {arXiv:2309.00669 [hep-ph]} \BibitemShut {NoStop}%
\bibitem [{\citenamefont {Acevedo}\ \emph {et~al.}(2023{\natexlab{b}})\citenamefont {Acevedo}, \citenamefont {Leane},\ and\ \citenamefont {Smirnov}}]{Acevedo:2023owd}%
  \BibitemOpen
  \bibfield  {author} {\bibinfo {author} {\bibfnamefont {J.~F.}\ \bibnamefont {Acevedo}}, \bibinfo {author} {\bibfnamefont {R.~K.}\ \bibnamefont {Leane}}, \ and\ \bibinfo {author} {\bibfnamefont {J.}~\bibnamefont {Smirnov}},\ }\href@noop {} {\  (\bibinfo {year} {2023}{\natexlab{b}})},\ \Eprint {http://arxiv.org/abs/2303.01516} {arXiv:2303.01516 [hep-ph]} \BibitemShut {NoStop}%
\bibitem [{\citenamefont {Sofue}(2013)}]{Sofue:2013kja}%
  \BibitemOpen
  \bibfield  {author} {\bibinfo {author} {\bibfnamefont {Y.}~\bibnamefont {Sofue}},\ }\href {\doibase 10.1093/pasj/65.6.118} {\bibfield  {journal} {\bibinfo  {journal} {Publ. Astron. Soc. Jap.}\ }\textbf {\bibinfo {volume} {65}},\ \bibinfo {pages} {118} (\bibinfo {year} {2013})},\ \Eprint {http://arxiv.org/abs/1307.8241} {arXiv:1307.8241 [astro-ph.GA]} \BibitemShut {NoStop}%
\bibitem [{\citenamefont {Navarro}\ \emph {et~al.}(1996)\citenamefont {Navarro}, \citenamefont {Frenk},\ and\ \citenamefont {White}}]{Navarro:1995iw}%
  \BibitemOpen
  \bibfield  {author} {\bibinfo {author} {\bibfnamefont {J.~F.}\ \bibnamefont {Navarro}}, \bibinfo {author} {\bibfnamefont {C.~S.}\ \bibnamefont {Frenk}}, \ and\ \bibinfo {author} {\bibfnamefont {S.~D.~M.}\ \bibnamefont {White}},\ }\href {\doibase 10.1086/177173} {\bibfield  {journal} {\bibinfo  {journal} {Astrophys. J.}\ }\textbf {\bibinfo {volume} {462}},\ \bibinfo {pages} {563} (\bibinfo {year} {1996})},\ \Eprint {http://arxiv.org/abs/astro-ph/9508025} {arXiv:astro-ph/9508025} \BibitemShut {NoStop}%
\bibitem [{\citenamefont {Generozov}\ \emph {et~al.}(2018)\citenamefont {Generozov}, \citenamefont {Stone}, \citenamefont {Metzger},\ and\ \citenamefont {Ostriker}}]{Generozov:2018niv}%
  \BibitemOpen
  \bibfield  {author} {\bibinfo {author} {\bibfnamefont {A.}~\bibnamefont {Generozov}}, \bibinfo {author} {\bibfnamefont {N.~C.}\ \bibnamefont {Stone}}, \bibinfo {author} {\bibfnamefont {B.~D.}\ \bibnamefont {Metzger}}, \ and\ \bibinfo {author} {\bibfnamefont {J.~P.}\ \bibnamefont {Ostriker}},\ }\href {\doibase 10.1093/mnras/sty1262} {\bibfield  {journal} {\bibinfo  {journal} {Mon. Not. Roy. Astron. Soc.}\ }\textbf {\bibinfo {volume} {478}},\ \bibinfo {pages} {4030} (\bibinfo {year} {2018})},\ \Eprint {http://arxiv.org/abs/1804.01543} {arXiv:1804.01543 [astro-ph.HE]} \BibitemShut {NoStop}%
\bibitem [{\citenamefont {Kroupa}\ \emph {et~al.}(2011)\citenamefont {Kroupa}, \citenamefont {Weidner}, \citenamefont {Pflamm-Altenburg}, \citenamefont {Thies}, \citenamefont {Dabringhausen}, \citenamefont {Marks},\ and\ \citenamefont {Maschberger}}]{Kroupa:2011aa}%
  \BibitemOpen
  \bibfield  {author} {\bibinfo {author} {\bibfnamefont {P.}~\bibnamefont {Kroupa}}, \bibinfo {author} {\bibfnamefont {C.}~\bibnamefont {Weidner}}, \bibinfo {author} {\bibfnamefont {J.}~\bibnamefont {Pflamm-Altenburg}}, \bibinfo {author} {\bibfnamefont {I.}~\bibnamefont {Thies}}, \bibinfo {author} {\bibfnamefont {J.}~\bibnamefont {Dabringhausen}}, \bibinfo {author} {\bibfnamefont {M.}~\bibnamefont {Marks}}, \ and\ \bibinfo {author} {\bibfnamefont {T.}~\bibnamefont {Maschberger}},\ }\href {\doibase 10.1007/978-94-007-5612-0{\_}4} {\  (\bibinfo {year} {2011}),\ 10.1007/978-94-007-5612-0{\_}4},\ \Eprint {http://arxiv.org/abs/1112.3340} {arXiv:1112.3340 [astro-ph.CO]} \BibitemShut {NoStop}%
\bibitem [{\citenamefont {Amaro-Seoane}(2019)}]{Amaro-Seoane:2019umn}%
  \BibitemOpen
  \bibfield  {author} {\bibinfo {author} {\bibfnamefont {P.}~\bibnamefont {Amaro-Seoane}},\ }\href {\doibase 10.1103/PhysRevD.99.123025} {\bibfield  {journal} {\bibinfo  {journal} {Phys. Rev. D}\ }\textbf {\bibinfo {volume} {99}},\ \bibinfo {pages} {123025} (\bibinfo {year} {2019})},\ \Eprint {http://arxiv.org/abs/1903.10871} {arXiv:1903.10871 [astro-ph.GA]} \BibitemShut {NoStop}%
\bibitem [{\citenamefont {Ariga}\ \emph {et~al.}(2019)\citenamefont {Ariga} \emph {et~al.}}]{FASER:2018eoc}%
  \BibitemOpen
  \bibfield  {author} {\bibinfo {author} {\bibfnamefont {A.}~\bibnamefont {Ariga}} \emph {et~al.} (\bibinfo {collaboration} {FASER}),\ }\href {\doibase 10.1103/PhysRevD.99.095011} {\bibfield  {journal} {\bibinfo  {journal} {Phys. Rev. D}\ }\textbf {\bibinfo {volume} {99}},\ \bibinfo {pages} {095011} (\bibinfo {year} {2019})},\ \Eprint {http://arxiv.org/abs/1811.12522} {arXiv:1811.12522 [hep-ph]} \BibitemShut {NoStop}%
\bibitem [{\citenamefont {Essig}\ \emph {et~al.}(2013)\citenamefont {Essig} \emph {et~al.}}]{Essig:2013lka}%
  \BibitemOpen
  \bibfield  {author} {\bibinfo {author} {\bibfnamefont {R.}~\bibnamefont {Essig}} \emph {et~al.},\ }in\ \href@noop {} {\emph {\bibinfo {booktitle} {{Snowmass 2013}: {Snowmass on the Mississippi}}}}\ (\bibinfo {year} {2013})\ \Eprint {http://arxiv.org/abs/1311.0029} {arXiv:1311.0029 [hep-ph]} \BibitemShut {NoStop}%
\bibitem [{\citenamefont {Chang}\ \emph {et~al.}(2017)\citenamefont {Chang}, \citenamefont {Essig},\ and\ \citenamefont {McDermott}}]{Chang:2016ntp}%
  \BibitemOpen
  \bibfield  {author} {\bibinfo {author} {\bibfnamefont {J.~H.}\ \bibnamefont {Chang}}, \bibinfo {author} {\bibfnamefont {R.}~\bibnamefont {Essig}}, \ and\ \bibinfo {author} {\bibfnamefont {S.~D.}\ \bibnamefont {McDermott}},\ }\href {\doibase 10.1007/JHEP01(2017)107} {\bibfield  {journal} {\bibinfo  {journal} {JHEP}\ }\textbf {\bibinfo {volume} {01}},\ \bibinfo {pages} {107} (\bibinfo {year} {2017})},\ \Eprint {http://arxiv.org/abs/1611.03864} {arXiv:1611.03864 [hep-ph]} \BibitemShut {NoStop}%
\bibitem [{\citenamefont {Pospelov}(2009)}]{Pospelov:2008zw}%
  \BibitemOpen
  \bibfield  {author} {\bibinfo {author} {\bibfnamefont {M.}~\bibnamefont {Pospelov}},\ }\href {\doibase 10.1103/PhysRevD.80.095002} {\bibfield  {journal} {\bibinfo  {journal} {Phys. Rev. D}\ }\textbf {\bibinfo {volume} {80}},\ \bibinfo {pages} {095002} (\bibinfo {year} {2009})},\ \Eprint {http://arxiv.org/abs/0811.1030} {arXiv:0811.1030 [hep-ph]} \BibitemShut {NoStop}%
\bibitem [{\citenamefont {Bennett}\ \emph {et~al.}(2021)\citenamefont {Bennett}, \citenamefont {Buldgen}, \citenamefont {De~Salas}, \citenamefont {Drewes}, \citenamefont {Gariazzo}, \citenamefont {Pastor},\ and\ \citenamefont {Wong}}]{Bennett:2020zkv}%
  \BibitemOpen
  \bibfield  {author} {\bibinfo {author} {\bibfnamefont {J.~J.}\ \bibnamefont {Bennett}}, \bibinfo {author} {\bibfnamefont {G.}~\bibnamefont {Buldgen}}, \bibinfo {author} {\bibfnamefont {P.~F.}\ \bibnamefont {De~Salas}}, \bibinfo {author} {\bibfnamefont {M.}~\bibnamefont {Drewes}}, \bibinfo {author} {\bibfnamefont {S.}~\bibnamefont {Gariazzo}}, \bibinfo {author} {\bibfnamefont {S.}~\bibnamefont {Pastor}}, \ and\ \bibinfo {author} {\bibfnamefont {Y.~Y.~Y.}\ \bibnamefont {Wong}},\ }\href {\doibase 10.1088/1475-7516/2021/04/073} {\bibfield  {journal} {\bibinfo  {journal} {JCAP}\ }\textbf {\bibinfo {volume} {04}},\ \bibinfo {pages} {073} (\bibinfo {year} {2021})},\ \Eprint {http://arxiv.org/abs/2012.02726} {arXiv:2012.02726 [hep-ph]} \BibitemShut {NoStop}%
\bibitem [{\citenamefont {Caputo}\ \emph {et~al.}(2026)\citenamefont {Caputo}, \citenamefont {Park},\ and\ \citenamefont {Yun}}]{Caputo:2025avc}%
  \BibitemOpen
  \bibfield  {author} {\bibinfo {author} {\bibfnamefont {A.}~\bibnamefont {Caputo}}, \bibinfo {author} {\bibfnamefont {J.}~\bibnamefont {Park}}, \ and\ \bibinfo {author} {\bibfnamefont {S.}~\bibnamefont {Yun}},\ }\href {\doibase 10.1103/m6zs-jxxp} {\bibfield  {journal} {\bibinfo  {journal} {Phys. Rev. D}\ }\textbf {\bibinfo {volume} {113}},\ \bibinfo {pages} {075001} (\bibinfo {year} {2026})},\ \Eprint {http://arxiv.org/abs/2511.15785} {arXiv:2511.15785 [hep-ph]} \BibitemShut {NoStop}%
\bibitem [{\citenamefont {Caputo}\ \emph {et~al.}(2021)\citenamefont {Caputo}, \citenamefont {Millar}, \citenamefont {O'Hare},\ and\ \citenamefont {Vitagliano}}]{Caputo:2021eaa}%
  \BibitemOpen
  \bibfield  {author} {\bibinfo {author} {\bibfnamefont {A.}~\bibnamefont {Caputo}}, \bibinfo {author} {\bibfnamefont {A.~J.}\ \bibnamefont {Millar}}, \bibinfo {author} {\bibfnamefont {C.~A.~J.}\ \bibnamefont {O'Hare}}, \ and\ \bibinfo {author} {\bibfnamefont {E.}~\bibnamefont {Vitagliano}},\ }\href {\doibase 10.1103/PhysRevD.104.095029} {\bibfield  {journal} {\bibinfo  {journal} {Phys. Rev. D}\ }\textbf {\bibinfo {volume} {104}},\ \bibinfo {pages} {095029} (\bibinfo {year} {2021})},\ \Eprint {http://arxiv.org/abs/2105.04565} {arXiv:2105.04565 [hep-ph]} \BibitemShut {NoStop}%
\bibitem [{\citenamefont {Carenza}\ \emph {et~al.}(2025)\citenamefont {Carenza}, \citenamefont {Ferreira},\ and\ \citenamefont {Nguyen}}]{Carenza:2025uwx}%
  \BibitemOpen
  \bibfield  {author} {\bibinfo {author} {\bibfnamefont {P.}~\bibnamefont {Carenza}}, \bibinfo {author} {\bibfnamefont {T.}~\bibnamefont {Ferreira}}, \ and\ \bibinfo {author} {\bibfnamefont {T.~T.~Q.}\ \bibnamefont {Nguyen}},\ }\href@noop {} {\  (\bibinfo {year} {2025})},\ \Eprint {http://arxiv.org/abs/2507.08932} {arXiv:2507.08932 [hep-ph]} \BibitemShut {NoStop}%
\bibitem [{\citenamefont {Frerick}\ \emph {et~al.}(2024)\citenamefont {Frerick}, \citenamefont {Jaeckel}, \citenamefont {Kahlhoefer},\ and\ \citenamefont {Schmidt-Hoberg}}]{Frerick:2023xnf}%
  \BibitemOpen
  \bibfield  {author} {\bibinfo {author} {\bibfnamefont {J.}~\bibnamefont {Frerick}}, \bibinfo {author} {\bibfnamefont {J.}~\bibnamefont {Jaeckel}}, \bibinfo {author} {\bibfnamefont {F.}~\bibnamefont {Kahlhoefer}}, \ and\ \bibinfo {author} {\bibfnamefont {K.}~\bibnamefont {Schmidt-Hoberg}},\ }\href {\doibase 10.1016/j.physletb.2023.138328} {\bibfield  {journal} {\bibinfo  {journal} {Phys. Lett. B}\ }\textbf {\bibinfo {volume} {848}},\ \bibinfo {pages} {138328} (\bibinfo {year} {2024})},\ \Eprint {http://arxiv.org/abs/2310.06017} {arXiv:2310.06017 [hep-ph]} \BibitemShut {NoStop}%
\bibitem [{\citenamefont {Garani}\ and\ \citenamefont {Palomares-Ruiz}(2022)}]{Garani:2021feo}%
  \BibitemOpen
  \bibfield  {author} {\bibinfo {author} {\bibfnamefont {R.}~\bibnamefont {Garani}}\ and\ \bibinfo {author} {\bibfnamefont {S.}~\bibnamefont {Palomares-Ruiz}},\ }\href {\doibase 10.1088/1475-7516/2022/05/042} {\bibfield  {journal} {\bibinfo  {journal} {JCAP}\ }\textbf {\bibinfo {volume} {05}},\ \bibinfo {pages} {042} (\bibinfo {year} {2022})},\ \Eprint {http://arxiv.org/abs/2104.12757} {arXiv:2104.12757 [hep-ph]} \BibitemShut {NoStop}%
\bibitem [{\citenamefont {Akerib}\ \emph {et~al.}(2022)\citenamefont {Akerib} \emph {et~al.}}]{Akerib:2022ort}%
  \BibitemOpen
  \bibfield  {author} {\bibinfo {author} {\bibfnamefont {D.~S.}\ \bibnamefont {Akerib}} \emph {et~al.},\ }in\ \href@noop {} {\emph {\bibinfo {booktitle} {{Snowmass 2021}}}}\ (\bibinfo {year} {2022})\ \Eprint {http://arxiv.org/abs/2203.08084} {arXiv:2203.08084 [hep-ex]} \BibitemShut {NoStop}%
\bibitem [{\citenamefont {Ajaj}\ \emph {et~al.}(2019)\citenamefont {Ajaj} \emph {et~al.}}]{DEAP:2019yzn}%
  \BibitemOpen
  \bibfield  {author} {\bibinfo {author} {\bibfnamefont {R.}~\bibnamefont {Ajaj}} \emph {et~al.} (\bibinfo {collaboration} {DEAP}),\ }\href {\doibase 10.1103/PhysRevD.100.022004} {\bibfield  {journal} {\bibinfo  {journal} {Phys. Rev. D}\ }\textbf {\bibinfo {volume} {100}},\ \bibinfo {pages} {022004} (\bibinfo {year} {2019})},\ \Eprint {http://arxiv.org/abs/1902.04048} {arXiv:1902.04048 [astro-ph.CO]} \BibitemShut {NoStop}%
\bibitem [{\citenamefont {Aprile}\ \emph {et~al.}(2021)\citenamefont {Aprile} \emph {et~al.}}]{XENON:2020gfr}%
  \BibitemOpen
  \bibfield  {author} {\bibinfo {author} {\bibfnamefont {E.}~\bibnamefont {Aprile}} \emph {et~al.} (\bibinfo {collaboration} {XENON}),\ }\href {\doibase 10.1103/PhysRevLett.126.091301} {\bibfield  {journal} {\bibinfo  {journal} {Phys. Rev. Lett.}\ }\textbf {\bibinfo {volume} {126}},\ \bibinfo {pages} {091301} (\bibinfo {year} {2021})},\ \Eprint {http://arxiv.org/abs/2012.02846} {arXiv:2012.02846 [hep-ex]} \BibitemShut {NoStop}%
\bibitem [{\citenamefont {Adhikari}\ \emph {et~al.}(2018)\citenamefont {Adhikari} \emph {et~al.}}]{Adhikari:2018ljm}%
  \BibitemOpen
  \bibfield  {author} {\bibinfo {author} {\bibfnamefont {G.}~\bibnamefont {Adhikari}} \emph {et~al.},\ }\href {\doibase 10.1038/s41586-018-0739-1} {\bibfield  {journal} {\bibinfo  {journal} {Nature}\ }\textbf {\bibinfo {volume} {564}},\ \bibinfo {pages} {83} (\bibinfo {year} {2018})},\ \bibinfo {note} {[Erratum: Nature 566, E2 (2019)]},\ \Eprint {http://arxiv.org/abs/1906.01791} {arXiv:1906.01791 [astro-ph.IM]} \BibitemShut {NoStop}%
\bibitem [{\citenamefont {Abdelhameed}\ \emph {et~al.}(2019)\citenamefont {Abdelhameed} \emph {et~al.}}]{CRESST:2019jnq}%
  \BibitemOpen
  \bibfield  {author} {\bibinfo {author} {\bibfnamefont {A.~H.}\ \bibnamefont {Abdelhameed}} \emph {et~al.} (\bibinfo {collaboration} {CRESST}),\ }\href {\doibase 10.1103/PhysRevD.100.102002} {\bibfield  {journal} {\bibinfo  {journal} {Phys. Rev. D}\ }\textbf {\bibinfo {volume} {100}},\ \bibinfo {pages} {102002} (\bibinfo {year} {2019})},\ \Eprint {http://arxiv.org/abs/1904.00498} {arXiv:1904.00498 [astro-ph.CO]} \BibitemShut {NoStop}%
\bibitem [{\citenamefont {Bernabei}\ \emph {et~al.}(2018)\citenamefont {Bernabei} \emph {et~al.}}]{Bernabei:2018yyw}%
  \BibitemOpen
  \bibfield  {author} {\bibinfo {author} {\bibfnamefont {R.}~\bibnamefont {Bernabei}} \emph {et~al.},\ }\href {\doibase 10.3390/universe4110116} {\bibfield  {journal} {\bibinfo  {journal} {Universe}\ }\textbf {\bibinfo {volume} {4}},\ \bibinfo {pages} {116} (\bibinfo {year} {2018})}\BibitemShut {NoStop}%
\bibitem [{\citenamefont {Agnes}\ \emph {et~al.}(2018)\citenamefont {Agnes} \emph {et~al.}}]{DarkSide:2018bpj}%
  \BibitemOpen
  \bibfield  {author} {\bibinfo {author} {\bibfnamefont {P.}~\bibnamefont {Agnes}} \emph {et~al.} (\bibinfo {collaboration} {DarkSide}),\ }\href {\doibase 10.1103/PhysRevLett.121.081307} {\bibfield  {journal} {\bibinfo  {journal} {Phys. Rev. Lett.}\ }\textbf {\bibinfo {volume} {121}},\ \bibinfo {pages} {081307} (\bibinfo {year} {2018})},\ \Eprint {http://arxiv.org/abs/1802.06994} {arXiv:1802.06994 [astro-ph.HE]} \BibitemShut {NoStop}%
\bibitem [{\citenamefont {Erickcek}\ \emph {et~al.}(2007)\citenamefont {Erickcek}, \citenamefont {Steinhardt}, \citenamefont {McCammon},\ and\ \citenamefont {McGuire}}]{Erickcek:2007jv}%
  \BibitemOpen
  \bibfield  {author} {\bibinfo {author} {\bibfnamefont {A.~L.}\ \bibnamefont {Erickcek}}, \bibinfo {author} {\bibfnamefont {P.~J.}\ \bibnamefont {Steinhardt}}, \bibinfo {author} {\bibfnamefont {D.}~\bibnamefont {McCammon}}, \ and\ \bibinfo {author} {\bibfnamefont {P.~C.}\ \bibnamefont {McGuire}},\ }\href {\doibase 10.1103/PhysRevD.76.042007} {\bibfield  {journal} {\bibinfo  {journal} {Phys. Rev. D}\ }\textbf {\bibinfo {volume} {76}},\ \bibinfo {pages} {042007} (\bibinfo {year} {2007})},\ \Eprint {http://arxiv.org/abs/0704.0794} {arXiv:0704.0794 [astro-ph]} \BibitemShut {NoStop}%
\bibitem [{\citenamefont {Gluscevic}\ and\ \citenamefont {Boddy}(2018)}]{Gluscevic:2017ywp}%
  \BibitemOpen
  \bibfield  {author} {\bibinfo {author} {\bibfnamefont {V.}~\bibnamefont {Gluscevic}}\ and\ \bibinfo {author} {\bibfnamefont {K.~K.}\ \bibnamefont {Boddy}},\ }\href {\doibase 10.1103/PhysRevLett.121.081301} {\bibfield  {journal} {\bibinfo  {journal} {Phys. Rev. Lett.}\ }\textbf {\bibinfo {volume} {121}},\ \bibinfo {pages} {081301} (\bibinfo {year} {2018})},\ \Eprint {http://arxiv.org/abs/1712.07133} {arXiv:1712.07133 [astro-ph.CO]} \BibitemShut {NoStop}%
\bibitem [{\citenamefont {Boddy}\ \emph {et~al.}(2018)\citenamefont {Boddy}, \citenamefont {Gluscevic}, \citenamefont {Poulin}, \citenamefont {Kovetz}, \citenamefont {Kamionkowski},\ and\ \citenamefont {Barkana}}]{Boddy:2018wzy}%
  \BibitemOpen
  \bibfield  {author} {\bibinfo {author} {\bibfnamefont {K.~K.}\ \bibnamefont {Boddy}}, \bibinfo {author} {\bibfnamefont {V.}~\bibnamefont {Gluscevic}}, \bibinfo {author} {\bibfnamefont {V.}~\bibnamefont {Poulin}}, \bibinfo {author} {\bibfnamefont {E.~D.}\ \bibnamefont {Kovetz}}, \bibinfo {author} {\bibfnamefont {M.}~\bibnamefont {Kamionkowski}}, \ and\ \bibinfo {author} {\bibfnamefont {R.}~\bibnamefont {Barkana}},\ }\href {\doibase 10.1103/PhysRevD.98.123506} {\bibfield  {journal} {\bibinfo  {journal} {Phys. Rev. D}\ }\textbf {\bibinfo {volume} {98}},\ \bibinfo {pages} {123506} (\bibinfo {year} {2018})},\ \Eprint {http://arxiv.org/abs/1808.00001} {arXiv:1808.00001 [astro-ph.CO]} \BibitemShut {NoStop}%
\bibitem [{\citenamefont {Nadler}\ \emph {et~al.}(2021)\citenamefont {Nadler} \emph {et~al.}}]{DES:2020fxi}%
  \BibitemOpen
  \bibfield  {author} {\bibinfo {author} {\bibfnamefont {E.~O.}\ \bibnamefont {Nadler}} \emph {et~al.} (\bibinfo {collaboration} {DES}),\ }\href {\doibase 10.1103/PhysRevLett.126.091101} {\bibfield  {journal} {\bibinfo  {journal} {Phys. Rev. Lett.}\ }\textbf {\bibinfo {volume} {126}},\ \bibinfo {pages} {091101} (\bibinfo {year} {2021})},\ \Eprint {http://arxiv.org/abs/2008.00022} {arXiv:2008.00022 [astro-ph.CO]} \BibitemShut {NoStop}%
\bibitem [{\citenamefont {Rogers}\ \emph {et~al.}(2022)\citenamefont {Rogers}, \citenamefont {Dvorkin},\ and\ \citenamefont {Peiris}}]{Rogers:2021byl}%
  \BibitemOpen
  \bibfield  {author} {\bibinfo {author} {\bibfnamefont {K.~K.}\ \bibnamefont {Rogers}}, \bibinfo {author} {\bibfnamefont {C.}~\bibnamefont {Dvorkin}}, \ and\ \bibinfo {author} {\bibfnamefont {H.~V.}\ \bibnamefont {Peiris}},\ }\href {\doibase 10.1103/PhysRevLett.128.171301} {\bibfield  {journal} {\bibinfo  {journal} {Phys. Rev. Lett.}\ }\textbf {\bibinfo {volume} {128}},\ \bibinfo {pages} {171301} (\bibinfo {year} {2022})},\ \Eprint {http://arxiv.org/abs/2111.10386} {arXiv:2111.10386 [astro-ph.CO]} \BibitemShut {NoStop}%
\bibitem [{\citenamefont {Aalbers}\ \emph {et~al.}(2023)\citenamefont {Aalbers} \emph {et~al.}}]{LZ:2022lsv}%
  \BibitemOpen
  \bibfield  {author} {\bibinfo {author} {\bibfnamefont {J.}~\bibnamefont {Aalbers}} \emph {et~al.} (\bibinfo {collaboration} {LZ}),\ }\href {\doibase 10.1103/PhysRevLett.131.041002} {\bibfield  {journal} {\bibinfo  {journal} {Phys. Rev. Lett.}\ }\textbf {\bibinfo {volume} {131}},\ \bibinfo {pages} {041002} (\bibinfo {year} {2023})},\ \Eprint {http://arxiv.org/abs/2207.03764} {arXiv:2207.03764 [hep-ex]} \BibitemShut {NoStop}%
\bibitem [{\citenamefont {Aprile}\ \emph {et~al.}(2020)\citenamefont {Aprile} \emph {et~al.}}]{XENON:2020kmp}%
  \BibitemOpen
  \bibfield  {author} {\bibinfo {author} {\bibfnamefont {E.}~\bibnamefont {Aprile}} \emph {et~al.} (\bibinfo {collaboration} {XENON}),\ }\href {\doibase 10.1088/1475-7516/2020/11/031} {\bibfield  {journal} {\bibinfo  {journal} {JCAP}\ }\textbf {\bibinfo {volume} {11}},\ \bibinfo {pages} {031} (\bibinfo {year} {2020})},\ \Eprint {http://arxiv.org/abs/2007.08796} {arXiv:2007.08796 [physics.ins-det]} \BibitemShut {NoStop}%
\bibitem [{\citenamefont {Meng}\ \emph {et~al.}(2021)\citenamefont {Meng} \emph {et~al.}}]{PandaX-4T:2021bab}%
  \BibitemOpen
  \bibfield  {author} {\bibinfo {author} {\bibfnamefont {Y.}~\bibnamefont {Meng}} \emph {et~al.} (\bibinfo {collaboration} {PandaX-4T}),\ }\href {\doibase 10.1103/PhysRevLett.127.261802} {\bibfield  {journal} {\bibinfo  {journal} {Phys. Rev. Lett.}\ }\textbf {\bibinfo {volume} {127}},\ \bibinfo {pages} {261802} (\bibinfo {year} {2021})},\ \Eprint {http://arxiv.org/abs/2107.13438} {arXiv:2107.13438 [hep-ex]} \BibitemShut {NoStop}%
\bibitem [{\citenamefont {Agnese}\ \emph {et~al.}(2019)\citenamefont {Agnese} \emph {et~al.}}]{SuperCDMS:2018gro}%
  \BibitemOpen
  \bibfield  {author} {\bibinfo {author} {\bibfnamefont {R.}~\bibnamefont {Agnese}} \emph {et~al.} (\bibinfo {collaboration} {SuperCDMS}),\ }\href {\doibase 10.1103/PhysRevD.99.062001} {\bibfield  {journal} {\bibinfo  {journal} {Phys. Rev. D}\ }\textbf {\bibinfo {volume} {99}},\ \bibinfo {pages} {062001} (\bibinfo {year} {2019})},\ \Eprint {http://arxiv.org/abs/1808.09098} {arXiv:1808.09098 [astro-ph.CO]} \BibitemShut {NoStop}%
\bibitem [{\citenamefont {Alfonso-Pita}\ \emph {et~al.}(2022)\citenamefont {Alfonso-Pita} \emph {et~al.}}]{Alfonso-Pita:2022akn}%
  \BibitemOpen
  \bibfield  {author} {\bibinfo {author} {\bibfnamefont {E.}~\bibnamefont {Alfonso-Pita}} \emph {et~al.},\ }in\ \href@noop {} {\emph {\bibinfo {booktitle} {{Snowmass 2021}}}}\ (\bibinfo {year} {2022})\ \Eprint {http://arxiv.org/abs/2207.12400} {arXiv:2207.12400 [physics.ins-det]} \BibitemShut {NoStop}%
\bibitem [{\citenamefont {Keith}\ \emph {et~al.}(2023)\citenamefont {Keith}, \citenamefont {Hooper},\ and\ \citenamefont {Linden}}]{Keith:2022xbd}%
  \BibitemOpen
  \bibfield  {author} {\bibinfo {author} {\bibfnamefont {C.}~\bibnamefont {Keith}}, \bibinfo {author} {\bibfnamefont {D.}~\bibnamefont {Hooper}}, \ and\ \bibinfo {author} {\bibfnamefont {T.}~\bibnamefont {Linden}},\ }\href {\doibase 10.1103/PhysRevD.107.103001} {\bibfield  {journal} {\bibinfo  {journal} {Phys. Rev. D}\ }\textbf {\bibinfo {volume} {107}},\ \bibinfo {pages} {103001} (\bibinfo {year} {2023})},\ \Eprint {http://arxiv.org/abs/2212.08080} {arXiv:2212.08080 [astro-ph.HE]} \BibitemShut {NoStop}%
\bibitem [{\citenamefont {Xu}\ and\ \citenamefont {Hooper}(2023)}]{Xu:2023zyz}%
  \BibitemOpen
  \bibfield  {author} {\bibinfo {author} {\bibfnamefont {F.}~\bibnamefont {Xu}}\ and\ \bibinfo {author} {\bibfnamefont {D.}~\bibnamefont {Hooper}},\ }\href@noop {} {\  (\bibinfo {year} {2023})},\ \Eprint {http://arxiv.org/abs/2308.15538} {arXiv:2308.15538 [astro-ph.HE]} \BibitemShut {NoStop}%
\bibitem [{\citenamefont {Kierans}(2020)}]{Kierans_2020}%
  \BibitemOpen
  \bibfield  {author} {\bibinfo {author} {\bibfnamefont {C.~A.}\ \bibnamefont {Kierans}},\ }in\ \href {\doibase 10.1117/12.2562352} {\emph {\bibinfo {booktitle} {Space Telescopes and Instrumentation 2020: Ultraviolet to Gamma Ray}}},\ \bibinfo {editor} {edited by\ \bibinfo {editor} {\bibfnamefont {J.-W.~A.}\ \bibnamefont {den Herder}}, \bibinfo {editor} {\bibfnamefont {K.}~\bibnamefont {Nakazawa}}, \ and\ \bibinfo {editor} {\bibfnamefont {S.}~\bibnamefont {Nikzad}}}\ (\bibinfo  {publisher} {SPIE},\ \bibinfo {year} {2020})\BibitemShut {NoStop}%
\bibitem [{\citenamefont {Tavani}\ \emph {et~al.}(2018)\citenamefont {Tavani} \emph {et~al.}}]{e-ASTROGAM:2017pxr}%
  \BibitemOpen
  \bibfield  {author} {\bibinfo {author} {\bibfnamefont {M.}~\bibnamefont {Tavani}} \emph {et~al.} (\bibinfo {collaboration} {e-ASTROGAM}),\ }\href {\doibase 10.1016/j.jheap.2018.07.001} {\bibfield  {journal} {\bibinfo  {journal} {JHEAp}\ }\textbf {\bibinfo {volume} {19}},\ \bibinfo {pages} {1} (\bibinfo {year} {2018})},\ \Eprint {http://arxiv.org/abs/1711.01265} {arXiv:1711.01265 [astro-ph.HE]} \BibitemShut {NoStop}%
\bibitem [{\citenamefont {Foster}\ \emph {et~al.}(2021)\citenamefont {Foster}, \citenamefont {Kongsore}, \citenamefont {Dessert}, \citenamefont {Park}, \citenamefont {Rodd}, \citenamefont {Cranmer},\ and\ \citenamefont {Safdi}}]{Foster:2021ngm}%
  \BibitemOpen
  \bibfield  {author} {\bibinfo {author} {\bibfnamefont {J.~W.}\ \bibnamefont {Foster}}, \bibinfo {author} {\bibfnamefont {M.}~\bibnamefont {Kongsore}}, \bibinfo {author} {\bibfnamefont {C.}~\bibnamefont {Dessert}}, \bibinfo {author} {\bibfnamefont {Y.}~\bibnamefont {Park}}, \bibinfo {author} {\bibfnamefont {N.~L.}\ \bibnamefont {Rodd}}, \bibinfo {author} {\bibfnamefont {K.}~\bibnamefont {Cranmer}}, \ and\ \bibinfo {author} {\bibfnamefont {B.~R.}\ \bibnamefont {Safdi}},\ }\href {\doibase 10.1103/PhysRevLett.127.051101} {\bibfield  {journal} {\bibinfo  {journal} {Phys. Rev. Lett.}\ }\textbf {\bibinfo {volume} {127}},\ \bibinfo {pages} {051101} (\bibinfo {year} {2021})},\ \Eprint {http://arxiv.org/abs/2102.02207} {arXiv:2102.02207 [astro-ph.CO]} \BibitemShut {NoStop}%
\bibitem [{\citenamefont {Schoedel}\ \emph {et~al.}(2023)\citenamefont {Schoedel} \emph {et~al.}}]{Schoedel:2023ott}%
  \BibitemOpen
  \bibfield  {author} {\bibinfo {author} {\bibfnamefont {R.}~\bibnamefont {Schoedel}} \emph {et~al.},\ }\href@noop {} {\  (\bibinfo {year} {2023})},\ \Eprint {http://arxiv.org/abs/2310.11912} {arXiv:2310.11912 [astro-ph.GA]} \BibitemShut {NoStop}%
\bibitem [{\citenamefont {Castro}\ \emph {et~al.}(2023)\citenamefont {Castro} \emph {et~al.}}]{Euclid:2023jih}%
  \BibitemOpen
  \bibfield  {author} {\bibinfo {author} {\bibfnamefont {T.}~\bibnamefont {Castro}} \emph {et~al.} (\bibinfo {collaboration} {Euclid}),\ }\href@noop {} {\  (\bibinfo {year} {2023})},\ \Eprint {http://arxiv.org/abs/2311.01465} {arXiv:2311.01465 [astro-ph.CO]} \BibitemShut {NoStop}%
\bibitem [{\citenamefont {Redondo}\ and\ \citenamefont {Postma}(2009)}]{Redondo:2008ec}%
  \BibitemOpen
  \bibfield  {author} {\bibinfo {author} {\bibfnamefont {J.}~\bibnamefont {Redondo}}\ and\ \bibinfo {author} {\bibfnamefont {M.}~\bibnamefont {Postma}},\ }\href {\doibase 10.1088/1475-7516/2009/02/005} {\bibfield  {journal} {\bibinfo  {journal} {JCAP}\ }\textbf {\bibinfo {volume} {02}},\ \bibinfo {pages} {005} (\bibinfo {year} {2009})},\ \Eprint {http://arxiv.org/abs/0811.0326} {arXiv:0811.0326 [hep-ph]} \BibitemShut {NoStop}%
\bibitem [{\citenamefont {Krnjaic}\ \emph {et~al.}(2023)\citenamefont {Krnjaic}, \citenamefont {Rocha},\ and\ \citenamefont {Sokolenko}}]{Krnjaic:2022wor}%
  \BibitemOpen
  \bibfield  {author} {\bibinfo {author} {\bibfnamefont {G.}~\bibnamefont {Krnjaic}}, \bibinfo {author} {\bibfnamefont {D.}~\bibnamefont {Rocha}}, \ and\ \bibinfo {author} {\bibfnamefont {A.}~\bibnamefont {Sokolenko}},\ }\href {\doibase 10.1103/PhysRevD.108.035047} {\bibfield  {journal} {\bibinfo  {journal} {Phys. Rev. D}\ }\textbf {\bibinfo {volume} {108}},\ \bibinfo {pages} {035047} (\bibinfo {year} {2023})},\ \Eprint {http://arxiv.org/abs/2210.06487} {arXiv:2210.06487 [hep-ph]} \BibitemShut {NoStop}%
\bibitem [{\citenamefont {Harris}\ \emph {et~al.}(2020)\citenamefont {Harris}, \citenamefont {Millman}, \citenamefont {van~der Walt}, \citenamefont {Gommers}, \citenamefont {Virtanen}, \citenamefont {Cournapeau}, \citenamefont {Wieser}, \citenamefont {Taylor}, \citenamefont {Berg}, \citenamefont {Smith}, \citenamefont {Kern}, \citenamefont {Picus}, \citenamefont {Hoyer}, \citenamefont {van Kerkwijk}, \citenamefont {Brett}, \citenamefont {Haldane}, \citenamefont {del Río}, \citenamefont {Wiebe}, \citenamefont {Peterson}, \citenamefont {Gérard-Marchant}, \citenamefont {Sheppard}, \citenamefont {Reddy}, \citenamefont {Weckesser}, \citenamefont {Abbasi}, \citenamefont {Gohlke},\ and\ \citenamefont {Oliphant}}]{Harris_2020}%
  \BibitemOpen
  \bibfield  {author} {\bibinfo {author} {\bibfnamefont {C.~R.}\ \bibnamefont {Harris}}, \bibinfo {author} {\bibfnamefont {K.~J.}\ \bibnamefont {Millman}}, \bibinfo {author} {\bibfnamefont {S.~J.}\ \bibnamefont {van~der Walt}}, \bibinfo {author} {\bibfnamefont {R.}~\bibnamefont {Gommers}}, \bibinfo {author} {\bibfnamefont {P.}~\bibnamefont {Virtanen}}, \bibinfo {author} {\bibfnamefont {D.}~\bibnamefont {Cournapeau}}, \bibinfo {author} {\bibfnamefont {E.}~\bibnamefont {Wieser}}, \bibinfo {author} {\bibfnamefont {J.}~\bibnamefont {Taylor}}, \bibinfo {author} {\bibfnamefont {S.}~\bibnamefont {Berg}}, \bibinfo {author} {\bibfnamefont {N.~J.}\ \bibnamefont {Smith}}, \bibinfo {author} {\bibfnamefont {R.}~\bibnamefont {Kern}}, \bibinfo {author} {\bibfnamefont {M.}~\bibnamefont {Picus}}, \bibinfo {author} {\bibfnamefont {S.}~\bibnamefont {Hoyer}}, \bibinfo {author} {\bibfnamefont {M.~H.}\ \bibnamefont {van Kerkwijk}}, \bibinfo {author} {\bibfnamefont {M.}~\bibnamefont {Brett}}, \bibinfo {author} {\bibfnamefont
  {A.}~\bibnamefont {Haldane}}, \bibinfo {author} {\bibfnamefont {J.~F.}\ \bibnamefont {del Río}}, \bibinfo {author} {\bibfnamefont {M.}~\bibnamefont {Wiebe}}, \bibinfo {author} {\bibfnamefont {P.}~\bibnamefont {Peterson}}, \bibinfo {author} {\bibfnamefont {P.}~\bibnamefont {Gérard-Marchant}}, \bibinfo {author} {\bibfnamefont {K.}~\bibnamefont {Sheppard}}, \bibinfo {author} {\bibfnamefont {T.}~\bibnamefont {Reddy}}, \bibinfo {author} {\bibfnamefont {W.}~\bibnamefont {Weckesser}}, \bibinfo {author} {\bibfnamefont {H.}~\bibnamefont {Abbasi}}, \bibinfo {author} {\bibfnamefont {C.}~\bibnamefont {Gohlke}}, \ and\ \bibinfo {author} {\bibfnamefont {T.~E.}\ \bibnamefont {Oliphant}},\ }\href {\doibase 10.1038/s41586-020-2649-2} {\bibfield  {journal} {\bibinfo  {journal} {Nature}\ }\textbf {\bibinfo {volume} {585}},\ \bibinfo {pages} {357–362} (\bibinfo {year} {2020})}\BibitemShut {NoStop}%
\bibitem [{\citenamefont {Virtanen}\ \emph {et~al.}(2020)\citenamefont {Virtanen} \emph {et~al.}}]{Virtanen:2019joe}%
  \BibitemOpen
  \bibfield  {author} {\bibinfo {author} {\bibfnamefont {P.}~\bibnamefont {Virtanen}} \emph {et~al.},\ }\href {\doibase 10.1038/s41592-019-0686-2} {\bibfield  {journal} {\bibinfo  {journal} {Nature Meth.}\ }\textbf {\bibinfo {volume} {17}},\ \bibinfo {pages} {261} (\bibinfo {year} {2020})},\ \Eprint {http://arxiv.org/abs/1907.10121} {arXiv:1907.10121 [cs.MS]} \BibitemShut {NoStop}%
\bibitem [{\citenamefont {Hunter}(2007)}]{HunterMatplotlib}%
  \BibitemOpen
  \bibfield  {author} {\bibinfo {author} {\bibfnamefont {J.~D.}\ \bibnamefont {Hunter}},\ }\href {\doibase 10.1109/MCSE.2007.55} {\bibfield  {journal} {\bibinfo  {journal} {Computing in Science \& Engineering}\ }\textbf {\bibinfo {volume} {9}},\ \bibinfo {pages} {90} (\bibinfo {year} {2007})}\BibitemShut {NoStop}%
\bibitem [{\citenamefont {{Kluyver}}\ \emph {et~al.}(2016)\citenamefont {{Kluyver}}, \citenamefont {{Ragan-Kelley}}, \citenamefont {{P{\'e}rez}}, \citenamefont {{Granger}}, \citenamefont {{Bussonnier}}, \citenamefont {{Frederic}}, \citenamefont {{Kelley}}, \citenamefont {{Hamrick}}, \citenamefont {{Grout}}, \citenamefont {{Corlay}}, \citenamefont {{Ivanov}}, \citenamefont {{Avila}}, \citenamefont {{Abdalla}}, \citenamefont {{Willing}},\ and\ \citenamefont {{Jupyter Development Team}}}]{2016ppap.book...87K}%
  \BibitemOpen
  \bibfield  {author} {\bibinfo {author} {\bibfnamefont {T.}~\bibnamefont {{Kluyver}}}, \bibinfo {author} {\bibfnamefont {B.}~\bibnamefont {{Ragan-Kelley}}}, \bibinfo {author} {\bibfnamefont {F.}~\bibnamefont {{P{\'e}rez}}}, \bibinfo {author} {\bibfnamefont {B.}~\bibnamefont {{Granger}}}, \bibinfo {author} {\bibfnamefont {M.}~\bibnamefont {{Bussonnier}}}, \bibinfo {author} {\bibfnamefont {J.}~\bibnamefont {{Frederic}}}, \bibinfo {author} {\bibfnamefont {K.}~\bibnamefont {{Kelley}}}, \bibinfo {author} {\bibfnamefont {J.}~\bibnamefont {{Hamrick}}}, \bibinfo {author} {\bibfnamefont {J.}~\bibnamefont {{Grout}}}, \bibinfo {author} {\bibfnamefont {S.}~\bibnamefont {{Corlay}}}, \bibinfo {author} {\bibfnamefont {P.}~\bibnamefont {{Ivanov}}}, \bibinfo {author} {\bibfnamefont {D.}~\bibnamefont {{Avila}}}, \bibinfo {author} {\bibfnamefont {S.}~\bibnamefont {{Abdalla}}}, \bibinfo {author} {\bibfnamefont {C.}~\bibnamefont {{Willing}}}, \ and\ \bibinfo {author} {\bibnamefont {{Jupyter Development Team}}},\ }in\ \href
  {\doibase 10.3233/978-1-61499-649-1-87} {\emph {\bibinfo {booktitle} {IOS Press}}}\ (\bibinfo {year} {2016})\ pp.\ \bibinfo {pages} {87--90}\BibitemShut {NoStop}%
\bibitem [{\citenamefont {Binosi}\ \emph {et~al.}(2009)\citenamefont {Binosi}, \citenamefont {Collins}, \citenamefont {Kaufhold},\ and\ \citenamefont {Theussl}}]{Binosi:2008ig}%
  \BibitemOpen
  \bibfield  {author} {\bibinfo {author} {\bibfnamefont {D.}~\bibnamefont {Binosi}}, \bibinfo {author} {\bibfnamefont {J.}~\bibnamefont {Collins}}, \bibinfo {author} {\bibfnamefont {C.}~\bibnamefont {Kaufhold}}, \ and\ \bibinfo {author} {\bibfnamefont {L.}~\bibnamefont {Theussl}},\ }\href {\doibase 10.1016/j.cpc.2009.02.020} {\bibfield  {journal} {\bibinfo  {journal} {Comput. Phys. Commun.}\ }\textbf {\bibinfo {volume} {180}},\ \bibinfo {pages} {1709} (\bibinfo {year} {2009})},\ \Eprint {http://arxiv.org/abs/0811.4113} {arXiv:0811.4113 [hep-ph]} \BibitemShut {NoStop}%
\bibitem [{\citenamefont {Rohatgi}(2022)}]{Rohatgi2022}%
  \BibitemOpen
  \bibfield  {author} {\bibinfo {author} {\bibfnamefont {A.}~\bibnamefont {Rohatgi}},\ }\href {https://automeris.io/WebPlotDigitizer} {\enquote {\bibinfo {title} {Webplotdigitizer: Version 4.6},}\ } (\bibinfo {year} {2022})\BibitemShut {NoStop}%
\end{thebibliography}%
\end{document}